\documentclass[twocolumn]{aastex631}
\usepackage{amsmath,amstext,amssymb}
\usepackage{float}
\hypersetup{
    colorlinks=true,
    linkcolor=blue,
    filecolor=magenta,      
    urlcolor=blue,
}

\graphicspath{{./}{figures/}}

\newcommand{\nimbus}{\texttt{nimbus}}

\begin{document}

\title{Inferring kilonova population properties with a hierarchical Bayesian framework \MakeUppercase{\romannumeral 1} : Non-detection methodology and single-event analyses}

\correspondingauthor{Siddharth R.\ Mohite}
\email{srmohite@uwm.edu}

\author[0000-0003-1356-7156]{Siddharth R.\ Mohite}
\altaffiliation{LSSTC Data Science Fellow}
\affiliation{Center for Gravitation, Cosmology and Astrophysics, Department of Physics, University of Wisconsin--Milwaukee, P.O.\ Box 413, Milwaukee, WI 53201, USA}
\affiliation{Center for Computational Astrophysics, Flatiron Institute, 162 5th Ave, New York, NY 10010, USA}

\author{Priyadarshini Rajkumar}
\affiliation{Department of Physics and Astronomy, Texas Tech University, Lubbock, TX 79409-1051, USA}

\author[0000-0003-3768-7515]{Shreya Anand}
\affil{Division of Physics, Mathematics and Astronomy, California Institute of Technology, Pasadena, CA 91125, USA}

\author[0000-0001-6295-2881]{David L.\ Kaplan}
\affiliation{Center for Gravitation, Cosmology and Astrophysics, Department of Physics, University of Wisconsin--Milwaukee, P.O.\ Box 413, Milwaukee, WI 53201, USA}

\author[0000-0002-8262-2924]{Michael W. Coughlin}
\affil{School of Physics and Astronomy, University of Minnesota, Minneapolis, Minnesota 55455, USA}

\author[0000-0002-3498-2167]{Ana Sagués-Carracedo}
\affil{The Oskar Klein Centre, Department of Physics, Stockholm University, AlbaNova, SE-106 91 Stockholm, Sweden}

\author[0000-0002-3836-7751]{Muhammed Saleem}
\affil{School of Physics and Astronomy, University of Minnesota, Minneapolis, Minnesota 55455, USA}

\author[0000-0003-3600-2406]{Jolien Creighton}
\affiliation{Center for Gravitation, Cosmology and Astrophysics, Department of Physics, University of Wisconsin--Milwaukee, P.O.\ Box 413, Milwaukee, WI 53201, USA}

\author[0000-0002-4611-9387]{Patrick~R.~Brady}
\affiliation{Center for Gravitation, Cosmology and Astrophysics, Department of Physics, University of Wisconsin--Milwaukee, P.O.\ Box 413, Milwaukee, WI 53201, USA}

\author[0000-0002-2184-6430]{Tom{\'a}s Ahumada}
\affil{Department of Astronomy, University of Maryland, College Park, MD 20742, USA}

\author[0000-0002-4694-7123]{Mouza Almualla}
\affil{Department of Physics, American University of Sharjah, PO Box 26666, Sharjah, UAE}

\author[0000-0002-8977-1498]{Igor Andreoni}
\affil{Division of Physics, Mathematics and Astronomy, California Institute of Technology, Pasadena, CA 91125, USA}

\author[0000-0002-8255-5127]{Mattia Bulla}
\affiliation{The Oskar Klein Centre, Department of Astronomy, Stockholm University, AlbaNova, SE-10691 Stockholm, Sweden}

\author[0000-0002-3168-0139]{Matthew J. Graham}
\affiliation{Division of Physics, Mathematics and Astronomy, California Institute of Technology, Pasadena, CA 91125, USA}

\author[0000-0002-5619-4938]{Mansi M. Kasliwal}
\affil{Division of Physics, Mathematics and Astronomy, California Institute of Technology, Pasadena, CA 91125, USA}

\author{Stephen Kaye}
\affiliation{Caltech Optical Observatories, California Institute of Technology, Pasadena, CA 91125, USA}

\author[0000-0003-2451-5482]{Russ R. Laher}
\affiliation{IPAC, California Institute of Technology, 1200 E. California Blvd, Pasadena, CA 91125, USA}

\author[0000-0002-1486-3582]{Kyung Min Shin}
\affiliation{California Institute of Technology, Pasadena, CA 91125, USA}

\author[0000-0003-4401-0430]{David L. Shupe}
\affiliation{IPAC, California Institute of Technology, 1200 E. California Blvd, Pasadena, CA 91125, USA}

\author[0000-0001-9898-5597]{Leo P. Singer}
\affiliation{Astrophysics Science Division, NASA Goddard Space Flight Center, MC 661, Greenbelt, MD 20771, USA}
\affiliation{Joint Space-Science Institute, University of Maryland, College Park, MD 20742, USA}



\begin{abstract}
We present \nimbus : a hierarchical Bayesian framework to infer the intrinsic luminosity parameters of kilonovae (KNe) associated with gravitational-wave (GW) events, based purely on non-detections. This framework makes use of GW 3-D distance information and electromagnetic upper limits from multiple surveys for multiple events, and self-consistently accounts for finite sky-coverage and probability of astrophysical origin. The framework is agnostic to the brightness evolution assumed and can account for multiple electromagnetic passbands simultaneously. Our analyses highlight the importance of accounting for model selection effects, especially in the context of non-detections. We show our methodology using a simple, two-parameter linear brightness model, taking the follow-up of GW190425 with the Zwicky Transient Facility (ZTF) as a single-event test case for two different prior choices of model parameters -- (i) uniform/uninformative priors and (ii) astrophysical priors based on surrogate models of Monte Carlo radiative transfer simulations of KNe.  We present results under the assumption that the KN is within the searched region to demonstrate functionality and the importance of prior choice. Our results show consistency with {\tt simsurvey} -- an astronomical survey simulation tool used previously in the literature to constrain the population of KNe. While our results based on uniform priors strongly constrain the parameter space, those based on astrophysical priors are largely uninformative, highlighting the need for deeper constraints. Future studies with multiple events having electromagnetic follow-up from multiple surveys should make it possible to constrain the KN population further.

\end{abstract}



\section{Introduction} \label{sec:intro}
Mergers of  neutron stars and neutron star-black hole binaries (BNS and NSBH) present unique opportunities to probe multi-messenger astrophysics \citep[e.g.,][]{Metzger_2019}. While they are among the best sources of gravitational-wave (GW) emission detectable by GW observatories \citep{GWTC1,GWTC2} such as Advanced LIGO and Advanced Virgo \citep{aLIGO2015,aVirgo2015}, their potential detection in the electro-magnetic (EM) spectrum by surveys around the world represents one of the most challenging searches for astrophysical transients. During the merger, significant amounts of neutron-star (NS) matter are ejected at sub-relativistic speeds due to either tidal or hydrodynamical forces; the radioactive decay of $r$-process elements synthesized in the neutron-rich merger ejecta powers a thermal ultraviolet, optical and near infrared transient, often referred to as a kilonova (KN) \citep{Li_1998,Rosswog2005,Metzger_2010,TaHo2013}. Despite their color- and luminosity-evolution being viewing-angle dependent \citep{Kasen2015, Bulla_2019, Kawaguchi2020, Korobkin2020, Zhu2021}, their (largely) isotropic emission makes KNe one of the promising targets for EM counterpart follow-up observations \citep{Roberts2011}. However, they can be short-lived, faint, and peak in the infrared, making detection difficult \citep{Kasen2015,Tanaka2016,Barnes2016,Metzger_2019,nakar2019electromagnetic}.


From an observational standpoint, the GW detection of the BNS merger GW170817 \citep{LVCGW170817} provided the first, and only, multi-messenger follow-up of a GW event to yield an associated KN (AT2017gfo) to date \citep{KNGW170817}.  Observations were recorded in the ultraviolet, optical and near-infrared \citep{2017gfoAUS, 2017gfoGeminiS, 2017gfoSwope, 2017gfoCowperthwaite, 2017gfoDrout, 2017gfoSwift, 2017gfoKasliwal, 2017gfoSpitzer, 2017gfoKilpatrick, 2017gfoMASTER, 2017gfoMcCully, 2017gfoNicholl, 2017gfoShappee, 2017gfoDECam, 2017gfoPian, 2017gfoSmartt, 2017gfolanthanides, 2017gfoJGEM}. These observations have highlighted the ability to test models of KNe and provide constraints on the ejecta mass and velocity \citep{Abbott2017_ejecta, 2017gfoCowperthwaite, Perego2017, 2017gfoPian, 2017gfoSmartt, Tanaka2017, Waxman2017, Coughlin_2019b, Kawaguchi2020, Heinzel2021,raaijmakers2021challenges}, $r$-process elemental abundances \citep{Cote2018, Hotokezaka2018, Radice2018, Tanaka2018, Hotokezaka2019,Siegel2019}, the NS equation of state \citep{FoHiNi2018,Coughlin_2018,RadiceDai2019, Hinderer2019, Breschi2021, Nicholl2021} and Hubble Constant \citep{Hotokezaka2018_HubbleConstant,DiCo2020, Dhawan2020}.


There are a plethora of studies in the literature that model the luminosity evolution of KNe (\citealt{2017gfoKasen,Coughlin_2018,Wollaeger2018, Bulla_2019,Kawaguchi2020}; also see references in \citealt{Metzger_2019}). 
Despite the detection of the KN from GW170817, there are significant uncertainties in the model parameter space \citep{Barnes2016, Rosswog2017, Zhu2018, 2017gfoSpitzer, Wu2019, Heinzel2021}. These uncertainties primarily stem from the range of ejecta masses expected from such mergers and the content of nuclear matter assumed in the models \citep{Barnes2020, Foucart2021, Kullmann2021, Just2021}. While uncertainties in the mass ejected from BNS systems have been shown to be driven mostly by the total mass and mass ratio of the system \citep{BausBaumJan2013,Hotokezaka2013,Koppel_2019,Kiuchi_2019}, those in models for NSBH systems are influenced by the mass ratio, BH spin and NS radius \citep{Etienne2009,Kyutoko2015,Kawaguchi_2016,FoHiNi2018,Zhu2020NSBH}.

The third observing run of Advanced LIGO and Virgo (O3\footnote{\url{https://gracedb.ligo.org/superevents/public/O3/}}), which lasted 11 months, yielded a total of 15 publicly announced NSBH and BNS candidates. Several teams, including Global Relay of Observatories Watching Transients Happen (GROWTH; \citealt{kasliwal2020kilonova}), Electromagnetic counterparts of gravitational wave sources at the Very Large Telescope (ENGRAVE; \citealt{Engrave2020}), Global Rapid Advanced Network Devoted to the Multi-messenger Addicts (GRANDMA; \citealt{Grandma2020}), Gravitational-wave Optical Transient Observer~(GOTO; \citealt{Goto2020}), All Sky Automated Survey for SuperNovae (ASAS-SN; \citealt{deJaeger2021_ASASN}), Asteroid Terrestrial Last Alert System (ATLAS; \citealt{ToDe2018}), Panoramic Survey Telescope and Rapid Response System (Pan-STARRS; \citealt{ChMa2016}), MASTER-Net \citep{2017gfoMASTER}, Dark Energy Survey Gravitational Wave Collaboration (DES-GW; \citealt{2017gfoDECam}) and Japanese collaboration for Gravitational wave ElectroMagnetic follow-up (J-GEM; \citealt{Sasada2021_JGEM}) conducted wide-field searches within the skymaps of BNS and NSBH candidates and pursued follow-up of interesting transient candidates found therein, but no plausible EM counterparts were found (e.g., \citealt{CoDi2019}).

Nevertheless, the apparent dearth of counterparts during all of O3 can illuminate our understanding of the intrinsic properties of KNe. On an individual GW event basis, observational upper limits can be used to constrain the KN emission from a potentially associated counterpart and infer properties of the binary \citep{Hosseinzadeh2019, Andreoni2020, AnandCoughlin_2020, MoSo2020}. Other works, e.g., \citet{CoDi2019,Lundquist2019,Goto2020,Grandma2020, kasliwal2020kilonova}, have demonstrated ways to synthesize survey observations for a suite of GW events to constrain the KN population as a whole. In particular, \citet{kasliwal2020kilonova} formulated a method for constraining the luminosity function of the KN population. Assuming a non-uniform distribution of KN initial luminosities between $-$10 and $-$20 absolute magnitude, their findings suggest that no more than 57\% (89\%) of KNe could be brighter than $-$16.6\,mag assuming flat (fading at 1\,mag\,day$^{-1}$) evolution \citep{kasliwal2020kilonova}.

In this paper, we present \nimbus \citep{nimbus_zenodo}: a hierarchical Bayesian framework to infer the intrinsic luminosity parameters of the population of KNe associated with GW events, based purely on non-detections. Key features of this framework include the simultaneous use of probabilistic source distance information from GW observations and corresponding upper limits from EM surveys, accounting for the fraction of the skymap searched by a given survey for each event, self-consistent inclusion of the probability of a GW event being of astrophysical origin ($p_{\rm astro}$) and the ability to model multi-band luminosity evolution. The framework is agnostic to the specific luminosity model used and thus can be used to constrain a wide variety of models in the literature. 

As a first example and proof of concept, we demonstrate realistic constraints possible on the KN emission from the past follow-up of the event GW190425 \citep{LVCGW190425} conducted with the Zwicky Transient Facility (ZTF) \citep{ZTF_GW190425}. ZTF is an optical time-domain survey, consisting of a CCD camera with a 47\,deg$^2$ field-of-view installed on the 48-inch Samuel Oschin Schmidt Telescope at the Palomar Observatory. Scanning the sky at an areal survey speed of $\sim$\,3750 square degrees per hour in three custom filters, ZTF-$g$, ZTF-$r$, ZTF-$i$; it reaches a median depth of 20.4\,mag in 30\,s exposures in its nominal nightly survey but can also conduct deeper target-of-opportunity followup of external events \citep{CoAh2019}; for a comprehensive review of the ZTF instrument, software, and survey see \citet{Bellm_2019ZTF, Masci_2019ZTF, Graham_2019, Dekany_ZTF_2020}.

Among the 13 events searched by ZTF in Advanced LIGO's third observing run \citep{kasliwal2020kilonova} that could have a probable EM counterpart, based on the probability of the system containing a NS i.e. p(BNS) or p(NSBH), GW190425 is so far the only significant binary merger event confirmed by LIGO and Virgo \citep{LVCGW190425} to likely be a BNS based on the posterior inference of its masses; therefore, our analysis herein focuses on this event alone. GW190425 was located at a distance of 159$^{+69}_{-71}$\,Mpc and its final 90\% credible localization spanned 8284\,deg$^2$ \citep{LVCGW190425}. For GW190425, ZTF observed $\sim$8000\,deg$^2$, corresponding to 45\% probability of the initial BAYESTAR skymap \citep{SiPr2016bayestar} which reduced to 21\% integrated probability within the 90\% credible region of the LALInference skymap \citep{Veitch2015} and attained a median depth of m$_{AB}\approx$21\,mag in $g$- and $r$-bands \citep{CoAh2019}. For the purpose of this analysis, we consider ZTF areal coverage within the entire LALInference skymap, which corresponds to 32\% probability. No KN was identified in the observed region of this event by ZTF or other optical telescopes \citep{gcn24167, gcn24172, gcn24187, gcn24190, gcn24191, gcn24197, gcn24210, gcn24224, gcn24227, gcn24285}, or for any other GW event followed-up with ZTF \citep{kasliwal2020kilonova}.

This paper is organized as follows. In Sec. \ref{Bayesian Inference}, we provide a detailed description of the Bayesian framework including a derivation of the model posterior and important aspects that impact the inference. We then present our main inference results on GW190425 using two different prior assumptions in Sec. \ref{sec:GW190425}. We also use this Section to compare our results with those obtained from {\tt simsurvey} \citep{Feindt_2019}, a simulation tool for astronomical surveys previously used in the literature to constrain KN luminosity distributions \citep{kasliwal2020kilonova}. We then conclude with a discussion of our results and future outlook in Sec. \ref{discussion}.

\section{Bayesian Framework} \label{Bayesian Inference}
In order to derive constraints on KN parameters, we make use of a hierarchical Bayesian statistical framework. Our goal is to find the posterior probability
distribution of the parameters of interest $\mathcal{\Vec{\theta}}$, given the data $\{d^{i}\}$. The derivation here follows analogous derivations of hierarchical population inference in GW literature \citep{FGMC2015, Gaebel2019, Mandel2019}.

\subsection{Model definitions} \label{model}
For this paper, we model the luminosity evolution of KNe using a two-parameter, linear family of light curves (as adopted in \citealt{kasliwal2020kilonova}). However, we will discuss extensions of our framework to other models as well. The absolute magnitude ($M$) in a given filter $\lambda$ is given as a linear function of time ($t$),
\begin{equation}
    M^{\lambda}(t) = M_{0} + \alpha \left(t - t_{0}\right)
    \label{eq:linear_evol}
\end{equation}

where $t_{0}$ is the initial time of the KN transient. We can see that the two parameters $\left(M_{0},\alpha \right)$, which represent an initial absolute magnitude and evolution rate respectively, completely define the evolution at all times. Therefore, for this simplistic parameterization, $\Vec{\theta} = \{M_0, \alpha\}$. We emphasize that our motivation to implement such a simple model is to demonstrate the framework and due to the fact that we rely on follow-up observations of KNe up to 3 days following the merger time, where such models are a relatively good fit to the data (see Sec. \ref{astro-prior} and Sec. \ref{discussion}). 

Before we begin with our derivation, we will state our notation as follows:
\begin{itemize}
    \item $N_{E}$ : Total number of events that were followed up, indexed by $i$.
    \item $N_{F}$ : Total number of fields-of-view for which EM observations have been recorded, indexed by $f$. For purposes of improved reference model subtraction, many optical/infrared surveys use discrete fields for observations rather than allowing  complete freedom (\citealt{Ghosh2017}, \citealt{Coughlin2018gwemopt} and references therein). However, this can be generalized to any discretization of the sky such as HEALPIX (Hierarchical Equal Area isoLatitude Pixelization\footnote{https://healpix.sourceforge.io/}; \citet{Gorski2005}) if needed.
    \item $N_{f}$ : Total number of observations for field $f$.
    \item $t^{f}_{j}$ : Time of observation, indexed by $j$ for each field $f$ over the duration of follow-up of the event. $j$ would run over the total number of observations for each field ($N_{f}$).
    \item $t_{0}$ : Initial time of the KN transient, which corresponds to the initial absolute magnitude $M_{0}$
    \item $\bar f$ : Index for fields not including the field $f$.
    \item $\bar F$ : Hypothesis that the KN is not in any of the observed fields.
    \item $A$ : Hypothesis that the event is of astrophysical origin.
    \item $T$ : Hypothesis that the event is of terrestrial origin (implying that the event is spurious).
    \item $P^{i}(A)$: The probability of the $i^{th}$ event being of astrophysical ($A$) origin. This is an estimate provided by the LIGO-Virgo-KAGRA collaboration for the associated GW event. It can either be a low-latency estimate or an update provided after a refined analysis.
    \item $\{d^{i}\}$ : The set of EM data associated with all events, indexed by $i$. We will further index this data by the field index $f$ and time of observation index $j$, in our derivation below. For this study, we take our data to be the set of limiting (apparent) magnitudes $\{m_{l}^{i,f,j}\}$ in each field at the given time of observation.
\end{itemize} 

\subsection{Derivation of the model posterior} \label{derivation}
We begin our derivation of the model posterior with the basic equation of Bayes' law:
\begin{equation}
    p(M_{0},\alpha | \{d^{i}\}) = \frac{p(\{d^{i}\} | M_{0},\alpha) p(M_{0},\alpha)}{p(\{d^{i}\})}
\end{equation}

where $p(\{d^{i}\} | M_{0},\alpha)$ is the likelihood, $p(M_{0},\alpha)$ is the prior distribution of the parameters $M_0$ and $\alpha$, and $p(\{d^{i}\})$ is the evidence. 
We carry out analyses based on different prior assumptions and show the effect it has on the posterior distribution of the KN parameters in Sec.~\ref{sec:GW190425}. The likelihood represents the probability density of observing the data ${d^{i}}$ given a model, for a set of events indexed by $i$, while the evidence is the probability of observing the data, marginalised over all parameters and serves as a normalization factor in the inference. Further, each event and its associated data are assumed to be independent. The likelihood $p(\{d^{i}\} | M_{0},\alpha)$ can then be written as a product over events.

\begin{equation}
    p(\{d^{i}\} | M_{0},\alpha) = \prod_{i=1}^{N_{E}} \Bigg{[} p(d^{i} | M_{0},\alpha) \Bigg{]}
\end{equation}
There are two possibilities for any given event -- either the event is astrophysical ($A$) or it is non-astrophysical/terrestrial ($T$). We note that the probability of the latter hypothesis is $(1 \, - \, P^{i}(A))$. We thus split the likelihood into two terms using the relative probabilities of each hypothesis,

\begin{eqnarray}
    p(\{d^{i}\} | M_{0},\alpha) &=& \prod_{i=1}^{N_{E}} \Bigg{[} p(d^{i} | M_{0},\alpha,A) P^{i}(A) \nonumber \\ 
    && +  p(d^{i} | T) (1 \, - \,P^{i}(A))\Bigg{]}.
\end{eqnarray}

The assumption in the last term in the parentheses is that the contribution to the likelihood cannot depend on the parameters of the KN model if the event is of terrestrial origin. This is straight-forward to check because in the case of a purely terrestrial event ($P^{i}(A) = 0$), we must recover the prior when performing inference.

We now use the fact that, for any event (indexed by $i$), EM observations are distributed over $N_{F}$ fields (indexed by $f$) at times of observation (indexed by $j$) such that every observation has associated limiting magnitudes ($m_{l}^{i,f,j}$). As stated above, we take our data $d^{i}$ for each event to be the set of these observed limiting (apparent) magnitudes $\{m_{l}^{i,f,j}\}$ i.e. $d^{i} \equiv \{m_{l}^{i,f,j}\}$. The likelihood thus becomes,
\begin{eqnarray}
     p(\{d^{i}\} | M_{0},\alpha) &=& \prod_{i=1}^{N_{E}} \Bigg{[} p(\{m_{l}^{i,f,j}\} | M_{0},\alpha,A) P^{i}(A) \nonumber\\
     && +  p(\{m_{l}^{i,f,j}\} | T) (1 \, - \,P^{i}(A))\Bigg{]}.
\end{eqnarray}

Furthermore, under the astrophysical hypothesis ($A$), the likelihood can be split into two more terms given that there are two possibilities for the KN event -- 
\begin{itemize}
    \item The KN is located within an observed field $f$ and consequently, not in any of the other fields ($\bar f$). In this case we need to find the probability that the KN is within a field $f$ i.e., $P(f)$ and sum the contributions to the likelihood from each field. The overall likelihood contribution from this hypothesis is
    \begin{eqnarray}
        \sum\limits_{f=1}^{N_{F}} \Big{(}p(\{m_{l}^{i,f}\} |  M_{0},\alpha,A,f) \prod\limits_{\bar f} p(\{m_{l}^{i,\bar f}\} |A,f) P(f)\Big{)} \nonumber
    \end{eqnarray}
    \item The KN event is not located in any observed field (hypothesis $\bar F$). This case has a probability equal to $(1-\sum\limits_{f=1}^{N_{F}} P(f))$. The overall likelihood contribution from this hypothesis is
    \begin{eqnarray}
        \prod\limits_{f} p(\{m_{l}^{i,f}\} |A,\bar F)\,\Big{(}1-\sum\limits_{f=1}^{N_{F}} P(f)\Big{)} \nonumber
    \end{eqnarray}
\end{itemize}

When information about a GW candidate event is released, it contains the 3D sky probability distribution of the location of the event, which includes the luminosity distance ($d_{L}$) to the source \citep{Singer2016,Singer2016Supp}. Using this, it is straightforward to compute the probability for a KN to be present in a given field. Referring to Eq.~3 in \citet{Singer2016}, the sum of probabilities over the sky is
\begin{equation}
    P(f) = \sum_{k=0}^{N_{pix}^f}  \rho_{k},
\end{equation}
where the sum is over the $N_{pix}^f$ pixels that are contained within field $f$ and $\rho_{k}$ is the probability of the event being in pixel $k$. The likelihood, written in terms of hypothesis contributions stated above, then becomes

\begin{eqnarray}
    p(\{d^{i}\} | M_{0},\alpha)  &=& \prod_{i=1}^{N_{E}} \Bigg{[}
       \Bigg{(} \sum_{f=1}^{N_{F}} \Big{(}p(\{m_{l}^{i,f}\} |
       M_{0},\alpha,A,f) \nonumber\\
    &&   \prod_{\bar f} p(\{m_{l}^{i,\bar f}\} |A,f) P(f)
       \Big{)} \nonumber \\ 
    && + \prod_{f} p(\{m_{l}^{i,f}\} |A,\bar F)
       \bigg{(}1-\sum_{f=1}^{N_{F}} P(f)\bigg{)} \Bigg{)} P^{i}(A)
       \nonumber \\
    && + \prod_{f} p(\{m_{l}^{i,f}\} | T) (1 \, - \,P^{i}(A))\Bigg{]},
\end{eqnarray}
where the second and third terms in the parentheses correspond to the hypotheses that the KN position is outside all the observed fields and that the event is terrestrial in nature, respectively.

Since the observations in each field will have observation times associated with them, each field observation would constrain the model independently. Thus, the likelihood term for each field can be written as a product over the number of observations corresponding to that field.

\begin{eqnarray}
    p(\{d^{i}\} | M_{0},\alpha) &=&  \prod_{i=1}^{N_{E}} \Bigg{[} \Bigg{(} \sum_{f=1}^{N_{F}} \Big{(}\prod_{j=1}^{N_{f}} p(m_{l}^{i,f,j} | M_{0},\alpha,A,f) \nonumber\\ 
    && \prod_{\substack{\bar f=1,\\ \bar f \neq f}}^{N_{F}} p(m_{l}^{i,\bar f} |A,f) P(f) \Big{)}\nonumber\\  
    && + \prod_{f} p(m_{l}^{i,f} |A,\bar F) \bigg{(}1-\sum_{f=1}^{N_{F}} P(f)\bigg{)} \Bigg{)} P^{i}(A)\nonumber\\ 
    && + \prod_{f} p(m_{l}^{i,f} | T) (1 \, - \,P^{i}(A))\Bigg{]}
\label{eq:likelihood_total}
\end{eqnarray}

We now focus on the first term in the likelihood for each field $p(m_{l}^{i,f,j} | M_{0},\alpha,A,f)$. In order to simplify this term and derive an expression for the same, we note that, in reality, a telescope measures an apparent magnitude ($m_{j}$) instead of an absolute magnitude. One can rewrite this likelihood term, using conditional probability, as an integral over the apparent magnitude of the KN event. 

\begin{eqnarray}
    p(m_{l}^{i,f,j} | M_{0},\alpha,A,f) = \nonumber\\
    \int_{-\infty}^{\infty} p(m_{l}^{i,f,j}| m_{j}) p(m_{j}| M_{0},\alpha,A,f) dm_{j}
    \label{eq:likelihood}
\end{eqnarray}

The relationship between the apparent magnitude ($m$), absolute magnitude ($M$) and luminosity distance ($d_{L}$) of an astrophysical source is given as 
\begin{equation}
    m = M + 5 \log_{10}\Bigg{(}\frac{d_{L}}{10 \mathrm{pc}}\Bigg{)}
\end{equation}

or equivalently,

\begin{equation}
    d_{L} = 10^{\Big{(}\frac{m - M}{5}\Big{)}} \Big{(}\frac{1}{10^{5}}\Big{)} \mathrm{Mpc}.
    \label{eq:distmod}
\end{equation}

The above formulae do not include the effects of extinction. We account for Milky Way extinction in our framework by appropriately modifying the limiting magnitudes for each field and filter. We make use of the {\tt dustmaps} package \citep{Dustmaps} and its implementation of the SFD dustmap \citep{SFD2011} to derive extinction values. Also, from Eq.~\ref{eq:distmod}, we can derive a limiting distance $d_{lim}^{i,f,j}$ for a corresponding limiting magnitude $m_{l}^{i,f,j}$. The intrinsic parameters of the KN $(M_{0},\alpha)$ along with the observation time ($t_{j}$) uniquely determine the absolute magnitude ($M_{j}(M_{0},\alpha)$) of the KN, at any given time. Eq.~\ref{eq:distmod} shows that for such a given absolute magnitude $M$, the apparent magnitude and distance are dependent variables that uniquely define each other. It is possible to relate the apparent magnitude distribution ($p(m_{j}|M_{0},\alpha,A,f)$) in the integral above to the marginal distance distribution ($p_{f}(d_{L})$) for each field $f$, which can be derived from the GW skymap, as 

\begin{equation}
    p(m_{j}|M_{0},\alpha,A,f) = p_{f}(d_{L}) \frac{d\,d_{L}}{dm_{j}} (M_{0},\alpha).
\end{equation}

We can thus rewrite the integral in Equation~\ref{eq:likelihood} as 

\begin{equation}
    p(m_{l}^{i,f,j} | M_{0},\alpha,A,f) =  \int_{0}^{\infty} p(m_{l}^{i,f,j}| m_{j}(d_{L},M_{0},\alpha)) p_{f}(d_{L}) d\,d_{L}.
    \label{eq:likelihood_app}
\end{equation}

Since this is a non-detection study, the only viable limiting magnitudes for non-detection are those that are strictly shallower (brighter) than the apparent magnitude from the KN model. We implement this by using a uniform distribution function for the conditional density $p(m_{l}^{i,f,j}| m_{j}(d,M_{0},\alpha))$ as,

\begin{equation}
     p(m_{l}^{i,f,j}| m_{j}(d_{L},M_{0},\alpha)) \propto \begin{cases}
     \frac{1}{(m_{j} - m_{l}^{i,f,j})} \,;\, m_{j} \leq m_{l}^{\mathrm{high}}\\
     \frac{1}{(m_{l}^{\mathrm{high}} - m_{l}^{\mathrm{low}})} \,;\, m_{j} > m_{l}^{\mathrm{high}}
     \end{cases}
    \label{eq:likelihood_app_uniform}
\end{equation}
where $m_{l}^{\mathrm{low}},\,m_{l}^{\mathrm{high}}$ are the lower and upper limits of the range of limiting magnitudes from the survey or distance information, respectively. See Sec.~\ref{normalization-choice} for a more detailed discussion on the choice of these limits. Further, it is important to account for the probabilistic nature of each limiting magnitude when considering the likelihood of any given model. We incorporate this requirement into our likelihood with a logistic function $\Phi(d_{L} - d_{lim}^{i,f,j})$. The logistic function ensures a smooth turnover in the likelihood between distances (apparent magnitudes) that pass the limiting distance (limiting magnitude) constraints and those that do not. From Equation~\ref{eq:distmod}, we can write the logistic function in terms of the distance as,

\begin{equation}
    \Phi(d_{L} - d_{lim}^{i,f,j}) = \frac{1}{1 + e^{-a(d_{L} - d_{lim}^{i,f,j} + b)}}.
    \label{eq:phi}
\end{equation}

We choose the constants $a$ and $b$ in Equation~\ref{eq:phi} based on errors in the limiting magnitude such that a 3-$\sigma$ error in $m_{l}^{i,f,j}$ corresponds to the distance at which the logistic function in Equation~\ref{eq:phi} is set to the cumulative probability weight of a Gaussian distribution beyond the lower 2-sigma limit ($\sim 2.3 \%$). Combining Eqs.~\ref{eq:likelihood_app_uniform} and \ref{eq:phi}, the likelihood term in Eq.~\ref{eq:likelihood_app} can be evaluated up to a normalization constant $k$.

Since the total likelihood in Eq.~\ref{eq:likelihood_total} is a sum of probability densities, care must be taken to normalize each term, corresponding to each hypothesis, separately. The constant $k$ can be derived by normalizing the likelihood term in Equation~\ref{eq:likelihood_app} between the appropriate limiting magnitude limits, or more directly between appropriate limiting distance limits:

\begin{equation}
    k = \frac{1}{\int_{m_{l}^{\rm low}}^{m_{l}^{\rm high}} p(m_{l}^{i,f,j} | M_{0},\alpha,A,f) dm_{l}^{i,f,j}}.
    \label{eq:norm}
\end{equation}
These limits can be chosen based on the extent of the marginal distance distribution for each field. We provide specific details of our assumptions for these limits in Sec.~\ref{sec:GW190425}.

This normalization also ensures that we account for selection effects based on the limiting magnitude limits of the survey. We defer the discussion of the impact this has on the inference to Sec.~\ref{normalization-choice}. The remaining terms in the field ($f$), non-field ($\bar f$) and terrestrial ($T$) hypotheses from Equation \ref{eq:likelihood_total} must be normalized. We assume that each of these terms follows a uniform distribution between the survey limits $m_{l}^{\rm low}$ -- $m_{l}^{\rm high}$. This assumption largely simplifies the form of the likelihood. While it is not necessary to assume such a form for each of these terms, and one can construct more complex distributions based on realistic data, this choice does not impact the inference because these distributions must necessarily be independent of the model parameters. This gives us a normalized density of

\begin{eqnarray}
    p(m_{l}^{i,\bar f} |A,f) & \, = & \, p(m_{l}^{i,f} |A,\bar F) \, = \, p(m_{l}^{i,f} | T) \nonumber\\
    && = \frac{1}{\Big{(} m_{l}^{\rm high} - m_{l}^{\rm low} \Big{)}}
\end{eqnarray}

This simplifies the likelihood in Equation~\ref{eq:likelihood_total} to give us a posterior

\begin{eqnarray}
    p(M_{0},\alpha | \{d^{i}\}) & \propto & \prod_{i=1}^{N_{E}}
        \Bigg{[} \Bigg{(} \sum_{f=1}^{N_{F}} \Big{(}\frac{\prod_{j}
        p(m_{l}^{i,f,j} | M_{0},\alpha,A,f)}{\prod_{j}
    p(m_{l}^{i,f,j} |A,\bar f)}\Big{)} P(f) \nonumber\\
    &&  +\bigg{(}1-\sum_{f=1}^{N_{F}} P(f)\bigg{)} \Bigg{)} P^{i}(A) \: \nonumber\\
    &&  + \: (1 \, - \,P^{i}(A))\Bigg{]} p(M_{0},\alpha).
\end{eqnarray}

\subsection{Impact of using survey limits and distance limits on inference} \label{normalization-choice}
We derived our model posterior for the framework in Sec.~\ref{derivation}. As seen in Equation~\ref{eq:norm}, an important quantity to compute for the posterior is the normalization factor $k$ which depends on the choice of upper limits $m_{l}^{\rm low}$, $m_{l}^{\rm high}$ and model parameters $(M_{0},\alpha)$. Such a factor is akin to accounting for selection effects (see \citealt{Mandel2019}) where one has to normalize the likelihood of observing a given model $(M_{0},\alpha)$ by the range of data supported by the model. From Eq.~\ref{eq:distmod}, it is possible to express upper limits for a model equivalently in terms of the apparent magnitude or the distance. Thus, a range in one of the quantities directly specifies a range in the other. The choice of which quantity to use to calculate, $m_{l}^{\rm low}$ and $m_{l}^{\rm high}$, significantly affects the result of the inference. There are two ways to select specific values for these normalizing parameters:

\begin{itemize}
    \item Survey Limits: A straight-forward method is to choose $m_{l}^{\rm low}$ and $m_{l}^{\rm high}$ directly from survey data when the telescope is observing. From Equation~\ref{eq:distmod}, this directly impacts the range of distances permitted for a given model $(M_{0},\alpha)$ and gives a different normalization value for each model. Such a method ensures that the normalization realistically accounts for model biases in the case of non-detection.
    \item Distance Limits: Alternatively, we can choose to use the distance posterior from the GW skymap data as our source of ground truth such that we calculate $m_{l}^{\rm low}$ and $m_{l}^{\rm high}$ based on the full range of possible distances\footnote{Computationally, we bound the distance between a realistic lower limit and the upper 5-$\sigma$ value from the distance posterior.}. This will change the values of $m_{l}^{\rm low}$ and $m_{l}^{\rm high}$ for each model. However, as the range in distance is the same for each model this ensures that the normalization factor is the same.
\end{itemize}

We note that our preferred results in this paper are those that use realistic survey upper limits. Unless stated otherwise, our reference to results in general will be with this choice. We present the differences that result from these two choices in Sec.~\ref{uniform-prior}.




\begin{figure*}[t]
    \plottwo{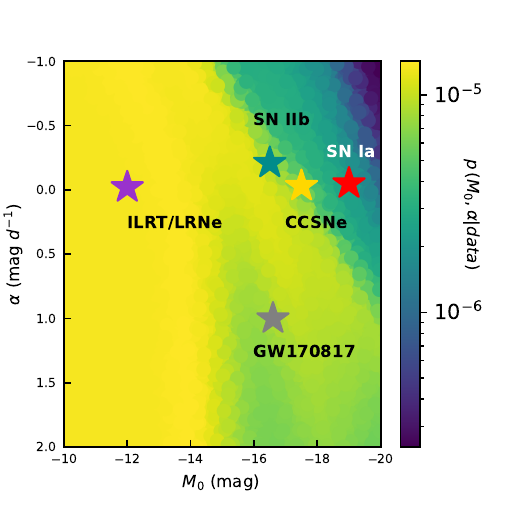}{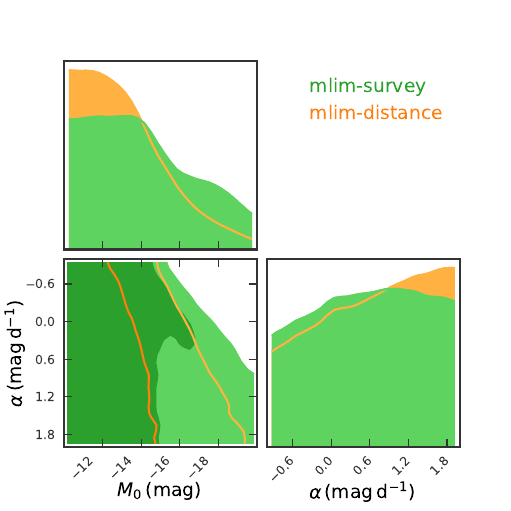}
    \caption{(left) 2-D posterior probability plot of KN model parameters (initial absolute magnitude $M_0$, evolution rate $\alpha$) with uniform priors, $M_{0} \sim \mathcal{U}(-20,-10)\, \mathrm{mag}, \alpha \sim \mathcal{U}(-1,2)\, \mathrm{mag \, day}^{-1}$, and normalization over realistic Survey Limits. Representative points are shown for the characteristic peak magnitude and rise rate (derived from the characteristic timescale) for some categories of transients (\citet{KasliwalThesis2011}, Fremling et al. in prep) -- SN Ia (red), Core Collapse SNe-CCSNe (gold), Faint, fast SNIIb (dark-cyan), ILRT/LRNe (dark-orchid) and GW170817 (grey). (right) Corner plot showing the 2-D and corresponding 1-D marginalized posterior distributions of the KN model parameters for the two different normalization schemes detailed in Sec.~\ref{normalization-choice} --- using realistic Survey Limits from ZTF data for GW190425 which are between 15--23 mag (green;\,{\tt mlim-survey}) and Distance Limits obtained for each model draw using 5$\sigma$ distance limits obtained from the skymap distance posterior for each field (orange;\,{\tt mlim-distance}). Contours indicate $68\%$ and $95\%$ confidence regions. Non-monotonic features in both panels (eg: $68\%$ green contour in the right panel) indicate the effects of model normalization. Details in Secs. \ref{normalization-choice},\ref{uniform-prior}.}
    \label{fig1}
\end{figure*}

\section{Kilonova Inference using GW190425} \label{sec:GW190425}
GW190425 was a highly significant ($p_{\rm astro} \sim 0.999$; \citealt{LVCGW190425}) GW event that was followed up by ZTF \citep{ZTF_GW190425} with an overall sky coverage of $\sim 32\%$ of the total skymap. Inferences on the component masses of the detected binary show it to be consistent with a BNS, although the possibility of either or both components being BHs cannot be ruled out from GW data alone. Here, we present results using the Bayesian framework \nimbus \, described here with upper limits from the ZTF follow-up of GW190425 to derive posterior constraints on KN parameters of the model light curve for BNS mergers. We note that unlike the band-specific linear evolution shown in Eq.~\ref{eq:linear_evol} we adopt a single ``average-band'' linear model with parameters ($M_{0},\alpha$) for our analyses presented here. This ''average-band'' model effectively assumes the same color evolution in all bands, allowing us to use ZTF observations in all filters for our analysis. This simplified model is conducive for testing the \nimbus\ framework as it significantly reduces the model parameter-space (since kilonova models predict a wide diversity in expected color evolution). For example, using this linear model fit, GW170817 has $M_{0}=-16.6$\,mag and $\alpha=1$\,mag day$^{-1}$ \citep{kasliwal2020kilonova}. Our analyses rely on two different prior distribution choices for ($M_{0},\alpha$) -- (i) uniform or agnostic priors (explained in Sec. \ref{uniform-prior}) and (ii) astrophysical priors motivated from theory and numerical modeling (explained in Sec. \ref{astro-prior}). We limit our analyses to use follow-up data up to 3 days from the trigger time since realistic models predict that most kilonovae will fade beyond the median ZTF limiting magnitude of 21\,m$_{\rm AB}$ (for this event) within 72\,hours after trigger time (see Sec.~\ref{discussion}).

\begin{figure*}[t]
    \plottwo{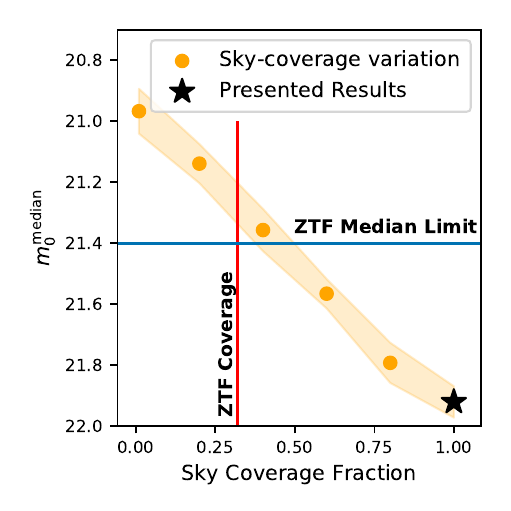}{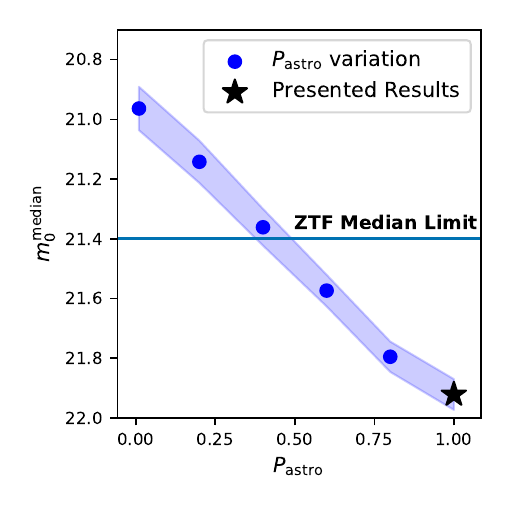}
    \caption{Variation in the median of the initial apparent magnitude distribution (assuming the source is at the mean luminosity distance from the 3-D skymap) as a function of sky-coverage (left) and probability of astrophysical origin $P_{\mathrm{astro}}$(right). As the sky-coverage/$P_{\mathrm{astro}}$ decreases (increases), constraints on the population parameter become weaker (stronger). Colored bands in both plots indicate $2-\sigma$ error regions from the interpolation of posterior probabilities as mentioned in Sec.\ref{sec:GW190425}. Horizontal lines at $m=21.4$\,mag in both plots indicate the median limiting magnitude across the 3 day ZTF observations for GW190425. Red vertical line in left plot indicates the actual sky coverage by ZTF.}
    \label{fig:skycov-pastro-variation}
\end{figure*}

In order to limit the effects of Milky Way extinction in the fields surveyed, we place a conservative threshold by excluding fields which have $E(B-V) > 2$\,mag. \emph{For the remainder of the paper, we make a simplifying assumption that the KN associated with GW190425 is located within the searched region\footnote{ \nimbus\ has the capability to accommodate for the excluded part of the skymap (see Fig.~\ref{fig:skycov-pastro-variation}).}} We ran simulations taking the full GW190425 skymap into account and found the results to be largely unconstraining; hence we adopt the above assumption in order to demonstrate the constraints possible with \nimbus\ in an ideal sky coverage scenario. Our constraints on KN model parameters, obtained using both prior choices stated above, are displayed in Table \ref{tab:table1}. In order to derive our constraints and plot our posterior probabilities in this paper, we make use of an interpolating or smoothing function such as the Gaussian Process module from {\tt scikit-learn} \citep{scikit-learn} to interpolate between our original samples from priors. We have checked that uncertainties from these interpolations (see Fig. \ref{fig:skycov-pastro-variation}) are within the statistical variations of the observed limiting magnitudes across the ZTF quadrants for each field of observation, where these variations are $>= 0.1\,\mathrm{mag}$ for a majority of fields. Sec.~\ref{uniform-prior} with uniform priors demonstrates the functionality of the framework. In particular, our results in this Section show the posterior constraints that are possible using the framework. In addition, we show how a general inference is sensitive to variations in sky coverage and $p_{\rm astro}$. We also use this Section to illustrate the differences in results based on the two different normalization choices as explained in Sec.~\ref{normalization-choice}: choosing the faint and bright apparent magnitude limits based on the observed range from ZTF (Survey Limits)  \textit{or} based on the minimum and maximum distances from the GW distance posterior (Distance Limits). Sec. \ref{astro-prior} demonstrates the ability to test astrophysical priors within the \nimbus \, framework. 

\subsection{Uniform priors}\label{uniform-prior}
We first implement the framework assuming uniform priors for our model parameters $M_{0} \sim \mathcal{U}(-20,-10) \, \mathrm{mag}, \alpha \sim \mathcal{U}(-1,2)\, \mathrm{mag \, day}^{-1}$. According to our convention (see Eq.\ref{eq:linear_evol}) a negative evolution rate would imply a rising light-curve for the kilonova while a positive one would be decaying. While $\alpha<0$ appears implausible based on our BNS KN model fits (see Sec.~\ref{astro-prior}), KNe from NSBH systems can exhibit a slow rise \citep{AnandCoughlin_2020} that could take $\gtrsim$2 days to peak. Within our time window of observations of 3 days, a NSBH KN model may be better approximated by a rising linear model than a fading one. Thus, we adopt a broad range for our evolution rate prior to accommodate rising to rapidly decaying KN models. Our prior on the initial magnitude is similarly broad, spanning a large fraction of the known transient phase-space \citep{KasliwalThesis2011}.
Such a prior is uninformative with respect to the realistic emission models of KNe that have implications for how ($M_{0},\alpha$) could be distributed. The posterior densities we obtain via the framework are shown in the left panel of Fig.~\ref{fig1}. Broadly, we see that, as expected for a non-detection, there is more posterior support for dimmer models (larger values of $M_{0}$ and $\alpha$) than brighter ones (smaller values of $M_{0}$ and $\alpha$). Another expected trend is the increase of support for brighter $M_{0}$ values with respect to the evolution rate as we vary $\alpha$ from $\sim -1$ to $1 \, \mathrm{mag \, day}^{-1}$. A significant part of the parameter space that belongs to bright and rising models ($M_{0} \lesssim -18$ and $\alpha \lesssim 0$) is disfavored over the rest of the models by factors of $\sim 10-100$. For most of the parameter space, where $M_{0} \gtrsim -15$, our results cannot place any constraints.

The right panel of Fig.~\ref{fig1} also compares the posterior constraints we obtain with two different normalization choices as explained in Sec.~\ref{normalization-choice}. The inference derived using realistic survey limits ({\tt mlim-survey}; green contours) rely on the range of limiting magnitudes obtained from ZTF during the follow-up of GW190425. For this event this corresponds to range limits of ($m_{\mathrm{low}}\,\approx~15,m_{\mathrm{high}}\,\approx~23$). On the other hand posteriors obtained using distance limits ({\tt mlim-distance}; orange contours) rely on the entire range of posterior distances from the 3-D skymap for the event. While constraints from the two choices are quite similar with respect to the evolution rate, the {\tt mlim-survey} method provides more support to models on the brighter end of the $M_{0}$ distribution compared to the {\tt mlim-distance} method. This is understandable since having a restricted range of limiting magnitudes from the survey reduces the range of viable distances for brighter models thereby providing a smaller parameter space that satisfies the likelihood. Accounting for this fact in the likelihood as a selection effect leads to an up-weighting of these brighter models with respect to the {\tt mlim-distance} method. This is also the reason we see non-monotonic features in the 2-D posterior distributions with the {\tt mlim-survey} method in both panels of Fig. \ref{fig1}. The effects of normalization arise the most for models that are at the marginal boundary with respect to the upper limits. As mentioned previously, our preferred results in this paper are those that use the {\tt mlim-survey} method.

As seen from Table \ref{tab:table1}, the $90 \%$ upper limit with a uniform prior is about $M_{0}^{90 \%} = -16.63$ \,mag.  We compare the result derived here to the probability of zero detections as a function of absolute magnitude shown in Fig.\,9 of \citet{kasliwal2020kilonova}, which indicates a $\sim 7\%$ probability for models with a similar initial absolute magnitude. Although the two separate constraints are consistent, there are significant differences between the two formalisms overall. The first is that our analysis assumes that an associated KN fell within the observed region. If we relax this assumption and use the $\sim$32\% sky coverage in total for this event by ZTF, our inference over the entire skymap region does not yield any meaningful constraints, as expected.

The second difference is that in this work, we infer the properties of a KN associated with GW190425, while the main result of \citet{kasliwal2020kilonova} places constraints on the luminosity function of the KN population as a whole. While strong constraints on the KN emission from a \textit{single} GW event are possible with a combination of complete sky coverage and deep upper limits, analyzing KN population properties importantly relies on \textit{several} events observed with decent sky coverage and depth. We note, however, that GW190425 likely contributes significantly to the constraints in \citet{kasliwal2020kilonova} given its nearby distance and ZTF median depth ($m\approx21$\,mag).

The $90\%$ limit stated above is also significant in that it represents the extrapolated peak magnitude of GW170817 with a average decay rate of $1 \, \mathrm{mag \, d^{-1}}$ \citep{kasliwal2020kilonova}. From the left panel in Fig.~\ref{fig1}, we can see that the data for the non-detection of a kilonova associated with GW190425 are still consistent with these parameters.

Fig. \ref{fig:skycov-pastro-variation} demonstrates how variations in sky coverage and $p_{\rm astro}$ can impact inferences in such a study in general. Our assumptions and derived constraints in this work correspond to ($p_{\rm astro} \sim 0.999$, sky coverage $= 100 \%$). However, as the sky coverage or astrophysical probability drops, the contribution of the survey upper limits to the posterior weakens and the constraints become more broad. In particular, below a certain value for these parameters, the median apparent magnitudes inferred for the KN are brighter than the median ZTF upper limits, pointing to the fact that at these values the alternative hypotheses - of either the KN being in a part of the sky where we do not have any observations or the KN event being a Terrestrial event - have more probability support. This result also demonstrates the functionality of the framework to account for arbitrary values of these probabilistic factors that impact inference.


\subsection{Comparison with \tt simsurvey}\label{simsurvey}
\begin{figure}
    \plotone{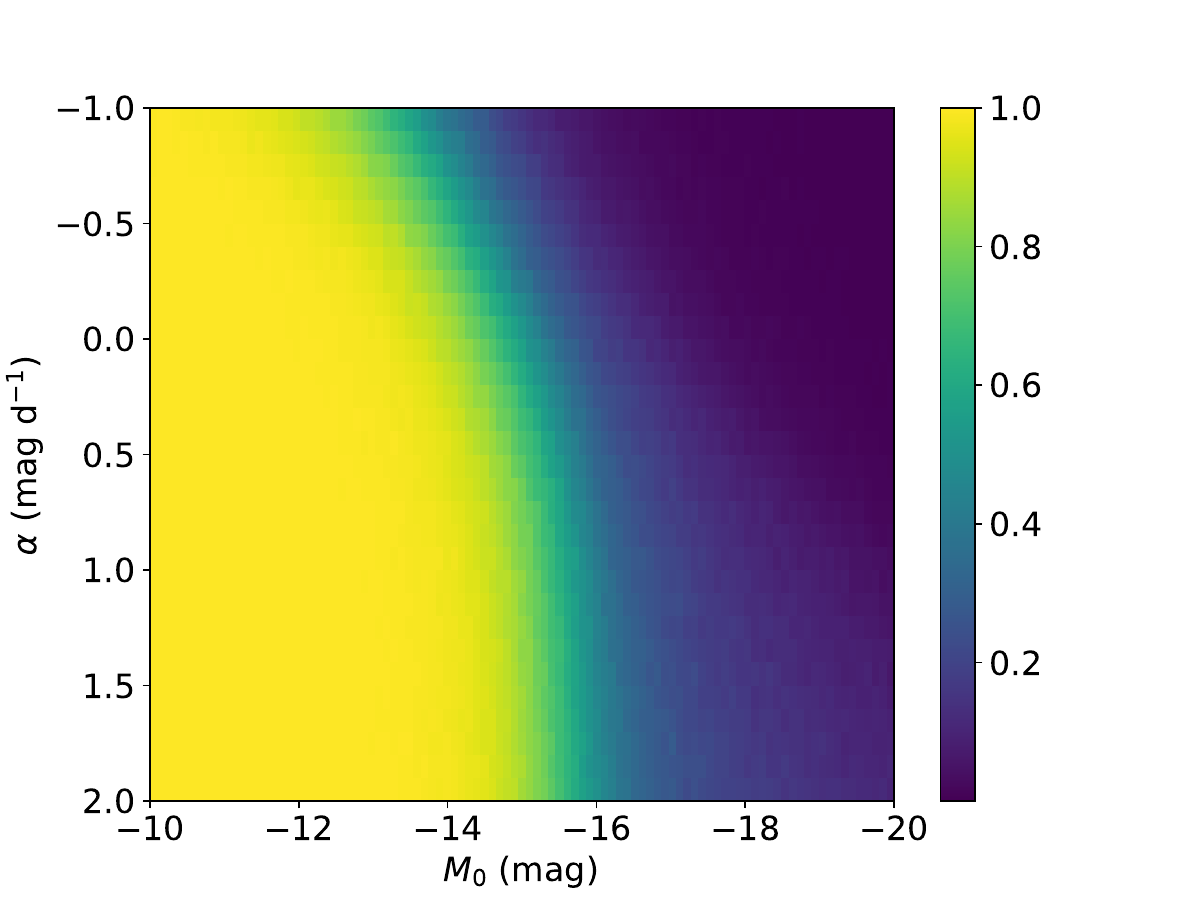}
    \caption{Non-detection probability 1-$\epsilon_i$, where $\epsilon_i$ is the recovery efficiency (number recovered / number injected) at a given absolute magnitude and evolution rate, for the grid of KNe simulated and recovered using {\tt simsurvey}. We simulate 100,000 KNe in each bin, with magnitudes ranging from -10.0 to -20.0\,mag and evolution rates ranging from -1.0 to 2.0\, mag day$^{-1}$.}
    \label{fig:InjectionRecovery}
\end{figure}

In order to benchmark our results against those from complementary methods in the literature, we compare the limits we obtain via \nimbus\ with those from the open-source simulator software {\tt simsurvey}\footnote{\url{https://github.com/ZwickyTransientFacility/simsurvey}} \citep{Feindt_2019}. {\tt simsurvey} simulates KN detections (or injections) based on survey limits for any given event \citep{Carracedo2021}. We estimate the efficiency (or probability) of detecting a KN of a given initial absolute magnitude and linear evolution rate by comparing detections with the total injections within the observed fields. The software takes as input the ZTF pointings and information (i.e., the observation time, limiting magnitude, filters, right ascension and declination for each field and CCD) for the first three days after the merger, and the 3D GW skymap for any given GW event. We simulate 100,000 KNe for a given absolute magnitude and evolution rate throughout the 3D GW probability region (see Fig.~\ref{fig:InjectionRecovery}). We assume a linear, colorless lightcurve model as stated in Sec. \ref{model}. Our detection criteria requires the KN to be detected at least once by ZTF: given actual detection experiments this is a necessary but likely insufficient criterion for identification, since both color information and evolution rate are needed to separate KNe from false positives \citep{Andreoni2020}. For example, the gamma-ray burst afterglows that have been discovered in the past with the ZTF Realtime Search and Triggering (ZTFReST; \citealt{ZTFRest2021}) pipeline have exhibited rapid evolution and reddening, requiring detections in both $g$- and $r$-bands, with $\geq$2 detections in either band for solid identification. Using {\tt simsurvey} we account for Milky Way extinction and exclude any KNe with $E(B-V)>2$\,mag. This process is repeated for a range of magnitudes (100 bins between $-$10 and $-$20\,mag) and evolution rates (31 bins from $-$1 and 2\,mags per day) resulting in a grid of efficiencies. This grid of efficiencies are then converted into non-detection probabilities (see Fig.~\ref{fig:InjectionRecovery}).

\begin{figure}[h]
    \plotone{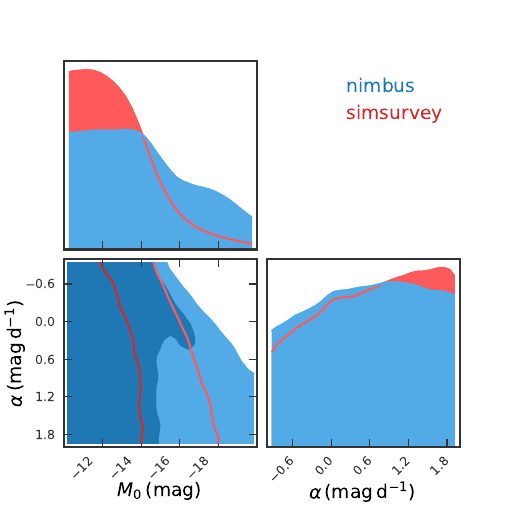}
    \caption{Comparison between \nimbus\ and {\tt simsurvey} with uniform priors. The corner plots compare the 2-D and corresponding 1-D marginalized posterior distributions for \nimbus\ (blue) against the normalized non-detection probabilities from {\tt simsurvey} (red). The 68\% and 95\% contours indicated on the plot demonstrate consistency between the two formalisms. We assume the same uniform priors in magnitude and evolution rate as for Fig.~\ref{fig1}.}
    \label{fig3}
\end{figure}

As this is a non-detection study, \nimbus\ generates posterior probabilities for models that are consistent with non-detection using observational upper limits; we compare the posterior support for models from \nimbus\ with the detection efficiency estimates for the same models from {\tt simsurvey}.
We normalize the non-detection probabilities in the {\tt simsurvey} model grid to sum to 1 in order to compare against \nimbus\ . In Fig.~\ref{fig3}, we show the 2-D and 1-D marginalized posterior distributions for the two light curve parameters from these formalisms. 


We note that for {\tt simsurvey}, the non-detection probability is calculated as $1.0 - \epsilon_i$, where $\epsilon_i$ is the recovery efficiency for a KN with a given absolute magnitude and evolution rate.
Therefore, it naturally follows that as the initial absolute magnitude gets dimmer, our constraints get progressively worse. Likewise, going from rising to fast-fading KN models, the constraints become weaker.

In general, these comparisons illustrate consistency between the constraints inferred by the two methods on the brighter edge of the initial magnitude distribution, as evidenced by the 2-D posterior in Fig.~\ref{fig3}.
We observe a large overlap in the 1-D marginalized posteriors for evolution rate for both formalisms. In both our hierarchical Bayesian formalism and the frequentist simulation-based approach, we observe that as the model evolution rate changes from $-1.0$\,mag day$^{-1}$ to $\sim 1.0$\,mag day$^{-1}$, constraints on KN models get progressively weaker. A rising KN is disfavored for values of $M_{0}$ on the brighter end of the initial magnitude range, as the transient would be brighter than these survey magnitudes. However, for evolution rates $>$1\,mag day$^{-1}$, the effects of normalization in \nimbus\ (see Sec. \ref{normalization-choice}) take into account that the survey limiting magnitudes lend more support to faster-decaying models, while the nightly limits themselves place nearly no constraints in this region of parameter space, leading to conservative estimates in the posterior curve relative to {\tt simsurvey}. The 1-D magnitude posteriors reveal that \nimbus\ has broader support for KN models of varying absolute magnitudes (plateauing around M$\gtrsim-$15) and more conservative constraints compared to {\tt simsurvey} for the brightest initial magnitudes.

Fundamentally, these two approaches are different but complementary. The {\tt simsurvey} analysis yields the probability of not detecting a KN with ($M_{0}$, $\alpha$) given the observations. \nimbus\ gives the posterior probability for a KN with ($M_{0}$, $\alpha$) that \textit{survives} the upper limits. Thus comparisons between these two approaches discussed here are analogous, but not exact. Note that while the results from {\tt simsurvey} here might seem similar to those obtained in Fig. \ref{fig1} using the {\tt mlim-distance} method of normalization, we emphasize that our preferred results using the {\tt mlim-survey} method are more realistic in that they use the actual observed range of ZTF limiting magnitudes from the follow-up of GW190425. As stated before, the {\tt mlim-distance} method uses the entire range of viable distances from the 3D GW skymap. One reason the {\tt simsurvey} results could be similar to this method is that the {\tt simsurvey} method also performs simulations in the entire region of the skymap based on the GW distance posterior. In order to better understand our results from the two formalisms, we also compared model probabilities using data from a single field of observation. Our results in this case show greater agreement indicating that the differences in the main results arise from the fundamentally different treatment of combining multiple fields with varying upper limits, luminosity distance distributions and different methods of model normalization.

\subsection{Astrophysical Priors} \label{astro-prior}

Variations in the masses, velocities, composition of the ejecta and inclination angle of the binary system result in different observed KN morphologies. BNS mass ejection mechanisms are categorized into two broad classes, i.e., dynamical ejecta and post-merger or wind ejecta \citep{nakar2019electromagnetic}. The tidal mass ejection occurring within $\sim$10 ms of the final inspiral stage is referred to as the dynamical ejecta. Bound NS material, which forms an accretion disk around the merger remnant, releases an outflow termed as the wind ejecta due to magnetically-driven, disk and neutrino winds. 

\begin{figure*}[t]
    \plottwo{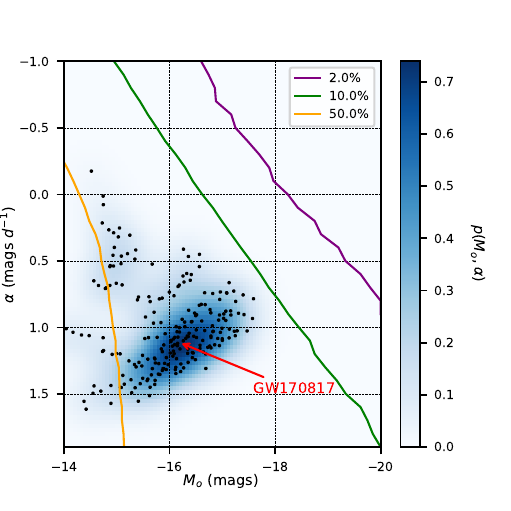}{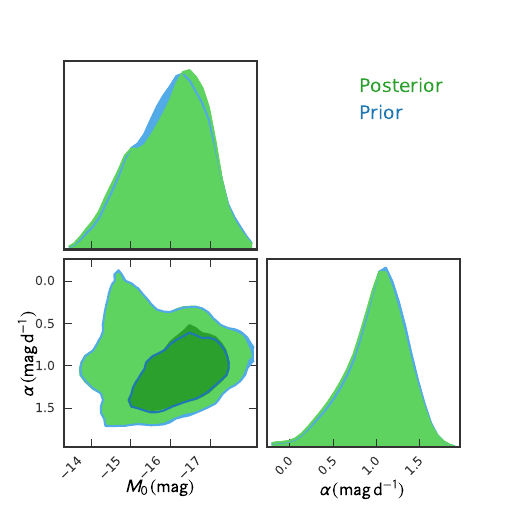}
    \caption{(left) Bolometric priors, informed by radiative-transport based KNe models \citep{Bulla_2019, DiCo2020}, showing regions of parameter space where particular luminosities and evolution rates are most probable or improbable for BNS KNe. To guide the eye, the best fit model for GW170817 is highlighted. The non-detection percentiles (solid lines) are calculated using {\tt simsurvey}, as discussed in Sec. \ref{astro-prior}. (right) Corner plot showing the 2-D and corresponding 1-D marginalized posterior distributions (green) using model based priors (blue).Contours indicate $68\%$ and $95\%$ confidence regions.}
    \label{fig4}
\end{figure*}

Using priors inspired from realistic astrophysical models of KNe based on simulations, we present our Bayesian constraints with GW190425 in Fig.~\ref{fig4}. These priors are derived from surrogate models~\citep{Coughlin_2018} trained on the outputs of the Monte Carlo Radiative-transfer code \textsc{possis} \citep{Bulla_2019}. Previous studies have underscored the importance of using astrophysical lightcurve priors in interpreting the emission from GW190425 \citep{barbieri2020distinguishing, Foley2020, Kyutoku2020, Dudi2021, Nicholl2021, raaijmakers2021challenges}.
Broadly speaking, the surrogate models, otherwise referred to as phenomenological models, use a machine learning technique to interpolate between data points. 
In this paper, we use a suite of 2D KN models assuming a three-component ejecta geometry, with dynamical ejecta split between equatorial lanthanide-rich and a polar lanthanide-poor components, and a spherical disk-wind ejecta component at lower velocities and with compositions intermediate to lanthanide-rich and lanthanide-poor \citep{DiCo2020}. 
The simulations cover four parameters: the inclination or the observer viewing angle ($\theta_{\rm obs}$), dynamical ejecta mass ($M_{\rm dyn}$), post-merger or wind ejecta mass ($M_{\rm wind}$), and half-opening angle for the lanthanide-rich dynamical ejecta component ($\phi$).

We assume a $\phi=30^{\circ}$ half opening angle and vary the other three parameters. Using the surrogate models, we predict BNS KN light curves for 10 viewing angles from a polar ($\theta_{\rm obs} = 0^{\circ}$) to an equatorial ($\theta_{\rm obs} = 90^{\circ}$) orientation, equally spaced in $\rm cos(\theta)$,
and for the following ejecta masses: $M_{\rm dyn}$ $\in [0.001,0.005,0.01,0.02]$ and $M_{\rm wind}$ $\in [0.01, 0.11 ]$ in steps of 0.02 $M_{\odot}$. In total, there are 240 BNS KN models. We then map from KN source properties (e.g., ejecta mass and inclination angle) to observables (peak magnitude and evolution rate) by performing lightcurve fits.


A typical simulated BNS KN rapidly rises to a maximum within a day or two and gradually decays. Since the decaying period dominates, we fit a linear model from the entire grid's median phase of the peak magnitude up to three days since the merger for $g$-, $r$- and $i$-band KN lightcurves. We omitted 20 models that had a mean squared error greater than 0.1 from this process. These omitted models specifically had low $M_{\rm dyn}$ but high $M_{\rm wind}$. 
For each combination of source parameters, we then combine $g$-, $r$- and $i$-band peak magnitudes and evolution rates based on the number of simulated KNe in \texttt{simsurvey} that fall in the ZTF observed region of any specific filter. For GW190425, we have about 2614, 2930, and 168 out of 100,000 simulated KNe that fall in the region observed with  $g$-, $r$- and $i$-band ZTF filters respectively. Gathering the grid-based values for peak absolute magnitudes and decay rates of KNe, we use kernel density estimation to construct a smooth probability density function and approximate the true distribution for the models considered. The left panel of Fig.~\ref{fig4} shows this grid-based prior model overlaid with {\tt simsurvey} non-detection probabilities.

Effectively, the astrophysical prior reduces a large portion of the parameter space previously considered by the uniform prior choice. The 2\%, 10\%, and 50\% non-detection percentiles estimated using {\tt simsurvey} are plotted over the astrophysical priors. Though the probable prior region is mostly encompassed by the 50\% curve, it lies entirely to the left of the 2\% and 10\% curves, indicating that our GW190425 ZTF observations are not deep enough to place stringent constraints over the astrophysical priors. This is reflected in the inference results from \nimbus\ (right panel of Fig.~\ref{fig4}) based on the astrophysical prior assumptions. Specifically, the  posterior and prior contour lines overlap considerably, showing that almost all of the KN models supported by the astrophysical prior survive the upper limits. In general, this implementation demonstrates the ability to use theory and simulation-based astrophysical models within the \nimbus \, framework and constrain them.

\begin{deluxetable}{ccc}
\tablecaption{$90\%$ upper limit of the kilonova initial absolute magnitude ($M_{0}$) for the different prior choices in Sec.\ref{sec:GW190425} }
\tablenum{1}
\label{tab:table1}
\tablehead{\colhead{Prior Choice} & \colhead{$M_{0}^{90 \%}$} \\ 
\colhead{} & \colhead{(mag)}} 

\startdata
Uniform-mlim & -16.63 \\
Uniform-distance & -15.01 \\
Astrophysical & -17.08\\
\enddata
\end{deluxetable}



\section{Discussion and Future Outlook} \label{discussion}
In this paper, we have presented a hierarchical Bayesian framework \nimbus\ that leverages data from non-detections of probable KNe. This Python package utilizes GW and EM follow-up information from each candidate event to provide posterior distributions for KN model parameters. The framework also accounts for the probability of
an event being astrophysical. Although the analysis presented here focuses on a single BNS event, GW190425, the framework has the capability to include multiple events. It is also straightforward to extend the framework to model the inference based on sub-populations of KN candidates such as BNS and NSBH mergers. We hope the multi-messenger astrophysics community finds use for and benefits from this package.

A current limitation of this study is that the framework does not account for events that have been detected in EM follow-up. In order to place stringent constraints on KN parameters, it would be ideal to include potential candidates for which there exists data from detected light curves. However, including information from detected events into the framework would involve non-trivial changes to the model likelihood and would necessitate an accurate understanding of survey selection effects. So far, GW170817 \citep{LVCGW170817} is the only GW event to have been associated with a KN counterpart \citep{KNGW170817}. Numerous studies in the literature \citep[e.g.,][]{2017gfoCowperthwaite,2017gfoDrout, Arcavi2018, Andreoni2020} have extrapolated follow-up data to arrive at an estimate of its initial absolute magnitude and decay rate. In particular, \citealt{kasliwal2020kilonova} compared their results to an extrapolated initial magnitude of $-16.6$ mag with a decay rate of $1$ mag per day. Our results do indicate non-negligible posterior support for such a model. Nevertheless, this represents a single point in model parameter space and we would instead require a number of detected events to inform the population of KNe. Further, restricting the study to non-detections is motivated by the fact that O3 did not yield any obvious EM counterparts. We defer the development of including detected events to a future study.

We presented results of our inference on GW190425 using two different prior choices for our model parameters (see Figs.~\ref{fig1} and \ref{fig4}). Our first choice, which is uniform in the parameters, is representative of an inference that is carried out with uninformative assumptions. Our second prior choice is based on surrogate models from Monte-Carlo radiative transfer simulations of KNe \citep{Coughlin_2018,Bulla_2019} and takes into account the effect of variations in ejecta masses and inclination angle on the resulting KN morphology. The inference using such a prior represents the possibility of testing realistic, physical models of KNe against upper limits obtained from surveys. While our implementation with uniform priors constrained the prior parameter space to a considerable extent and shows consistency with previous efforts \citep{kasliwal2020kilonova}, the posterior results based on surrogate KNe models are largely uninformative with respect to the prior. Overall, these results show how priors on model parameters can influence the constraints obtained and the need to examine results in light of the prior distribution.

One of the assumptions we have made in presenting our results above is that the KN counterpart to GW190425 is localized within the surveyed region of the skymap. GW190425, as mentioned previously in Sec.~\ref{sec:GW190425}, had an overall sky coverage by ZTF of $\sim 32\%$. Given that a significant fraction of the skymap is not surveyed and therefore would result in uninformative constraints, we made this assumption to demonstrate the utility of the Bayesian framework. In LIGO-Virgo-KAGRA's fourth observing run, we expect that for $\sim$8-10\% of BNS and NSBH systems discovered, ZTF will be able to observe $>$90\% of the localization \citep{PetrovSinger2021}, and hence our above assumption would hold reasonably true in those particular cases.

Furthermore, throughout this study we have assumed a kilonova luminosity evolution model that is linear in time. A linear model only needs two parameters to define it and our goal in this paper has been to demonstrate framework functionality at the cost of model accuracy. Such a simplistic choice might not be representative of realistic evolution models (see \citealt{Metzger_2019}) that depend on more complex parameters related to the binary system. In principle, it should be feasible to include arbitrary models for luminosity evolution since the framework only expects a function that returns predictions for the absolute magnitude of the KN as a function of time. For all priors applied, we consider only the first three days of evolution of the KN (and therefore the first three days of observations after the merger time of GW190425). This choice is motivated by the fact that ZTF is unlikely to detect a KN at the distance of GW190425 in the $g$- and $r$-bands after three days post-merger. More specifically, at four days, all KN models in our set have an apparent magnitude in the $g$- and $r$-bands fainter than the median depth of ZTF in this study ($\sim$\,21\,mag).

In this work we also neglected color evolution in our studies of kilonova non-detections. In addition to differentiating observations in different filters, in the future we intend to account for the K-correction effect on kilonova color evolution which is especially relevant for cosmological sources. We will implement this feature in \nimbus\ following the existing implementation in {\tt simsurvey}.

We highlight here that due to the adaptability of \nimbus\ to various lightcurve models and a hierarchical framework, it could even be used to jointly constrain the properties of a potential kilonova and short GRB optical afterglow associated with the GW event (as in \citealt{DiCo2020}, Pang et al. in prep) based on the rapid optical follow-up performed by various facilities.

In order to establish consistency with existing results in the literature, we compared our results to those from the simulator software {\tt simsurvey}. As shown in Fig.~\ref{fig3}, the two formalisms are largely consistent although some qualitative differences exist. In the future, with observations and upper limits from more events, it will be possible to test for further consistency between frameworks investigating KN populations.


In our specific implementation with astrophysical priors in Sec.~\ref{uniform-prior}, we used a prior on the KN luminosity parameters, i.e. the initial absolute magnitude and evolution rate that depends on intrinsic parameters such as the dynamical or wind ejecta masses. Our Bayesian approach makes it straightforward to convert our posteriors on the luminosity parameters into constraints on these intrinsic parameters. Alternatively, since the framework is agnostic to the KN model used, it should be possible to directly use priors on the physical parameters that govern the light-curve morphology. In such a case, the inference would directly constrain parameters such as the ejecta mass from the binary merger, although the computational feasibility of such an implementation needs to be investigated. 
The use of these astrophysical priors is based on including variations in the observer viewing (inclination) angle and its effect on each KN model. Non-trivial couplings between the observer angle and the signal-to-noise ratio of the GW signal can lead to some selection bias. To mitigate this effect in the future, we will select skymaps from a realistic distribution of GW signals detected by LIGO \citep{PetrovSinger2021} which will inform the distribution of observer angles for our kilonova models.

Looking forward to O4 and beyond, we expect that \nimbus\ will be an important framework for analyzing joint EM-GW observations. \citealt{PetrovSinger2021} predict a median of $\sim$35 BNS yr$^{-1}$ with O4 sensitivity and roughly double the number during O5. Thus, with several tens of EM follow-ups of BNS events from O4 and O5, we will use \nimbus\ to place stringent population-level luminosity function constraints based on non-detections. For the well-localized ($\lesssim100$\,deg$^2$) and nearby ($\lesssim200$\,Mpc) events (predicted to range from 0-13\,BNS mergers; \citealt{PetrovSinger2021}) detected by LIGO in O4 for which we have excellent optical sky coverage, \nimbus\ is ideally placed to constrain the intrinsic kilonova properties which can translate to constraints on binary system parameters such as mass ratio and NS radius in the face of non-detection. A similar analysis of NSBH mergers is also feasible, though ejecta mass yield sensitively depends on the mass ratio of the system (e.g. \citealt{KrugerFoucart2020}). As a follow-up study, we hope to explore the scientific merit of conducting EM-GW follow-ups with the Vera C. Rubin Observatory, assuming the cadence and filter strategy for KN identification outlined in \citep{Andreoni2021_VRO}, using the \nimbus\ framework.

Constraining the ejecta masses of the KN population could potentially provide us better insights into the amount of $r$-process material contributed to the formation of KNe \citep{Hotokezaka2018}. It will also help in understanding the relationship and breaking the degeneracy that exists between binary parameters (equation of state, spin and mass ratio) \citep{FoHiNi2018, Hinderer2019,  RadiceDai2019, Zhu2020NSBH}, ejecta mass and KN light curve morphology. \citep{Coughlin_2019b, Hotokezaka2019, Breschi2021, raaijmakers2021challenges}. The future GW and EM multi-messenger landscape will provide the opportunity to explore this further.  
\begin{acknowledgements}
Based on observations obtained with the Samuel Oschin Telescope 48-inch and the 60-inch Telescope at the Palomar Observatory as part of the Zwicky Transient Facility project. ZTF is supported by the National Science Foundation under Grant No. AST-1440341 and a collaboration including Caltech, IPAC, the Weizmann Institute for Science, the Oskar Klein Center at Stockholm University, the University of Maryland, the University of Washington, Deutsches Elektronen-Synchrotron and Humboldt University, Los Alamos National Laboratories, the TANGO Consortium of Taiwan, the University of Wisconsin at Milwaukee, and Lawrence Berkeley National Laboratories. Operations are conducted by COO, IPAC, and UW. This work was supported by the GROWTH (Global Relay of Observatories Watching Transients Happen) project funded by the National Science Foundation under PIRE Grant No 1545949. GROWTH is a collaborative project among California Institute of Technology (USA), University of Maryland College Park (USA), University of Wisconsin Milwaukee (USA), Texas Tech University (USA), San Diego State University (USA), University of Washington (USA), Los Alamos National Laboratory (USA), Tokyo Institute of Technology (Japan), National Central University (Taiwan), Indian Institute of Astrophysics (India), Indian Institute of Technology Bombay (India), Weizmann Institute of Science (Israel), The Oskar Klein Centre at Stockholm University (Sweden), Humboldt University (Germany), Liverpool John Moores University (UK) and University of Sydney (Australia). SRM thanks the LSSTC Data Science Fellowship Program, which is funded by LSSTC, NSF Cybertraining Grant \#1829740, the Brinson Foundation, and the Moore Foundation; his participation in the program has benefited this work. SRM and JC acknowledge support from the NSF grant NSF PHY \#1912649. We are grateful for computational resources provided by the Leonard E Parker Center for Gravitation, Cosmology and Astrophysics at the University of Wisconsin-Milwaukee. MMK acknowledges generous support from the David and Lucille Packard Foundation. M.C. and M.S. acknowledges support from the National Science Foundation with grant number PHY-2010970. S.A. acknowledges support from the GROWTH PIRE Grant No 1545949. A.S.C acknowledges support from the G.R.E.A.T research environment, funded by {\em Vetenskapsr\aa det},  the Swedish Research Council, project number 2016-06012. M.B. acknowledges support from the Swedish Research Council (Reg. no. 2020-03330). We thank the reviewer whose comments and suggestions helped improve and clarify this manuscript.
\end{acknowledgements}

\software{{\tt ipython} \citep{ipython}, {\tt jupyter} \citep{jupyter}, {\tt matplotlib} \citep{matplotlib}, {\tt python} \citep{python3}, {\tt NumPy} \citep{numpy}, {\tt scikit-learn} \citep{scikit-learn}, {\tt scipy} \citep{scipy}}

\facilities{LIGO, ZTF/PO:1.2m}

\bibliography{references}

\begin{thebibliography}{}
\expandafter\ifx\csname natexlab\endcsname\relax\def\natexlab#1{#1}\fi

\bibitem[{Aasi {et~al.}(2015)Aasi, Abbott, Abbott, Abbott, Abernathy, Ackley,
  Adams, Adams, Addesso, Adhikari, Adya, Affeldt, Aggarwal, Aguiar, Ain, Ajith,
  Alemic, Allen, Amariutei, Anderson, Anderson, Arai, Araya, Arceneaux, Areeda,
  Ashton, Ast, Aston, Aufmuth, Aulbert, Aylott, Babak, Baker, Ballmer,
  Barayoga, Barbet, Barclay, Barish, Barker, Barr, Barsotti, Bartlett, Barton,
  Bartos, Bassiri, Batch, Baune, Behnke, Bell, Bell, Benacquista, Bergman,
  Bergmann, Berry, Betzwieser, Bhagwat, Bhandare, Bilenko, Billingsley, Birch,
  Biscans, Biwer, Blackburn, Blackburn, Blair, Blair, Bock, Bodiya, Bojtos,
  Bond, Bork, Born, Bose, Brady, Braginsky, Brau, Bridges, Brinkmann, Brooks,
  Brown, Brown, Brown, Buchman, Buikema, Buonanno, Cadonati, Bustillo, Camp,
  Cannon, Cao, Capano, Caride, Caudill, Cavagli{\`{a}}, Cepeda, Chakraborty,
  Chalermsongsak, Chamberlin, Chao, Charlton, Chen, Cho, Cho, Chow,
  Christensen, Chu, Chung, Ciani, Clara, Clark, Collette, Cominsky, Constancio,
  Cook, Corbitt, Cornish, Corsi, Costa, Coughlin, Countryman, Couvares, Coward,
  Cowart, Coyne, Coyne, Craig, Creighton, Creighton, Cripe, Crowder, Cumming,
  Cunningham, Cutler, Dahl, Canton, Damjanic, Danilishin, Danzmann, Dartez,
  Dave, Daveloza, Davies, Daw, DeBra, Pozzo, Denker, Dent, Dergachev, DeRosa,
  DeSalvo, Dhurandhar, D{\'D}{{\i}}az, Palma, Dojcinoski, Dominguez, Donovan,
  Dooley, Doravari, Douglas, Downes, Driggers, Du, Dwyer, Eberle, Edo, Edwards,
  Edwards, Effler, Eggenstein, Ehrens, Eichholz, Eikenberry, Essick, Etzel,
  Evans, Evans, Factourovich, Fairhurst, Fan, Fang, Farr, Farr, Favata, Fays,
  Fehrmann, Fejer, Feldbaum, Ferreira, Fisher, Frei, Freise, Frey, Fricke,
  Fritschel, Frolov, Fuentes-Tapia, Fulda, Fyffe, Gair, Gaonkar, Gehrels,
  Gergely, Giaime, Giardina, Gleason, Goetz, Goetz, Gondan, Gonz{\'{a}}lez,
  Gordon, Gorodetsky, Gossan, Go{\ss}ler, Gräf, Graff, Grant, Gras, Gray,
  Greenhalgh, Gretarsson, Grote, Grunewald, Guido, Guo, Gushwa, Gustafson,
  Gustafson, Hacker, Hall, Hammond, Hanke, Hanks, Hanna, Hannam, Hanson,
  Hardwick, Harry, Harry, Hart, Hartman, Haster, Haughian, Hee, Heintze,
  Heinzel, Hendry, Heng, Heptonstall, Heurs, Hewitson, Hild, Hoak, Hodge,
  Hollitt, Holt, Hopkins, Hosken, Hough, Houston, Howell, Hu, Huerta, Hughey,
  Husa, Huttner, Huynh, Huynh-Dinh, Idrisy, Indik, Ingram, Inta, Islas, Isler,
  Isogai, Iyer, Izumi, Jacobson, Jang, Jawahar, Ji, Jim{\'{e}}nez-Forteza,
  Johnson, Jones, Jones, Ju, Haris, Kalogera, Kandhasamy, Kang, Kanner,
  Katsavounidis, Katzman, Kaufer, Kaufer, Kaur, Kawabe, Kawazoe, Keiser,
  Keitel, Kelley, Kells, Keppel, Key, Khalaidovski, Khalili, Khazanov, Kim,
  Kim, Kim, Kim, Kim, King, King, Kinzel, Kissel, Klimenko, Kline, Koehlenbeck,
  Kokeyama, Kondrashov, Korobko, Korth, Kozak, Kringel, Krishnan, Krueger,
  Kuehn, Kumar, Kumar, Kuo, Landry, Lantz, Larson, Lasky, Lazzarini, Lazzaro,
  Le, Leaci, Leavey, Lebigot, Lee, Lee, Lee, Leong, Levin, Levine, Lewis, Li,
  Libbrecht, Libson, Lin, Littenberg, Lockerbie, Lockett, Logue, Lombardi,
  Lormand, Lough, Lubinski, Lück, Lundgren, Lynch, Ma, Macarthur, MacDonald,
  Machenschalk, MacInnis, Macleod, Maga{\~{n}}a-Sandoval, Magee, Mageswaran,
  Maglione, Mailand, Mandel, Mandic, Mangano, Mansell, M{\'{a}}rka,
  M{\'{a}}rka, Markosyan, Maros, Martin, Martin, Martynov, Marx, Mason,
  Massinger, Matichard, Matone, Mavalvala, Mazumder, Mazzolo, McCarthy,
  McClelland, McCormick, McGuire, McIntyre, McIver, McLin, McWilliams, Meadors,
  Meinders, Melatos, Mendell, Mercer, Meshkov, Messenger, Meyers, Miao,
  Middleton, Mikhailov, Miller, Miller, Millhouse, Ming, Mirshekari, Mishra,
  Mitra, Mitrofanov, Mitselmakher, Mittleman, Moe, Mohanty, Mohapatra, Moore,
  Moraru, Moreno, Morriss, Mossavi, Mow-Lowry, Mueller, Mueller, Mukherjee,
  Mullavey, Munch, Murphy, Murray, Mytidis, Nash, Nayak, Necula, Nedkova,
  Newton, Nguyen, Nielsen, Nissanke, Nitz, Nolting, Normandin, Nuttall,
  Ochsner, O'Dell, Oelker, Ogin, Oh, Oh, Ohme, Oppermann, Oram, O'Reilly,
  Ortega, O'Shaughnessy, Osthelder, Ott, Ottaway, Ottens, Overmier, Owen,
  Padilla, Pai, Pai, Palashov, Pal-Singh, Pan, Pankow, Pannarale, Pant, Papa,
  Paris, Patrick, Pedraza, Pekowsky, Pele, Penn, Perreca, Phelps, Pierro,
  Pinto, Pitkin, Poeld, Post, Poteomkin, Powell, Prasad, Predoi, Premachandra,
  Prestegard, Price, Principe, Privitera, Prix, Prokhorov, Puncken, Pürrer,
  Qin, Quetschke, Quintero, Quiroga, Quitzow-James, Raab, Rabeling, Radkins,
  Raffai, Raja, Rajalakshmi, Rakhmanov, Ramirez, Raymond, Reed, Reid, Reitze,
  Reula, Riles, Robertson, Robie, Rollins, Roma, Romano, Romanov, Romie, Rowan,
  Rüdiger, Ryan, Sachdev, Sadecki, Sadeghian, Saleem, Salemi, Sammut,
  Sandberg, Sanders, Sannibale, Santiago-Prieto, Sathyaprakash, Saulson,
  Savage, Sawadsky, Scheuer, Schilling, Schmidt, Schnabel, Schofield,
  Schreiber, Schuette, Schutz, Scott, Scott, Sellers, Sengupta, Sergeev, Serna,
  Sevigny, Shaddock, Shahriar, Shaltev, Shao, Shapiro, Shawhan, Shoemaker,
  Sidery, Siemens, Sigg, Silva, Simakov, Singer, Singer, Singh, Sintes,
  Slagmolen, Smith, Smith, Smith, Smith-Lefebvre, Son, Sorazu, Souradeep,
  Staley, Stebbins, Steinke, Steinlechner, Steinlechner, Steinmeyer, Stephens,
  Steplewski, Stevenson, Stone, Strain, Strigin, Sturani, Stuver, Summerscales,
  Sutton, Szczepanczyk, Szeifert, Talukder, Tanner, T{\'{a}}pai, Tarabrin,
  Taracchini, Taylor, Tellez, Theeg, Thirugnanasambandam, Thomas, Thomas,
  Thorne, Thorne, Thrane, Tiwari, Tomlinson, Torres, Torrie, Traylor, Tse,
  Tshilumba, Ugolini, Unnikrishnan, Urban, Usman, Vahlbruch, Vajente, Valdes,
  Vallisneri, van Veggel, Vass, Vaulin, Vecchio, Veitch, Veitch, Venkateswara,
  Vincent-Finley, Vitale, Vo, Vorvick, Vousden, Vyatchanin, Wade, Wade, Wade,
  Walker, Wallace, Walsh, Wang, Wang, Wang, Ward, Warner, Was, Weaver, Weinert,
  Weinstein, Weiss, Welborn, Wen, Wessels, Westphal, Wette, Whelan, Whitcomb,
  White, Whiting, Wilkinson, Williams, Williams, Williamson, Willis, Willke,
  Wimmer, Winkler, Wipf, Wittel, Woan, Worden, Xie, Yablon, Yakushin, Yam,
  Yamamoto, Yancey, Yang, Zanolin, Zhang, Zhang, Zhang, Zhang, Zhao, Zhou, Zhu,
  Zucker, Zuraw, \& Zweizig}]{aLIGO2015}
Aasi, J., Abbott, B.~P., Abbott, R., {et~al.} 2015, Classical and Quantum
  Gravity, 32, 074001

\bibitem[{Abbott {et~al.}(2017{\natexlab{a}})Abbott, Abbott, Abbott, Acernese,
  Ackley, Adams, Adams, Addesso, Adhikari, Adya, Affeldt, Afrough, Agarwal,
  Agathos, Agatsuma, Aggarwal, Aguiar, Aiello, Ain, Ajith, Allen, Allen,
  Allocca, Altin, Amato, Ananyeva, Anderson, Anderson, Angelova, Antier,
  Appert, Arai, Araya, Areeda, Arnaud, Arun, Ascenzi, Ashton, Ast, Aston,
  Astone, Atallah, Aufmuth, Aulbert, AultONeal, Austin, Avila-Alvarez, Babak,
  Bacon, Bader, Bae, Baker, Baldaccini, Ballardin, Banagiri, Barayoga, Barclay,
  Barish, Barker, Barkett, Barone, Barr, Barsotti, Barsuglia, Barta, Bartlett,
  Bartos, Bassiri, Basti, Batch, Bawaj, Bayley, Bazzan, B{\'{e}}csy, Beer,
  Bejger, Belahcene, Bell, Bergmann, Bernuzzi, Bero, Berry, Bersanetti,
  Bertolini, Betzwieser, Bhagwat, Bhandare, Bilenko, Billingsley, Billman,
  Birch, Birney, Birnholtz, Biscans, Biscoveanu, Bisht, Bitossi, Biwer,
  Bizouard, Blackburn, Blackman, Blair, Blair, Blair, Bloemen, Bock, Bode,
  Boer, Bogaert, Bohe, Bondu, Bonilla, Bonnand, Boom, Bork, Boschi, Bose,
  Bossie, Bouffanais, Bozzi, Bradaschia, Brady, Branchesi, Brau, Briant,
  Brillet, Brinkmann, Brisson, Brockill, Broida, Brooks, Brown, Brunett,
  Buchanan, Buikema, Bulik, Bulten, Buonanno, Buskulic, Buy, Byer, Cabero,
  Cadonati, Cagnoli, Cahillane, Bustillo, Callister, Calloni, Camp, Canepa,
  Canizares, Cannon, Cao, Cao, Capano, Capocasa, Carbognani, Caride, Carney,
  Diaz, Casentini, Caudill, Cavagli{\`{a}}, Cavalier, Cavalieri, Cella, Cepeda,
  Cerd{\'{a}}-Dur{\'{a}}n, Cerretani, Cesarini, Chamberlin, Chan, Chao,
  Charlton, Chase, Chassande-Mottin, Chatterjee, Chatziioannou, Cheeseboro,
  Chen, Chen, Chen, Cheng, Chia, Chincarini, Chiummo, Chmiel, Cho, Cho, Chow,
  Christensen, Chu, Chua, Chua, Chung, Chung, Ciani, Ciolfi, Cirelli, Cirone,
  Clara, Clark, Clearwater, Cleva, Cocchieri, Coccia, Cohadon, Cohen, Colla,
  Collette, Cominsky, Jr., Conti, Cooper, Corban, Corbitt,
  Cordero-Carri{\'{o}}n, Corley, Cornish, Corsi, Cortese, Costa, Coughlin,
  Coughlin, Coulon, Countryman, Couvares, Covas, Cowan, Coward, Cowart, Coyne,
  Coyne, Creighton, Creighton, Cripe, Crowder, Cullen, Cumming, Cunningham,
  Cuoco, Canton, D{\'{a}}lya, Danilishin, D'Antonio, Danzmann, Dasgupta, Costa,
  Dattilo, Dave, Davier, Davis, Daw, Day, De, DeBra, Degallaix, Laurentis,
  Del{\'{e}}glise, Pozzo, Demos, Denker, Dent, Pietri, Dergachev, Rosa, DeRosa,
  Rossi, DeSalvo, de~Varona, Devenson, Dhurandhar, D{\'{\i}}az, Dietrich,
  Fiore, Giovanni, Girolamo, Lieto, Pace, Palma, Renzo, Doctor, Dolique,
  Donovan, Dooley, Doravari, Dorrington, Douglas, {\'{A}}lvarez, Downes, Drago,
  Dreissigacker, Driggers, Du, Ducrot, Dupej, Dwyer, Edo, Edwards, Effler,
  Eggenstein, Ehrens, Eichholz, Eikenberry, Eisenstein, Essick, Estevez,
  Etienne, Etzel, Evans, Evans, Factourovich, Fafone, Fair, Fairhurst, Fan,
  Farinon, Farr, Farr, Fauchon-Jones, Favata, Fays, Fee, Fehrmann, Feicht,
  Fejer, Fernandez-Galiana, Ferrante, Ferreira, Ferrini, Fidecaro, Finstad,
  Fiori, Fiorucci, Fishbach, Fisher, Fitz-Axen, Flaminio, Fletcher, Fong, Font,
  Forsyth, Forsyth, Fournier, Frasca, Frasconi, Frei, Freise, Frey, Frey,
  Fries, Fritschel, Frolov, Fulda, Fyffe, Gabbard, Gadre, Gaebel, Gair,
  Gammaitoni, Ganija, Gaonkar, Garcia-Quiros, Garufi, Gateley, Gaudio, Gaur,
  Gayathri, Gehrels, Gemme, Genin, Gennai, George, George, Gergely, Germain,
  Ghonge, Ghosh, Ghosh, Ghosh, Giaime, Giardina, Giazotto, Gill, Glover, Goetz,
  Goetz, Gomes, Goncharov, Gonz{\'{a}}lez, Castro, Gopakumar, Gorodetsky,
  Gossan, Gosselin, Gouaty, Grado, Graef, Granata, Grant, Gras, Gray, Greco,
  Green, Gretarsson, Groot, Grote, Grunewald, Gruning, Guidi, Guo, Gupta,
  Gupta, Gushwa, Gustafson, Gustafson, Halim, Hall, Hall, Hamilton, Hammond,
  Haney, Hanke, Hanks, Hanna, Hannam, Hannuksela, Hanson, Hardwick, Harms,
  Harry, Harry, Hart, Haster, Haughian, Healy, Heidmann, Heintze, Heitmann,
  Hello, Hemming, Hendry, Heng, Hennig, Heptonstall, Heurs, Hild, Hinderer,
  Hoak, Hofman, Holt, Holz, Hopkins, Horst, Hough, Houston, Howell, Hreibi, Hu,
  Huerta, Huet, Hughey, Husa, Huttner, Huynh-Dinh, Indik, Inta, Intini, Isa,
  Isac, Isi, Iyer, Izumi, Jacqmin, Jani, Jaranowski, Jawahar,
  Jim{\'{e}}nez-Forteza, Johnson, Johnson-McDaniel, Jones, Jones, Jonker, Ju,
  Junker, Kalaghatgi, Kalogera, Kamai, Kandhasamy, Kang, Kanner, Kapadia,
  Karki, Karvinen, Kasprzack, Kastaun, Katolik, Katsavounidis, Katzman, Kaufer,
  Kawabe, Kawaguchi, K{\'{e}}f{\'{e}}lian, Keitel, Kemball, Kennedy, Kent, Key,
  Khalili, Khan, Khan, Khan, Khazanov, Kijbunchoo, Kim, Kim, Kim, Kim, Kim,
  Kim, Kimbrell, King, King, Kinley-Hanlon, Kirchhoff, Kissel, Kleybolte,
  Klimenko, Knowles, Koch, Koehlenbeck, Koley, Kondrashov, Kontos, Korobko,
  Korth, Kowalska, Kozak, Krämer, Kringel, Kr{\'{o}}lak, Kuehn, Kumar, Kumar,
  Kumar, Kuo, Kutynia, Kwang, Lackey, Lai, Landry, Lang, Lange, Lantz, Lanza,
  Larson, Lartaux-Vollard, Lasky, Laxen, Lazzarini, Lazzaro, Leaci, Leavey,
  Lee, Lee, Lee, Lee, Lee, Lehmann, Lenon, Leonardi, Leroy, Letendre, Levin,
  Li, Linker, Littenberg, Liu, Liu, Lo, Lockerbie, London, Lord, Lorenzini,
  Loriette, Lormand, Losurdo, Lough, Lousto, Lovelace, Lück, Lumaca, Lundgren,
  Lynch, Ma, Macas, Macfoy, Machenschalk, MacInnis, Macleod, Hernandez,
  Maga{\~{n}}a-Sandoval, Zertuche, Magee, Majorana, Maksimovic, Man, Mandic,
  Mangano, Mansell, Manske, Mantovani, Marchesoni, Marion, M{\'{a}}rka,
  M{\'{a}}rka, Markakis, Markosyan, Markowitz, Maros, Marquina, Martelli,
  Martellini, Martin, Martin, Martynov, Mason, Massera, Masserot, Massinger,
  Masso-Reid, Mastrogiovanni, Matas, Matichard, Matone, Mavalvala, Mazumder,
  McCarthy, McClelland, McCormick, McCuller, McGuire, McIntyre, McIver,
  McManus, McNeill, McRae, McWilliams, Meacher, Meadors, Mehmet, Meidam,
  Mejuto-Villa, Melatos, Mendell, Mercer, Merilh, Merzougui, Meshkov,
  Messenger, Messick, Metzdorff, Meyers, Miao, Michel, Middleton, Mikhailov,
  Milano, Miller, Miller, Miller, Millhouse, Milovich-Goff, Minazzoli,
  Minenkov, Ming, Mishra, Mitra, Mitrofanov, Mitselmakher, Mittleman, Moffa,
  Moggi, Mogushi, Mohan, Mohapatra, Montani, Moore, Moraru, Moreno, Morriss,
  Mours, Mow-Lowry, Mueller, Muir, Mukherjee, Mukherjee, Mukherjee, Mukund,
  Mullavey, Munch, Mu{\~{n}}iz, Muratore, Murray, Napier, Nardecchia,
  Naticchioni, Nayak, Neilson, Nelemans, Nelson, Nery, Neunzert, Nevin,
  Newport, Newton, Ng, Nguyen, Nichols, Nielsen, Nissanke, Nitz, Noack, Nocera,
  Nolting, North, Nuttall, Oberling, O'Dea, Ogin, Oh, Oh, Ohme, Okada, Oliver,
  Oppermann, Oram, O'Reilly, Ormiston, Ortega, O'Shaughnessy, Ossokine,
  Ottaway, Overmier, Owen, Pace, Page, Page, Pai, Pai, Palamos, Palashov,
  Palomba, Pal-Singh, Pan, Pan, Pang, Pang, Pankow, Pannarale, Pant, Paoletti,
  Paoli, Papa, Parida, Parker, Pascucci, Pasqualetti, Passaquieti, Passuello,
  Patil, Patricelli, Pearlstone, Pedraza, Pedurand, Pekowsky, Pele, Penn,
  Perez, Perreca, Perri, Pfeiffer, Phelps, Piccinni, Pichot, Piergiovanni,
  Pierro, Pillant, Pinard, Pinto, Pirello, Pitkin, Poe, Poggiani, Popolizio,
  Porter, Post, Powell, Prasad, Pratt, Pratten, Predoi, Prestegard, Prijatelj,
  Principe, Privitera, Prodi, Prokhorov, Puncken, Punturo, Puppo, Pürrer, Qi,
  Quetschke, Quintero, Quitzow-James, Rabeling, Radkins, Raffai, Raja, Rajan,
  Rajbhandari, Rakhmanov, Ramirez, Ramos-Buades, Rapagnani, Raymond, Razzano,
  Read, Regimbau, Rei, Reid, Reitze, Ren, Reyes, Ricci, Ricker, Rieger, Riles,
  Rizzo, Robertson, Robie, Robinet, Rocchi, Rolland, Rollins, Roma, Romano,
  Romel, Romie, Rosi{\'{n}}ska, Ross, Rowan, Rüdiger, Ruggi, Rutins, Ryan,
  Sachdev, Sadecki, Sadeghian, Sakellariadou, Salconi, Saleem, Salemi,
  Samajdar, Sammut, Sampson, Sanchez, Sanchez, Sanchis-Gual, Sandberg, Sanders,
  Sassolas, Sauter, Savage, Sawadsky, Schale, Scheel, Scheuer, Schmidt,
  Schmidt, Schnabel, Schofield, Schönbeck, Schreiber, Schuette, Schulte,
  Schutz, Schwalbe, Scott, Scott, Seidel, Sellers, Sengupta, Sentenac, Sequino,
  Sergeev, Shaddock, Shaffer, Shah, Shahriar, Shaner, Shao, Shapiro, Shawhan,
  Sheperd, Shoemaker, Shoemaker, Siellez, Siemens, Sieniawska, Sigg, Silva,
  Singer, Singh, Singhal, Sintes, Slagmolen, Smith, Smith, Smith, Somala, Son,
  Sonnenberg, Sorazu, Sorrentino, Souradeep, Spencer, Srivastava, Staats,
  Staley, Steinke, Steinlechner, Steinlechner, Steinmeyer, Stevenson, Stone,
  Stops, Strain, Stratta, Strigin, Strunk, Sturani, Stuver, Summerscales, Sun,
  Sunil, Suresh, Sutton, Swinkels, Szczepa{\'{n}}czyk, Tacca, Tait, Talbot,
  Talukder, Tanner, T{\'{a}}pai, Taracchini, Tasson, Taylor, Taylor, Tewari,
  Theeg, Thies, Thomas, Thomas, Thomas, Thorne, Thrane, Tiwari, Tiwari,
  Tokmakov, Toland, Tonelli, Tornasi, Torres-Forn{\'{e}}, Torrie, Töyrä,
  Travasso, Traylor, Trinastic, Tringali, Trozzo, Tsang, Tse, Tso, Tsukada,
  Tsuna, Tuyenbayev, Ueno, Ugolini, Unnikrishnan, Urban, Usman, Vahlbruch,
  Vajente, Valdes, van Bakel, van Beuzekom, van~den Brand, Broeck, Vander-Hyde,
  van~der Schaaf, van Heijningen, van Veggel, Vardaro, Varma, Vass,
  Vas{\'{u}}th, Vecchio, Vedovato, Veitch, Veitch, Venkateswara, Venugopalan,
  Verkindt, Vetrano, Vicer{\'{e}}, Viets, Vinciguerra, Vine, Vinet, Vitale, Vo,
  Vocca, Vorvick, Vyatchanin, Wade, Wade, Wade, Walet, Walker, Wallace, Walsh,
  Wang, Wang, Wang, Wang, Wang, Ward, Warner, Was, Watchi, Weaver, Wei,
  Weinert, Weinstein, Weiss, Wen, Wessel, We{\ss}els, Westerweck, Westphal,
  Wette, Whelan, Whiting, Whittle, Wilken, Williams, Williams, Williamson,
  Willis, Willke, Wimmer, Winkler, Wipf, Wittel, Woan, Woehler, Wofford, Wong,
  Worden, Wright, Wu, Wysocki, Xiao, Yamamoto, Yancey, Yang, Yap, Yazback, Yu,
  Yu, Yvert, Zadro{\.{z}}ny, Zanolin, Zelenova, Zendri, Zevin, Zhang, Zhang,
  Zhang, Zhang, Zhao, Zhou, Zhou, Zhu, Zhu, Zimmerman, Zucker, \&
  and}]{Abbott2017_ejecta}
Abbott, B.~P., Abbott, R., Abbott, T.~D., {et~al.} 2017{\natexlab{a}}, The
  Astrophysical Journal, 850, L39

\bibitem[{Abbott {et~al.}(2017{\natexlab{b}})Abbott, Abbott, Abbott, Acernese,
  Ackley, Adams, Adams, Addesso, Adhikari, Adya, Affeldt, Afrough, Agarwal,
  Agathos, Agatsuma, Aggarwal, Aguiar, Aiello, Ain, Ajith, Allen, Allen,
  Allocca, Altin, Amato, Ananyeva, Anderson, Anderson, Angelova, Antier,
  Appert, Arai, Araya, Areeda, Arnaud, Arun, Ascenzi, Ashton, Ast, Aston,
  Astone, Atallah, Aufmuth, Aulbert, AultONeal, Austin, Avila-Alvarez, Babak,
  Bacon, Bader, Bae, Bailes, Baker, Baldaccini, Ballardin, Ballmer, Banagiri,
  Barayoga, Barclay, Barish, Barker, Barkett, Barone, Barr, Barsotti,
  Barsuglia, Barta, Barthelmy, Bartlett, Bartos, Bassiri, Basti, Batch, Bawaj,
  Bayley, Bazzan, B\'ecsy, Beer, Bejger, Belahcene, Bell, Berger, Bergmann,
  Bernuzzi, Bero, Berry, Bersanetti, Bertolini, Betzwieser, Bhagwat, Bhandare,
  Bilenko, Billingsley, Billman, Birch, Birney, Birnholtz, Biscans, Biscoveanu,
  Bisht, Bitossi, Biwer, Bizouard, Blackburn, Blackman, Blair, Blair, Blair,
  Bloemen, Bock, Bode, Boer, Bogaert, Bohe, Bondu, Bonilla, Bonnand, Boom,
  Bork, Boschi, Bose, Bossie, Bouffanais, Bozzi, Bradaschia, Brady, Branchesi,
  Brau, Briant, Brillet, Brinkmann, Brisson, Brockill, Broida, Brooks, Brown,
  Brown, Brunett, Buchanan, Buikema, Bulik, Bulten, Buonanno, Buskulic, Buy,
  Byer, Cabero, Cadonati, Cagnoli, Cahillane, Calder\'on~Bustillo, Callister,
  Calloni, Camp, Canepa, Canizares, Cannon, Cao, Cao, Capano, Capocasa,
  Carbognani, Caride, Carney, Carullo, Casanueva~Diaz, Casentini, Caudill,
  Cavagli\`a, Cavalier, Cavalieri, Cella, Cepeda, Cerd\'a-Dur\'an, Cerretani,
  Cesarini, Chamberlin, Chan, Chao, Charlton, Chase, Chassande-Mottin,
  Chatterjee, Chatziioannou, Cheeseboro, Chen, Chen, Chen, Cheng, Chia,
  Chincarini, Chiummo, Chmiel, Cho, Cho, Chow, Christensen, Chu, Chua, Chua,
  Chung, Chung, Ciani, Ciolfi, Cirelli, Cirone, Clara, Clark, Clearwater,
  Cleva, Cocchieri, Coccia, Cohadon, Cohen, Colla, Collette, Cominsky,
  Constancio, Conti, Cooper, Corban, Corbitt, Cordero-Carri\'on, Corley,
  Cornish, Corsi, Cortese, Costa, Coughlin, Coughlin, Coulon, Countryman,
  Couvares, Covas, Cowan, Coward, Cowart, Coyne, Coyne, Creighton, Creighton,
  Cripe, Crowder, Cullen, Cumming, Cunningham, Cuoco, Dal~Canton, D\'alya,
  Danilishin, D'Antonio, Danzmann, Dasgupta, Da~Silva~Costa, Dattilo, Dave,
  Davier, Davis, Daw, Day, De, DeBra, Degallaix, De~Laurentis, Del\'eglise,
  Del~Pozzo, Demos, Denker, Dent, De~Pietri, Dergachev, De~Rosa, DeRosa,
  De~Rossi, DeSalvo, de~Varona, Devenson, Dhurandhar, D\'{\i}az, Dietrich,
  Di~Fiore, Di~Giovanni, Di~Girolamo, Di~Lieto, Di~Pace, Di~Palma, Di~Renzo,
  Doctor, Dolique, Donovan, Dooley, Doravari, Dorrington, Douglas,
  Dovale~\'Alvarez, Downes, Drago, Dreissigacker, Driggers, Du, Ducrot, Dudi,
  Dupej, Dwyer, Edo, Edwards, Effler, Eggenstein, Ehrens, Eichholz, Eikenberry,
  Eisenstein, Essick, Estevez, Etienne, Etzel, Evans, Evans, Factourovich,
  Fafone, Fair, Fairhurst, Fan, Farinon, Farr, Farr, Fauchon-Jones, Favata,
  Fays, Fee, Fehrmann, Feicht, Fejer, Fernandez-Galiana, Ferrante, Ferreira,
  Ferrini, Fidecaro, Finstad, Fiori, Fiorucci, Fishbach, Fisher, Fitz-Axen,
  Flaminio, Fletcher, Fong, Font, Forsyth, Forsyth, Fournier, Frasca, Frasconi,
  Frei, Freise, Frey, Frey, Fries, Fritschel, Frolov, Fulda, Fyffe, Gabbard,
  Gadre, Gaebel, Gair, Gammaitoni, Ganija, Gaonkar, Garcia-Quiros, Garufi,
  Gateley, Gaudio, Gaur, Gayathri, Gehrels, Gemme, Genin, Gennai, George,
  George, Gergely, Germain, Ghonge, Ghosh, Ghosh, Ghosh, Giaime, Giardina,
  Giazotto, Gill, Glover, Goetz, Goetz, Gomes, Goncharov, Gonz\'alez,
  Gonzalez~Castro, Gopakumar, Gorodetsky, Gossan, Gosselin, Gouaty, Grado,
  Graef, Granata, Grant, Gras, Gray, Greco, Green, Gretarsson, Groot, Grote,
  Grunewald, Gruning, Guidi, Guo, Gupta, Gupta, Gushwa, Gustafson, Gustafson,
  Halim, Hall, Hall, Hamilton, Hammond, Haney, Hanke, Hanks, Hanna, Hannam,
  Hannuksela, Hanson, Hardwick, Harms, Harry, Harry, Hart, Haster, Haughian,
  Healy, Heidmann, Heintze, Heitmann, Hello, Hemming, Hendry, Heng, Hennig,
  Heptonstall, Heurs, Hild, Hinderer, Ho, Hoak, Hofman, Holt, Holz, Hopkins,
  Horst, Hough, Houston, Howell, Hreibi, Hu, Huerta, Huet, Hughey, Husa,
  Huttner, Huynh-Dinh, Indik, Inta, Intini, Isa, Isac, Isi, Iyer, Izumi,
  Jacqmin, Jani, Jaranowski, Jawahar, Jim\'enez-Forteza, Johnson,
  Johnson-McDaniel, Jones, Jones, Jonker, Ju, Junker, Kalaghatgi, Kalogera,
  Kamai, Kandhasamy, Kang, Kanner, Kapadia, Karki, Karvinen, Kasprzack,
  Kastaun, Katolik, Katsavounidis, Katzman, Kaufer, Kawabe, K\'ef\'elian,
  Keitel, Kemball, Kennedy, Kent, Key, Khalili, Khan, Khan, Khan, Khazanov,
  Kijbunchoo, Kim, Kim, Kim, Kim, Kim, Kim, Kimbrell, King, King,
  Kinley-Hanlon, Kirchhoff, Kissel, Kleybolte, Klimenko, Knowles, Koch,
  Koehlenbeck, Koley, Kondrashov, Kontos, Korobko, Korth, Kowalska, Kozak,
  Kr\"amer, Kringel, Krishnan, Kr\'olak, Kuehn, Kumar, Kumar, Kumar, Kuo,
  Kutynia, Kwang, Lackey, Lai, Landry, Lang, Lange, Lantz, Lanza, Larson,
  Lartaux-Vollard, Lasky, Laxen, Lazzarini, Lazzaro, Leaci, Leavey, Lee, Lee,
  Lee, Lee, Lee, Lehmann, Lenon, Leon, Leonardi, Leroy, Letendre, Levin, Li,
  Linker, Littenberg, Liu, Liu, Lo, Lockerbie, London, Lord, Lorenzini,
  Loriette, Lormand, Losurdo, Lough, Lousto, Lovelace, L\"uck, Lumaca,
  Lundgren, Lynch, Ma, Macas, Macfoy, Machenschalk, MacInnis, Macleod, Maga\~na
  Hernandez, Maga\~na Sandoval, Maga\~na Zertuche, Magee, Majorana, Maksimovic,
  Man, Mandic, Mangano, Mansell, Manske, Mantovani, Marchesoni, Marion,
  M\'arka, M\'arka, Markakis, Markosyan, Markowitz, Maros, Marquina, Marsh,
  Martelli, Martellini, Martin, Martin, Martynov, Marx, Mason, Massera,
  Masserot, Massinger, Masso-Reid, Mastrogiovanni, Matas, Matichard, Matone,
  Mavalvala, Mazumder, McCarthy, McClelland, McCormick, McCuller, McGuire,
  McIntyre, McIver, McManus, McNeill, McRae, McWilliams, Meacher, Meadors,
  Mehmet, Meidam, Mejuto-Villa, Melatos, Mendell, Mercer, Merilh, Merzougui,
  Meshkov, Messenger, Messick, Metzdorff, Meyers, Miao, Michel, Middleton,
  Mikhailov, Milano, Miller, Miller, Miller, Millhouse, Milovich-Goff,
  Minazzoli, Minenkov, Ming, Mishra, Mitra, Mitrofanov, Mitselmakher,
  Mittleman, Moffa, Moggi, Mogushi, Mohan, Mohapatra, Molina, Montani, Moore,
  Moraru, Moreno, Morisaki, Morriss, Mours, Mow-Lowry, Mueller, Muir,
  Mukherjee, Mukherjee, Mukherjee, Mukund, Mullavey, Munch, Mu\~niz, Muratore,
  Murray, Nagar, Napier, Nardecchia, Naticchioni, Nayak, Neilson, Nelemans,
  Nelson, Nery, Neunzert, Nevin, Newport, Newton, Ng, Nguyen, Nguyen, Nichols,
  Nielsen, Nissanke, Nitz, Noack, Nocera, Nolting, North, Nuttall, Oberling,
  O'Dea, Ogin, Oh, Oh, Ohme, Okada, Oliver, Oppermann, Oram, O'Reilly,
  Ormiston, Ortega, O'Shaughnessy, Ossokine, Ottaway, Overmier, Owen, Pace,
  Page, Page, Pai, Pai, Palamos, Palashov, Palomba, Pal-Singh, Pan, Pan, Pang,
  Pang, Pankow, Pannarale, Pant, Paoletti, Paoli, Papa, Parida, Parker,
  Pascucci, Pasqualetti, Passaquieti, Passuello, Patil, Patricelli, Pearlstone,
  Pedraza, Pedurand, Pekowsky, Pele, Penn, Perez, Perreca, Perri, Pfeiffer,
  Phelps, Piccinni, Pichot, Piergiovanni, Pierro, Pillant, Pinard, Pinto,
  Pirello, Pitkin, Poe, Poggiani, Popolizio, Porter, Post, Powell, Prasad,
  Pratt, Pratten, Predoi, Prestegard, Prijatelj, Principe, Privitera, Prix,
  Prodi, Prokhorov, Puncken, Punturo, Puppo, P\"urrer, Qi, Quetschke, Quintero,
  Quitzow-James, Raab, Rabeling, Radkins, Raffai, Raja, Rajan, Rajbhandari,
  Rakhmanov, Ramirez, Ramos-Buades, Rapagnani, Raymond, Razzano, Read,
  Regimbau, Rei, Reid, Reitze, Ren, Reyes, Ricci, Ricker, Rieger, Riles, Rizzo,
  Robertson, Robie, Robinet, Rocchi, Rolland, Rollins, Roma, Romano, Romano,
  Romel, Romie, Rosi\ifmmode~\acute{n}\else \'{n}\fi{}ska, Ross, Rowan,
  R\"udiger, Ruggi, Rutins, Ryan, Sachdev, Sadecki, Sadeghian, Sakellariadou,
  Salconi, Saleem, Salemi, Samajdar, Sammut, Sampson, Sanchez, Sanchez,
  Sanchis-Gual, Sandberg, Sanders, Sassolas, Sathyaprakash, Saulson, Sauter,
  Savage, Sawadsky, Schale, Scheel, Scheuer, Schmidt, Schmidt, Schnabel,
  Schofield, Sch\"onbeck, Schreiber, Schuette, Schulte, Schutz, Schwalbe,
  Scott, Scott, Seidel, Sellers, Sengupta, Sentenac, Sequino, Sergeev,
  Shaddock, Shaffer, Shah, Shahriar, Shaner, Shao, Shapiro, Shawhan, Sheperd,
  Shoemaker, Shoemaker, Siellez, Siemens, Sieniawska, Sigg, Silva, Singer,
  Singh, Singhal, Sintes, Slagmolen, Smith, Smith, Smith, Somala, Son,
  Sonnenberg, Sorazu, Sorrentino, Souradeep, Spencer, Srivastava, Staats,
  Staley, Steinke, Steinlechner, Steinlechner, Steinmeyer, Stevenson, Stone,
  Stops, Strain, Stratta, Strigin, Strunk, Sturani, Stuver, Summerscales, Sun,
  Sunil, Suresh, Sutton, Swinkels, Szczepa\ifmmode~\acute{n}\else
  \'{n}\fi{}czyk, Tacca, Tait, Talbot, Talukder, Tanner, T\'apai, Taracchini,
  Tasson, Taylor, Taylor, Tewari, Theeg, Thies, Thomas, Thomas, Thomas, Thorne,
  Thorne, Thrane, Tiwari, Tiwari, Tokmakov, Toland, Tonelli, Tornasi,
  Torres-Forn\'e, Torrie, T\"oyr\"a, Travasso, Traylor, Trinastic, Tringali,
  Trozzo, Tsang, Tse, Tso, Tsukada, Tsuna, Tuyenbayev, Ueno, Ugolini,
  Unnikrishnan, Urban, Usman, Vahlbruch, Vajente, Valdes, Vallisneri, van
  Bakel, van Beuzekom, van~den Brand, Van Den~Broeck, Vander-Hyde, van~der
  Schaaf, van Heijningen, van Veggel, Vardaro, Varma, Vass, Vas\'uth, Vecchio,
  Vedovato, Veitch, Veitch, Venkateswara, Venugopalan, Verkindt, Vetrano,
  Vicer\'e, Viets, Vinciguerra, Vine, Vinet, Vitale, Vo, Vocca, Vorvick,
  Vyatchanin, Wade, Wade, Wade, Walet, Walker, Wallace, Walsh, Wang, Wang,
  Wang, Wang, Wang, Ward, Warner, Was, Watchi, Weaver, Wei, Weinert, Weinstein,
  Weiss, Wen, Wessel, We\ss{}els, Westerweck, Westphal, Wette, Whelan,
  Whitcomb, Whiting, Whittle, Wilken, Williams, Williams, Williamson, Willis,
  Willke, Wimmer, Winkler, Wipf, Wittel, Woan, Woehler, Wofford, Wong, Worden,
  Wright, Wu, Wysocki, Xiao, Yamamoto, Yancey, Yang, Yap, Yazback, Yu, Yu,
  Yvert, Zadro\ifmmode~\dot{z}\else \.{z}\fi{}ny, Zanolin, Zelenova, Zendri,
  Zevin, Zhang, Zhang, Zhang, Zhang, Zhao, Zhou, Zhou, Zhu, Zhu, Zimmerman,
  Zucker, \& Zweizig}]{LVCGW170817}
---. 2017{\natexlab{b}}, Phys. Rev. Lett., 119, 161101

\bibitem[{{Abbott} {et~al.}(2017){Abbott}, {Abbott}, {Abbott}, {Acernese},
  {Ackley}, {Adams}, {Adams}, {Addesso}, {Adhikari}, {Adya}, {Affeldt},
  {Afrough}, {Agarwal}, {Agathos}, {Agatsuma}, {Aggarwal}, {Aguiar}, {Aiello},
  {Ain}, {Ajith}, {Allen}, {Allen}, {Allocca}, {Altin}, {Amato}, {Ananyeva},
  {Anderson}, {Anderson}, {Angelova}, {Antier}, {Appert}, {Arai}, {Araya},
  {Areeda}, {Arnaud}, {Arun}, {Ascenzi}, {Ashton}, {Ast}, {Aston}, {Astone},
  {Atallah}, {Aufmuth}, {Aulbert}, {AultONeal}, {Austin}, {Avila-Alvarez},
  {Babak}, {Bacon}, {Bader}, {Bae}, {Baker}, {Baldaccini}, {Ballardin},
  {Ballmer}, {Banagiri}, {Barayoga}, {Barclay}, {Barish}, {Barker}, {Barkett},
  {Barone}, {Barr}, {Barsotti}, {Barsuglia}, {Barta}, {Barthelmy}, {Bartlett},
  {Bartos}, {Bassiri}, {Basti}, {Batch}, {Bawaj}, {Bayley}, {Bazzan},
  {B{\'e}csy}, {Beer}, {Bejger}, {Belahcene}, {Bell}, {Berger}, {Bergmann},
  {Bero}, {Berry}, {Bersanetti}, {Bertolini}, {Betzwieser}, {Bhagwat},
  {Bhandare}, {Bilenko}, {Billingsley}, {Billman}, {Birch}, {Birney},
  {Birnholtz}, {Biscans}, {Biscoveanu}, {Bisht}, {Bitossi}, {Biwer},
  {Bizouard}, {Blackburn}, {Blackman}, {Blair}, {Blair}, {Blair}, {Bloemen},
  {Bock}, {Bode}, {Boer}, {Bogaert}, {Bohe}, {Bondu}, {Bonilla}, {Bonnand},
  {Boom}, {Bork}, {Boschi}, {Bose}, {Bossie}, {Bouffanais}, {Bozzi},
  {Bradaschia}, {Brady}, {Branchesi}, {Brau}, {Briant}, {Brillet}, {Brinkmann},
  {Brisson}, {Brockill}, {Broida}, {Brooks}, {Brown}, {Brown}, {Brunett},
  {Buchanan}, {Buikema}, {Bulik}, {Bulten}, {Buonanno}, {Buskulic}, {Buy},
  {Byer}, {Cabero}, {Cadonati}, {Cagnoli}, {Cahillane}, {Calder{\'o}n
  Bustillo}, {Callister}, {Calloni}, {Camp}, {Canepa}, {Canizares}, {Cannon},
  {Cao}, {Cao}, {Capano}, {Capocasa}, {Carbognani}, {Caride}, {Carney},
  {Casanueva Diaz}, {Casentini}, {Caudill}, {Cavagli{\`a}}, {Cavalier},
  {Cavalieri}, {Cella}, {Cepeda}, {Cerd{\'a}-Dur{\'a}n}, {Cerretani},
  {Cesarini}, {Chamberlin}, {Chan}, {Chao}, {Charlton}, {Chase},
  {Chassande-Mottin}, {Chatterjee}, {Chatziioannou}, {Cheeseboro}, {Chen},
  {Chen}, {Chen}, {Cheng}, {Chia}, {Chincarini}, {Chiummo}, {Chmiel}, {Cho},
  {Cho}, {Chow}, {Christensen}, {Chu}, {Chua}, {Chua}, {Chung}, {Chung},
  {Ciani}, {Ciolfi}, {Cirelli}, {Cirone}, {Clara}, {Clark}, {Clearwater},
  {Cleva}, {Cocchieri}, {Coccia}, {Cohadon}, {Cohen}, {Colla}, {Collette},
  {Cominsky}, {Constancio}, {Conti}, {Cooper}, {Corban}, {Corbitt},
  {Cordero-Carri{\'o}n}, {Corley}, {Cornish}, {Corsi}, {Cortese}, {Costa},
  {Coughlin}, {Coughlin}, {Coulon}, {Countryman}, {Couvares}, {Covas}, {Cowan},
  {Coward}, {Cowart}, {Coyne}, {Coyne}, {Creighton}, {Creighton}, {Cripe},
  {Crowder}, {Cullen}, {Cumming}, {Cunningham}, {Cuoco}, {Dal Canton},
  {D{\'a}lya}, {Danilishin}, {D'Antonio}, {Danzmann}, {Dasgupta}, {Da Silva
  Costa}, {Dattilo}, {Dave}, {Davier}, {Davis}, {Daw}, {Day}, {De}, {DeBra},
  {Degallaix}, {De Laurentis}, {Del{\'e}glise}, {Del Pozzo}, {Demos}, {Denker},
  {Dent}, {De Pietri}, {Dergachev}, {De Rosa}, {DeRosa}, {De Rossi}, {DeSalvo},
  {de Varona}, {Devenson}, {Dhurandhar}, {D{\'\i}az}, {Di Fiore}, {Di
  Giovanni}, {Di Girolamo}, {Di Lieto}, {Di Pace}, {Di Palma}, {Di Renzo},
  {Doctor}, {Dolique}, {Donovan}, {Dooley}, {Doravari}, {Dorrington},
  {Douglas}, {Dovale {\'A}lvarez}, {Downes}, {Drago}, {Dreissigacker},
  {Driggers}, {Du}, {Ducrot}, {Dupej}, {Dwyer}, {Edo}, {Edwards}, {Effler},
  {Ehrens}, {Eichholz}, {Eikenberry}, {Eisenstein}, {Essick}, {Estevez},
  {Etienne}, {Etzel}, {Evans}, {Evans}, {Factourovich}, {Fafone}, {Fair},
  {Fairhurst}, {Fan}, {Farinon}, {Farr}, {Farr}, {Fauchon-Jones}, {Favata},
  {Fays}, {Fee}, {Fehrmann}, {Feicht}, {Fejer}, {Fernandez-Galiana},
  {Ferrante}, {Ferreira}, {Ferrini}, {Fidecaro}, {Finstad}, {Fiori},
  {Fiorucci}, {Fishbach}, {Fisher}, {Fitz-Axen}, {Flaminio}, {Fletcher},
  {Fong}, {Font}, {Forsyth}, {Forsyth}, {Fournier}, {Frasca}, {Frasconi},
  {Frei}, {Freise}, {Frey}, {Frey}, {Fries}, {Fritschel}, {Frolov}, {Fulda},
  {Fyffe}, {Gabbard}, {Gadre}, {Gaebel}, {Gair}, {Gammaitoni}, {Ganija},
  {Gaonkar}, {Garcia-Quiros}, {Garufi}, {Gateley}, {Gaudio}, {Gaur},
  {Gayathri}, {Gehrels}, {Gemme}, {Genin}, {Gennai}, {George}, {George},
  {Gergely}, {Germain}, {Ghonge}, {Ghosh}, {Ghosh}, {Ghosh}, {Giaime},
  {Giardina}, {Giazotto}, {Gill}, {Glover}, {Goetz}, {Goetz}, {Gomes},
  {Goncharov}, {Gonz{\'a}lez}, {Gonzalez Castro}, {Gopakumar}, {Gorodetsky},
  {Gossan}, {Gosselin}, {Gouaty}, {Grado}, {Graef}, {Granata}, {Grant}, {Gras},
  {Gray}, {Greco}, {Green}, {Gretarsson}, {Griswold}, {Groot}, {Grote},
  {Grunewald}, {Gruning}, {Guidi}, {Guo}, {Gupta}, {Gupta}, {Gushwa},
  {Gustafson}, {Gustafson}, {Halim}, {Hall}, {Hall}, {Hamilton}, {Hammond},
  {Haney}, {Hanke}, {Hanks}, {Hanna}, {Hannam}, {Hannuksela}, {Hanson},
  {Hardwick}, {Harms}, {Harry}, {Harry}, {Hart}, {Haster}, {Haughian}, {Healy},
  {Heidmann}, {Heintze}, {Heitmann}, {Hello}, {Hemming}, {Hendry}, {Heng},
  {Hennig}, {Heptonstall}, {Heurs}, {Hild}, {Hinderer}, {Hoak}, {Hofman},
  {Holt}, {Holz}, {Hopkins}, {Horst}, {Hough}, {Houston}, {Howell}, {Hreibi},
  {Hu}, {Huerta}, {Huet}, {Hughey}, {Husa}, {Huttner}, {Huynh-Dinh}, {Indik},
  {Inta}, {Intini}, {Isa}, {Isac}, {Isi}, {Iyer}, {Izumi}, {Jacqmin}, {Jani},
  {Jaranowski}, {Jawahar}, {Jim{\'e}nez-Forteza}, {Johnson}, {Jones}, {Jones},
  {Jonker}, {Ju}, {Junker}, {Kalaghatgi}, {Kalogera}, {Kamai}, {Kandhasamy},
  {Kang}, {Kanner}, {Kapadia}, {Karki}, {Karvinen}, {Kasprzack}, {Katolik},
  {Katsavounidis}, {Katzman}, {Kaufer}, {Kawabe}, {K{\'e}f{\'e}lian}, {Keitel},
  {Kemball}, {Kennedy}, {Kent}, {Key}, {Khalili}, {Khan}, {Khan}, {Khan},
  {Khazanov}, {Kijbunchoo}, {Kim}, {Kim}, {Kim}, {Kim}, {Kim}, {Kim},
  {Kimbrell}, {King}, {King}, {Kinley-Hanlon}, {Kirchhoff}, {Kissel},
  {Kleybolte}, {Klimenko}, {Knowles}, {Koch}, {Koehlenbeck}, {Koley},
  {Kondrashov}, {Kontos}, {Korobko}, {Korth}, {Kowalska}, {Kozak},
  {Kr{\"a}mer}, {Kringel}, {Krishnan}, {Kr{\'o}lak}, {Kuehn}, {Kumar}, {Kumar},
  {Kumar}, {Kuo}, {Kutynia}, {Kwang}, {Lackey}, {Lai}, {Landry}, {Lang},
  {Lange}, {Lantz}, {Lanza}, {Larson}, {Lartaux-Vollard}, {Lasky}, {Laxen},
  {Lazzarini}, {Lazzaro}, {Leaci}, {Leavey}, {Lee}, {Lee}, {Lee}, {Lee}, {Lee},
  {Lehmann}, {Lenon}, {Leonardi}, {Leroy}, {Letendre}, {Levin}, {Li}, {Linker},
  {Littenberg}, {Liu}, {Lo}, {Lockerbie}, {London}, {Lord}, {Lorenzini},
  {Loriette}, {Lormand}, {Losurdo}, {Lough}, {Lousto}, {Lovelace}, {L{\"u}ck},
  {Lumaca}, {Lundgren}, {Lynch}, {Ma}, {Macas}, {Macfoy}, {Machenschalk},
  {MacInnis}, {Macleod}, {Maga{\~n}a Hernandez}, {Maga{\~n}a-Sandoval},
  {Maga{\~n}a Zertuche}, {Magee}, {Majorana}, {Maksimovic}, {Man}, {Mandic},
  {Mangano}, {Mansell}, {Manske}, {Mantovani}, {Marchesoni}, {Marion},
  {M{\'a}rka}, {M{\'a}rka}, {Markakis}, {Markosyan}, {Markowitz}, {Maros},
  {Marquina}, {Marsh}, {Martelli}, {Martellini}, {Martin}, {Martin},
  {Martynov}, {Mason}, {Massera}, {Masserot}, {Massinger}, {Masso-Reid},
  {Mastrogiovanni}, {Matas}, {Matichard}, {Matone}, {Mavalvala}, {Mazumder},
  {McCarthy}, {McClelland}, {McCormick}, {McCuller}, {McGuire}, {McIntyre},
  {McIver}, {McManus}, {McNeill}, {McRae}, {McWilliams}, {Meacher}, {Meadors},
  {Mehmet}, {Meidam}, {Mejuto-Villa}, {Melatos}, {Mendell}, {Mercer}, {Merilh},
  {Merzougui}, {Meshkov}, {Messenger}, {Messick}, {Metzdorff}, {Meyers},
  {Miao}, {Michel}, {Middleton}, {Mikhailov}, {Milano}, {Miller}, {Miller},
  {Miller}, {Millhouse}, {Milovich-Goff}, {Minazzoli}, {Minenkov}, {Ming},
  {Mishra}, {Mitra}, {Mitrofanov}, {Mitselmakher}, {Mittleman}, {Moffa},
  {Moggi}, {Mogushi}, {Mohan}, {Mohapatra}, {Montani}, {Moore}, {Moraru},
  {Moreno}, {Morriss}, {Mours}, {Mow-Lowry}, {Mueller}, {Muir}, {Mukherjee},
  {Mukherjee}, {Mukherjee}, {Mukund}, {Mullavey}, {Munch}, {Mu{\~n}iz},
  {Muratore}, {Murray}, {Napier}, {Nardecchia}, {Naticchioni}, {Nayak},
  {Neilson}, {Nelemans}, {Nelson}, {Nery}, {Neunzert}, {Nevin}, {Newport},
  {Newton}, {Ng}, {Nguyen}, {Nguyen}, {Nichols}, {Nielsen}, {Nissanke}, {Nitz},
  {Noack}, {Nocera}, {Nolting}, {North}, {Nuttall}, {Oberling}, {O'Dea},
  {Ogin}, {Oh}, {Oh}, {Ohme}, {Okada}, {Oliver}, {Oppermann}, {Oram},
  {O'Reilly}, {Ormiston}, {Ortega}, {O'Shaughnessy}, {Ossokine}, {Ottaway},
  {Overmier}, {Owen}, {Pace}, {Page}, {Page}, {Pai}, {Pai}, {Palamos},
  {Palashov}, {Palomba}, {Pal-Singh}, {Pan}, {Pan}, {Pang}, {Pang}, {Pankow},
  {Pannarale}, {Pant}, {Paoletti}, {Paoli}, {Papa}, {Parida}, {Parker},
  {Pascucci}, {Pasqualetti}, {Passaquieti}, {Passuello}, {Patil}, {Patricelli},
  {Pearlstone}, {Pedraza}, {Pedurand}, {Pekowsky}, {Pele}, {Penn}, {Perez},
  {Perreca}, {Perri}, {Pfeiffer}, {Phelps}, {Piccinni}, {Pichot},
  {Piergiovanni}, {Pierro}, {Pillant}, {Pinard}, {Pinto}, {Pirello}, {Pitkin},
  {Poe}, {Poggiani}, {Popolizio}, {Porter}, {Post}, {Powell}, {Prasad},
  {Pratt}, {Pratten}, {Predoi}, {Prestegard}, {Price}, {Prijatelj}, {Principe},
  {Privitera}, {Prodi}, {Prokhorov}, {Puncken}, {Punturo}, {Puppo},
  {P{\"u}rrer}, {Qi}, {Quetschke}, {Quintero}, {Quitzow-James}, {Raab},
  {Rabeling}, {Radkins}, {Raffai}, {Raja}, {Rajan}, {Rajbhandari}, {Rakhmanov},
  {Ramirez}, {Ramos-Buades}, {Rapagnani}, {Raymond}, {Razzano}, {Read},
  {Regimbau}, {Rei}, {Reid}, {Reitze}, {Ren}, {Reyes}, {Ricci}, {Ricker},
  {Rieger}, {Riles}, {Rizzo}, {Robertson}, {Robie}, {Robinet}, {Rocchi},
  {Rolland}, {Rollins}, {Roma}, {Romano}, {Romel}, {Romie}, {Rosi{\'n}ska},
  {Ross}, {Rowan}, {R{\"u}diger}, {Ruggi}, {Rutins}, {Ryan}, {Sachdev},
  {Sadecki}, {Sadeghian}, {Sakellariadou}, {Salconi}, {Saleem}, {Salemi},
  {Samajdar}, {Sammut}, {Sampson}, {Sanchez}, {Sanchez}, {Sanchis-Gual},
  {Sandberg}, {Sanders}, {Sassolas}, {Sathyaprakash}, {Saulson}, {Sauter},
  {Savage}, {Sawadsky}, {Schale}, {Scheel}, {Scheuer}, {Schmidt}, {Schmidt},
  {Schnabel}, {Schofield}, {Sch{\"o}nbeck}, {Schreiber}, {Schuette}, {Schulte},
  {Schutz}, {Schwalbe}, {Scott}, {Scott}, {Seidel}, {Sellers}, {Sengupta},
  {Sentenac}, {Sequino}, {Sergeev}, {Shaddock}, {Shaffer}, {Shah}, {Shahriar},
  {Shaner}, {Shao}, {Shapiro}, {Shawhan}, {Sheperd}, {Shoemaker}, {Shoemaker},
  {Siellez}, {Siemens}, {Sieniawska}, {Sigg}, {Silva}, {Singer}, {Singh},
  {Singhal}, {Sintes}, {Slagmolen}, {Smith}, {Smith}, {Smith}, {Somala}, {Son},
  {Sonnenberg}, {Sorazu}, {Sorrentino}, {Souradeep}, {Spencer}, {Srivastava},
  {Staats}, {Staley}, {Steinke}, {Steinlechner}, {Steinlechner}, {Steinmeyer},
  {Stevenson}, {Stone}, {Stops}, {Strain}, {Stratta}, {Strigin}, {Strunk},
  {Sturani}, {Stuver}, {Summerscales}, {Sun}, {Sunil}, {Suresh}, {Sutton},
  {Swinkels}, {Szczepa{\'n}czyk}, {Tacca}, {Tait}, {Talbot}, {Talukder},
  {Tanner}, {T{\'a}pai}, {Taracchini}, {Tasson}, {Taylor}, {Taylor}, {Tewari},
  {Theeg}, {Thies}, {Thomas}, {Thomas}, {Thomas}, {Thorne}, {Thorne}, {Thrane},
  {Tiwari}, {Tiwari}, {Tokmakov}, {Toland}, {Tonelli}, {Tornasi},
  {Torres-Forn{\'e}}, {Torrie}, {T{\"o}yr{\"a}}, {Travasso}, {Traylor},
  {Trinastic}, {Tringali}, {Trozzo}, {Tsang}, {Tse}, {Tso}, {Tsukada}, {Tsuna},
  {Tuyenbayev}, {Ueno}, {Ugolini}, {Unnikrishnan}, {Urban}, {Usman},
  {Vahlbruch}, {Vajente}, {Valdes}, {van Bakel}, {van Beuzekom}, {van den
  Brand}, {Van Den Broeck}, {Vander-Hyde}, {van der Schaaf}, {van Heijningen},
  {van Veggel}, {Vardaro}, {Varma}, {Vass}, {Vas{\'u}th}, {Vecchio},
  {Vedovato}, {Veitch}, {Veitch}, {Venkateswara}, {Venugopalan}, {Verkindt},
  {Vetrano}, {Vicer{\'e}}, {Viets}, {Vinciguerra}, {Vine}, {Vinet}, {Vitale},
  {Vo}, {Vocca}, {Vorvick}, {Vyatchanin}, {Wade}, {Wade}, {Wade}, {Walet},
  {Walker}, {Wallace}, {Walsh}, {Wang}, {Wang}, {Wang}, {Wang}, {Wang}, {Ward},
  {Warner}, {Was}, {Watchi}, {Weaver}, {Wei}, {Weinert}, {Weinstein}, {Weiss},
  {Wen}, {Wessel}, {Wessels}, {Westerweck}, {Westphal}, {Wette}, {Whelan},
  {Whitcomb}, {Whiting}, {Whittle}, {Wilken}, {Williams}, {Williams},
  {Williamson}, {Willis}, {Willke}, {Wimmer}, {Winkler}, {Wipf}, {Wittel},
  {Woan}, {Woehler}, {Wofford}, {Wong}, {Worden}, {Wright}, {Wu}, {Wysocki},
  {Xiao}, {Yamamoto}, {Yancey}, {Yang}, {Yap}, {Yazback}, {Yu}, {Yu}, {Yvert},
  {Zadro{\.z}ny}, {Zanolin}, {Zelenova}, {Zendri}, {Zevin}, {Zhang}, {Zhang},
  {Zhang}, {Zhang}, {Zhao}, {Zhou}, {Zhou}, {Zhu}, {Zhu}, {Zimmerman},
  {Zucker}, {Zweizig}, {LIGO Scientific Collaboration}, {Virgo Collaboration},
  {Wilson-Hodge}, {Bissaldi}, {Blackburn}, {Briggs}, {Burns}, {Cleveland},
  {Connaughton}, {Gibby}, {Giles}, {Goldstein}, {Hamburg}, {Jenke}, {Hui},
  {Kippen}, {Kocevski}, {McBreen}, {Meegan}, {Paciesas}, {Poolakkil}, {Preece},
  {Racusin}, {Roberts}, {Stanbro}, {Veres}, {von Kienlin}, {GBM}, {Savchenko},
  {Ferrigno}, {Kuulkers}, {Bazzano}, {Bozzo}, {Brandt}, {Chenevez},
  {Courvoisier}, {Diehl}, {Domingo}, {Hanlon}, {Jourdain}, {Laurent}, {Lebrun},
  {Lutovinov}, {Martin-Carrillo}, {Mereghetti}, {Natalucci}, {Rodi}, {Roques},
  {Sunyaev}, {Ubertini}, {INTEGRAL}, {Aartsen}, {Ackermann}, {Adams},
  {Aguilar}, {Ahlers}, {Ahrens}, {Samarai}, {Altmann}, {Andeen}, {Anderson},
  {Ansseau}, {Anton}, {Arg{\"u}elles}, {Auffenberg}, {Axani}, {Bagherpour},
  {Bai}, {Barron}, {Barwick}, {Baum}, {Bay}, {Beatty}, {Becker Tjus},
  {Bernardini}, {Besson}, {Binder}, {Bindig}, {Blaufuss}, {Blot}, {Bohm},
  {B{\"o}rner}, {Bos}, {Bose}, {B{\"o}ser}, {Botner}, {Bourbeau}, {Bourbeau},
  {Bradascio}, {Braun}, {Brayeur}, {Brenzke}, {Bretz}, {Bron},
  {Brostean-Kaiser}, {Burgman}, {Carver}, {Casey}, {Casier}, {Cheung},
  {Chirkin}, {Christov}, {Clark}, {Classen}, {Coenders}, {Collin}, {Conrad},
  {Cowen}, {Cross}, {Day}, {de Andr{\'e}}, {De Clercq}, {DeLaunay},
  {Dembinski}, {De Ridder}, {Desiati}, {de Vries}, {de Wasseige}, {de With},
  {DeYoung}, {D{\'\i}az-V{\'e}lez}, {di Lorenzo}, {Dujmovic}, {Dumm},
  {Dunkman}, {Dvorak}, {Eberhardt}, {Ehrhardt}, {Eichmann}, {Eller}, {Evenson},
  {Fahey}, {Fazely}, {Felde}, {Filimonov}, {Finley}, {Flis}, {Franckowiak},
  {Friedman}, {Fuchs}, {Gaisser}, {Gallagher}, {Gerhardt}, {Ghorbani}, {Giang},
  {Glauch}, {Gl{\"u}senkamp}, {Goldschmidt}, {Gonzalez}, {Grant}, {Griffith},
  {Haack}, {Hallgren}, {Halzen}, {Hanson}, {Hebecker}, {Heereman}, {Helbing},
  {Hellauer}, {Hickford}, {Hignight}, {Hill}, {Hoffman}, {Hoffmann},
  {Hokanson-Fasig}, {Hoshina}, {Huang}, {Huber}, {Hultqvist}, {H{\"u}nnefeld},
  {In}, {Ishihara}, {Jacobi}, {Japaridze}, {Jeong}, {Jero}, {Jones},
  {Kalaczynski}, {Kang}, {Kappes}, {Karg}, {Karle}, {Kauer}, {Keivani},
  {Kelley}, {Kheirandish}, {Kim}, {Kim}, {Kintscher}, {Kiryluk}, {Kittler},
  {Klein}, {Kohnen}, {Koirala}, {Kolanoski}, {K{\"o}pke}, {Kopper}, {Kopper},
  {Koschinsky}, {Koskinen}, {Kowalski}, {Krings}, {Kroll}, {Kr{\"u}ckl},
  {Kunnen}, {Kunwar}, {Kurahashi}, {Kuwabara}, {Kyriacou}, {Labare},
  {Lanfranchi}, {Larson}, {Lauber}, {Lesiak-Bzdak}, {Leuermann}, {Liu}, {Lu},
  {L{\"u}nemann}, {Luszczak}, {Madsen}, {Maggi}, {Mahn}, {Mancina}, {Maruyama},
  {Mase}, {Maunu}, {McNally}, {Meagher}, {Medici}, {Meier}, {Menne}, {Merino},
  {Meures}, {Miarecki}, {Micallef}, {Moment{\'e}}, {Montaruli}, {Moore},
  {Moulai}, {Nahnhauer}, {Nakarmi}, {Naumann}, {Neer}, {Niederhausen},
  {Nowicki}, {Nygren}, {Obertacke Pollmann}, {Olivas}, {O'Murchadha},
  {Palczewski}, {Pandya}, {Pankova}, {Peiffer}, {Pepper}, {P{\'e}rez de los
  Heros}, {Pieloth}, {Pinat}, {Price}, {Przybylski}, {Raab}, {R{\"a}del},
  {Rameez}, {Rawlins}, {Rea}, {Reimann}, {Relethford}, {Relich}, {Resconi},
  {Rhode}, {Richman}, {Robertson}, {Rongen}, {Rott}, {Ruhe}, {Ryckbosch},
  {Rysewyk}, {S{\"a}lzer}, {Sanchez Herrera}, {Sandrock}, {Sandroos},
  {Santander}, {Sarkar}, {Sarkar}, {Satalecka}, {Schlunder}, {Schmidt},
  {Schneider}, {Schoenen}, {Sch{\"o}neberg}, {Schumacher}, {Seckel},
  {Seunarine}, {Soedingrekso}, {Soldin}, {Song}, {Spiczak}, {Spiering},
  {Stachurska}, {Stamatikos}, {Stanev}, {Stasik}, {Stettner}, {Steuer},
  {Stezelberger}, {Stokstad}, {St{\"o}ssl}, {Strotjohann}, {Stuttard},
  {Sullivan}, {Sutherland}, {Taboada}, {Tatar}, {Tenholt}, {Ter-Antonyan},
  {Terliuk}, {Te{\v{s}}i{\'c}}, {Tilav}, {Toale}, {Tobin}, {Toscano}, {Tosi},
  {Tselengidou}, {Tung}, {Turcati}, {Turley}, {Ty}, {Unger}, {Usner},
  {Vandenbroucke}, {Van Driessche}, {van Eijndhoven}, {Vanheule}, {van Santen},
  {Vehring}, {Vogel}, {Vraeghe}, {Walck}, {Wallace}, {Wallraff}, {Wandler},
  {Wandkowsky}, {Waza}, {Weaver}, {Weiss}, {Wendt}, {Werthebach}, {Whelan},
  {Wiebe}, {Wiebusch}, {Wille}, {Williams}, {Wills}, {Wolf}, {Wood}, {Woolsey},
  {Woschnagg}, {Xu}, {Xu}, {Xu}, {Yanez}, {Yodh}, {Yoshida}, {Yuan}, {Zoll},
  {IceCube Collaboration}, {Balasubramanian}, {Mate}, {Bhalerao},
  {Bhattacharya}, {Vibhute}, {Dewangan}, {Rao}, {Vadawale}, {AstroSat Cadmium
  Zinc Telluride Imager Team}, {Svinkin}, {Hurley}, {Aptekar}, {Frederiks},
  {Golenetskii}, {Kozlova}, {Lysenko}, {Oleynik}, {Tsvetkova}, {Ulanov},
  {Cline}, {IPN Collaboration}, {Li}, {Xiong}, {Zhang}, {Lu}, {Song}, {Cao},
  {Chang}, {Chen}, {Chen}, {Chen}, {Chen}, {Chen}, {Chen}, {Cui}, {Cui},
  {Deng}, {Dong}, {Du}, {Fu}, {Gao}, {Gao}, {Gao}, {Ge}, {Gu}, {Guan}, {Guo},
  {Han}, {Hu}, {Huang}, {Huo}, {Jia}, {Jiang}, {Jiang}, {Jin}, {Jin}, {Li},
  {Li}, {Li}, {Li}, {Li}, {Li}, {Li}, {Li}, {Li}, {Li}, {Li}, {Liang}, {Liao},
  {Liu}, {Liu}, {Liu}, {Liu}, {Liu}, {Liu}, {Liu}, {Lu}, {Lu}, {Luo}, {Ma},
  {Meng}, {Nang}, {Nie}, {Ou}, {Qu}, {Sai}, {Sun}, {Tan}, {Tao}, {Tao}, {Tuo},
  {Wang}, {Wang}, {Wang}, {Wang}, {Wang}, {Wen}, {Wu}, {Wu}, {Xiao}, {Xu},
  {Xu}, {Yan}, {Yang}, {Yang}, {Yang}, {Zhang}, {Zhang}, {Zhang}, {Zhang},
  {Zhang}, {Zhang}, {Zhang}, {Zhang}, {Zhang}, {Zhang}, {Zhang}, {Zhang},
  {Zhang}, {Zhang}, {Zhang}, {Zhang}, {Zhang}, {Zhang}, {Zhao}, {Zhao}, {Zhao},
  {Zheng}, {Zhu}, {Zhu}, {Zou}, {Insight-HXMT Collaboration}, {Albert},
  {Andr{\'e}}, {Anghinolfi}, {Ardid}, {Aubert}, {Aublin}, {Avgitas}, {Baret},
  {Barrios-Mart{\'\i}}, {Basa}, {Belhorma}, {Bertin}, {Biagi}, {Bormuth},
  {Bourret}, {Bouwhuis}, {Br{\^a}nza{\textcommabelow s}}, {Bruijn}, {Brunner},
  {Busto}, {Capone}, {Caramete}, {Carr}, {Celli}, {Cherkaoui El Moursli},
  {Chiarusi}, {Circella}, {Coelho}, {Coleiro}, {Coniglione}, {Costantini},
  {Coyle}, {Creusot}, {D{\'\i}az}, {Deschamps}, {De Bonis}, {Distefano}, {Di
  Palma}, {Domi}, {Donzaud}, {Dornic}, {Drouhin}, {Eberl}, {El Bojaddaini}, {El
  Khayati}, {Els{\"a}sser}, {Enzenh{\"o}fer}, {Ettahiri}, {Fassi}, {Felis},
  {Fusco}, {Gay}, {Giordano}, {Glotin}, {Gr{\'e}goire}, {Ruiz}, {Graf},
  {Hallmann}, {van Haren}, {Heijboer}, {Hello}, {Hern{\'a}ndez-Rey},
  {H{\"o}ssl}, {Hofest{\"a}dt}, {Hugon}, {Illuminati}, {James}, {de Jong},
  {Jongen}, {Kadler}, {Kalekin}, {Katz}, {Kiessling}, {Kouchner}, {Kreter},
  {Kreykenbohm}, {Kulikovskiy}, {Lachaud}, {Lahmann}, {Lef{\`e}vre}, {Leonora},
  {Lotze}, {Loucatos}, {Marcelin}, {Margiotta}, {Marinelli},
  {Mart{\'\i}nez-Mora}, {Mele}, {Melis}, {Michael}, {Migliozzi}, {Moussa},
  {Navas}, {Nezri}, {Organokov}, {P{\u{a}}v{\u{a}}la{\textcommabelow s}},
  {Pellegrino}, {Perrina}, {Piattelli}, {Popa}, {Pradier}, {Quinn}, {Racca},
  {Riccobene}, {S{\'a}nchez-Losa}, {Salda{\~n}a}, {Salvadori}, {Samtleben},
  {Sanguineti}, {Sapienza}, {Sieger}, {Spurio}, {Stolarczyk}, {Taiuti},
  {Tayalati}, {Trovato}, {Turpin}, {T{\"o}nnis}, {Vallage}, {Van Elewyck},
  {Versari}, {Vivolo}, {Vizzoca}, {Wilms}, {Zornoza}, {Z{\'u}{\~n}iga},
  {ANTARES Collaboration}, {Beardmore}, {Breeveld}, {Burrows}, {Cenko},
  {Cusumano}, {D'A{\`\i}}, {de Pasquale}, {Emery}, {Evans}, {Giommi},
  {Gronwall}, {Kennea}, {Krimm}, {Kuin}, {Lien}, {Marshall}, {Melandri},
  {Nousek}, {Oates}, {Osborne}, {Pagani}, {Page}, {Palmer}, {Perri}, {Siegel},
  {Sbarufatti}, {Tagliaferri}, {Tohuvavohu}, {Swift Collaboration}, {Tavani},
  {Verrecchia}, {Bulgarelli}, {Evangelista}, {Pacciani}, {Feroci}, {Pittori},
  {Giuliani}, {Del Monte}, {Donnarumma}, {Argan}, {Trois}, {Ursi}, {Cardillo},
  {Piano}, {Longo}, {Lucarelli}, {Munar-Adrover}, {Fuschino}, {Labanti},
  {Marisaldi}, {Minervini}, {Fioretti}, {Parmiggiani}, {Gianotti}, {Trifoglio},
  {Di Persio}, {Antonelli}, {Barbiellini}, {Caraveo}, {Cattaneo}, {Costa},
  {Colafrancesco}, {D'Amico}, {Ferrari}, {Morselli}, {Paoletti}, {Picozza},
  {Pilia}, {Rappoldi}, {Soffitta}, {Vercellone}, {AGILE Team}, {Foley},
  {Coulter}, {Kilpatrick}, {Drout}, {Piro}, {Shappee}, {Siebert}, {Simon},
  {Ulloa}, {Kasen}, {Madore}, {Murguia-Berthier}, {Pan}, {Prochaska},
  {Ramirez-Ruiz}, {Rest}, {Rojas-Bravo}, {1M2H Team}, {Berger},
  {Soares-Santos}, {Annis}, {Alexander}, {Allam}, {Balbinot}, {Blanchard},
  {Brout}, {Butler}, {Chornock}, {Cook}, {Cowperthwaite}, {Diehl},
  {Drlica-Wagner}, {Drout}, {Durret}, {Eftekhari}, {Finley}, {Fong}, {Frieman},
  {Fryer}, {Garc{\'\i}a-Bellido}, {Gruendl}, {Hartley}, {Herner}, {Kessler},
  {Lin}, {Lopes}, {Louren{\c{c}}o}, {Margutti}, {Marshall}, {Matheson},
  {Medina}, {Metzger}, {Mu{\~n}oz}, {Muir}, {Nicholl}, {Nugent}, {Palmese},
  {Paz-Chinch{\'o}n}, {Quataert}, {Sako}, {Sauseda}, {Schlegel}, {Scolnic},
  {Secco}, {Smith}, {Sobreira}, {Villar}, {Vivas}, {Wester}, {Williams},
  {Yanny}, {Zenteno}, {Zhang}, {Abbott}, {Banerji}, {Bechtol},
  {Benoit-L{\'e}vy}, {Bertin}, {Brooks}, {Buckley-Geer}, {Burke}, {Capozzi},
  {Carnero Rosell}, {Carrasco Kind}, {Castander}, {Crocce}, {Cunha},
  {D'Andrea}, {da Costa}, {Davis}, {DePoy}, {Desai}, {Dietrich}, {Eifler},
  {Fernandez}, {Flaugher}, {Fosalba}, {Gaztanaga}, {Gerdes}, {Giannantonio},
  {Goldstein}, {Gruen}, {Gschwend}, {Gutierrez}, {Honscheid}, {James},
  {Jeltema}, {Johnson}, {Johnson}, {Kent}, {Krause}, {Kron}, {Kuehn}, {Lahav},
  {Lima}, {Maia}, {March}, {Martini}, {McMahon}, {Menanteau}, {Miller},
  {Miquel}, {Mohr}, {Nichol}, {Ogando}, {Plazas}, {Romer}, {Roodman}, {Rykoff},
  {Sanchez}, {Scarpine}, {Schindler}, {Schubnell}, {Sevilla-Noarbe}, {Sheldon},
  {Smith}, {Smith}, {Stebbins}, {Suchyta}, {Swanson}, {Tarle}, {Thomas},
  {Troxel}, {Tucker}, {Vikram}, {Walker}, {Wechsler}, {Weller}, {Carlin},
  {Gill}, {Li}, {Marriner}, {Neilsen}, {Dark Energy Camera GW-EM
  Collaboration}, {DES Collaboration}, {Haislip}, {Kouprianov}, {Reichart},
  {Sand}, {Tartaglia}, {Valenti}, {Yang}, {DLT40 Collaboration}, {Benetti},
  {Brocato}, {Campana}, {Cappellaro}, {Covino}, {D'Avanzo}, {D'Elia}, {Getman},
  {Ghirlanda}, {Ghisellini}, {Limatola}, {Nicastro}, {Palazzi}, {Pian},
  {Piranomonte}, {Possenti}, {Rossi}, {Salafia}, {Tomasella}, {Amati},
  {Antonelli}, {Bernardini}, {Bufano}, {Capaccioli}, {Casella}, {Dadina}, {De
  Cesare}, {Di Paola}, {Giuffrida}, {Giunta}, {Israel}, {Lisi}, {Maiorano},
  {Mapelli}, {Masetti}, {Pescalli}, {Pulone}, {Salvaterra}, {Schipani},
  {Spera}, {Stamerra}, {Stella}, {Testa}, {Turatto}, {Vergani}, {Aresu},
  {Bachetti}, {Buffa}, {Burgay}, {Buttu}, {Caria}, {Carretti}, {Casasola},
  {Castangia}, {Carboni}, {Casu}, {Concu}, {Corongiu}, {Deiana}, {Egron},
  {Fara}, {Gaudiomonte}, {Gusai}, {Ladu}, {Loru}, {Leurini}, {Marongiu},
  {Melis}, {Melis}, {Migoni}, {Milia}, {Navarrini}, {Orlati}, {Ortu}, {Palmas},
  {Pellizzoni}, {Perrodin}, {Pisanu}, {Poppi}, {Righini}, {Saba}, {Serra},
  {Serrau}, {Stagni}, {Surcis}, {Vacca}, {Vargiu}, {Hunt}, {Jin}, {Klose},
  {Kouveliotou}, {Mazzali}, {M{\o}ller}, {Nava}, {Piran}, {Selsing}, {Vergani},
  {Wiersema}, {Toma}, {Higgins}, {Mundell}, {di Serego Alighieri}, {G{\'o}tz},
  {Gao}, {Gomboc}, {Kaper}, {Kobayashi}, {Kopac}, {Mao}, {Starling}, {Steele},
  {van der Horst}, {GRAWITA: GRAvitational Wave Inaf TeAm}, {Acero}, {Atwood},
  {Baldini}, {Barbiellini}, {Bastieri}, {Berenji}, {Bellazzini}, {Bissaldi},
  {Blandford}, {Bloom}, {Bonino}, {Bottacini}, {Bregeon}, {Buehler}, {Buson},
  {Cameron}, {Caputo}, {Caraveo}, {Cavazzuti}, {Chekhtman}, {Cheung}, {Chiang},
  {Ciprini}, {Cohen-Tanugi}, {Cominsky}, {Costantin}, {Cuoco}, {D'Ammando}, {de
  Palma}, {Digel}, {Di Lalla}, {Di Mauro}, {Di Venere}, {Dubois}, {Fegan},
  {Focke}, {Franckowiak}, {Fukazawa}, {Funk}, {Fusco}, {Gargano}, {Gasparrini},
  {Giglietto}, {Giordano}, {Giroletti}, {Glanzman}, {Green}, {Grondin},
  {Guillemot}, {Guiriec}, {Harding}, {Horan}, {J{\'o}hannesson}, {Kamae},
  {Kensei}, {Kuss}, {La Mura}, {Latronico}, {Lemoine-Goumard}, {Longo},
  {Loparco}, {Lovellette}, {Lubrano}, {Magill}, {Maldera}, {Manfreda},
  {Mazziotta}, {McEnery}, {Meyer}, {Michelson}, {Mirabal}, {Monzani},
  {Moretti}, {Morselli}, {Moskalenko}, {Negro}, {Nuss}, {Ojha}, {Omodei},
  {Orienti}, {Orlando}, {Palatiello}, {Paliya}, {Paneque}, {Pesce-Rollins},
  {Piron}, {Porter}, {Principe}, {Rain{\`o}}, {Rando}, {Razzano}, {Razzaque},
  {Reimer}, {Reimer}, {Reposeur}, {Rochester}, {Saz Parkinson}, {Sgr{\`o}},
  {Siskind}, {Spada}, {Spandre}, {Suson}, {Takahashi}, {Tanaka}, {Thayer},
  {Thayer}, {Thompson}, {Tibaldo}, {Torres}, {Torresi}, {Troja}, {Venters},
  {Vianello}, {Zaharijas}, {Fermi Large Area Telescope Collaboration},
  {Allison}, {Bannister}, {Dobie}, {Kaplan}, {Lenc}, {Lynch}, {Murphy},
  {Sadler}, {Australia Telescope Compact Array}, {Hotan}, {James}, {Oslowski},
  {Raja}, {Shannon}, {Whiting}, {Australian SKA Pathfinder}, {Arcavi},
  {Howell}, {McCully}, {Hosseinzadeh}, {Hiramatsu}, {Poznanski}, {Barnes},
  {Zaltzman}, {Vasylyev}, {Maoz}, {Las Cumbres Observatory Group}, {Cooke},
  {Bailes}, {Wolf}, {Deller}, {Lidman}, {Wang}, {Gendre}, {Andreoni}, {Ackley},
  {Pritchard}, {Bessell}, {Chang}, {M{\"o}ller}, {Onken}, {Scalzo},
  {Ridden-Harper}, {Sharp}, {Tucker}, {Farrell}, {Elmer}, {Johnston},
  {Venkatraman Krishnan}, {Keane}, {Green}, {Jameson}, {Hu}, {Ma}, {Sun}, {Wu},
  {Wang}, {Shang}, {Hu}, {Ashley}, {Yuan}, {Li}, {Tao}, {Zhu}, {Zhang},
  {Suntzeff}, {Zhou}, {Yang}, {Orange}, {Morris}, {Cucchiara}, {Giblin},
  {Klotz}, {Staff}, {Thierry}, {Schmidt}, {OzGrav}, {(Deeper}, {Wider},
  {program}, {AST3}, {CAASTRO Collaborations}, {Tanvir}, {Levan}, {Cano}, {de
  Ugarte-Postigo}, {Gonz{\'a}lez-Fern{\'a}ndez}, {Greiner}, {Hjorth}, {Irwin},
  {Kr{\"u}hler}, {Mandel}, {Milvang-Jensen}, {O'Brien}, {Rol}, {Rosetti},
  {Rosswog}, {Rowlinson}, {Steeghs}, {Th{\"o}ne}, {Ulaczyk}, {Watson}, {Bruun},
  {Cutter}, {Figuera Jaimes}, {Fujii}, {Fruchter}, {Gompertz}, {Jakobsson},
  {Hodosan}, {J{\`e}rgensen}, {Kangas}, {Kann}, {Rabus}, {Schr{\o}der},
  {Stanway}, {Wijers}, {VINROUGE Collaboration}, {Lipunov}, {Gorbovskoy},
  {Kornilov}, {Tyurina}, {Balanutsa}, {Kuznetsov}, {Vlasenko}, {Podesta},
  {Lopez}, {Podesta}, {Levato}, {Saffe}, {Mallamaci}, {Budnev}, {Gress},
  {Kuvshinov}, {Gorbunov}, {Vladimirov}, {Zimnukhov}, {Gabovich}, {Yurkov},
  {Sergienko}, {Rebolo}, {Serra-Ricart}, {Tlatov}, {Ishmuhametova}, {MASTER
  Collaboration}, {Abe}, {Aoki}, {Aoki}, {Asakura}, {Baar}, {Barway}, {Bond},
  {Doi}, {Finet}, {Fujiyoshi}, {Furusawa}, {Honda}, {Itoh}, {Kanda},
  {Kawabata}, {Kawabata}, {Kim}, {Koshida}, {Kuroda}, {Lee}, {Liu},
  {Matsubayashi}, {Miyazaki}, {Morihana}, {Morokuma}, {Motohara}, {Murata},
  {Nagai}, {Nagashima}, {Nagayama}, {Nakaoka}, {Nakata}, {Ohsawa}, {Ohshima},
  {Ohta}, {Okita}, {Saito}, {Saito}, {Sako}, {Sekiguchi}, {Sumi}, {Tajitsu},
  {Takahashi}, {Takayama}, {Tamura}, {Tanaka}, {Tanaka}, {Terai}, {Tominaga},
  {Tristram}, {Uemura}, {Utsumi}, {Yamaguchi}, {Yasuda}, {Yoshida}, {Zenko},
  {J-GEM}, {Adams}, {Anupama}, {Bally}, {Barway}, {Bellm}, {Blagorodnova},
  {Cannella}, {Chandra}, {Chatterjee}, {Clarke}, {Cobb}, {Cook}, {Copperwheat},
  {De}, {Emery}, {Feindt}, {Foster}, {Fox}, {Frail}, {Fremling}, {Frohmaier},
  {Garcia}, {Ghosh}, {Giacintucci}, {Goobar}, {Gottlieb}, {Grefenstette},
  {Hallinan}, {Harrison}, {Heida}, {Helou}, {Ho}, {Horesh}, {Hotokezaka}, {Ip},
  {Itoh}, {Jacobs}, {Jencson}, {Kasen}, {Kasliwal}, {Kassim}, {Kim}, {Kiran},
  {Kuin}, {Kulkarni}, {Kupfer}, {Lau}, {Madsen}, {Mazzali}, {Miller},
  {Miyasaka}, {Mooley}, {Myers}, {Nakar}, {Ngeow}, {Nugent}, {Ofek},
  {Palliyaguru}, {Pavana}, {Perley}, {Peters}, {Pike}, {Piran}, {Qi}, {Quimby},
  {Rana}, {Rosswog}, {Rusu}, {Sadler}, {Van Sistine}, {Sollerman}, {Xu}, {Yan},
  {Yatsu}, {Yu}, {Zhang}, {Zhao}, {GROWTH}, {JAGWAR}, {Caltech-NRAO},
  {TTU-NRAO}, {NuSTAR Collaborations}, {Chambers}, {Huber}, {Schultz},
  {Bulger}, {Flewelling}, {Magnier}, {Lowe}, {Wainscoat}, {Waters}, {Willman},
  {Pan-STARRS}, {Ebisawa}, {Hanyu}, {Harita}, {Hashimoto}, {Hidaka}, {Hori},
  {Ishikawa}, {Isobe}, {Iwakiri}, {Kawai}, {Kawai}, {Kawamuro}, {Kawase},
  {Kitaoka}, {Makishima}, {Matsuoka}, {Mihara}, {Morita}, {Morita}, {Nakahira},
  {Nakajima}, {Nakamura}, {Negoro}, {Oda}, {Sakamaki}, {Sasaki}, {Serino},
  {Shidatsu}, {Shimomukai}, {Sugawara}, {Sugita}, {Sugizaki}, {Tachibana},
  {Takao}, {Tanimoto}, {Tomida}, {Tsuboi}, {Tsunemi}, {Ueda}, {Ueno}, {Yamada},
  {Yamaoka}, {Yamauchi}, {Yatabe}, {Yoneyama}, {Yoshii}, {MAXI Team}, {Coward},
  {Crisp}, {Macpherson}, {Andreoni}, {Laugier}, {Noysena}, {Klotz}, {Gendre},
  {Thierry}, {Turpin}, {Consortium}, {Im}, {Choi}, {Kim}, {Yoon}, {Lim}, {Lee},
  {Lee}, {Kim}, {Ko}, {Joe}, {Kwon}, {Kim}, {Lim}, {Choi}, {KU Collaboration},
  {Fynbo}, {Malesani}, {Xu}, {Optical Telescope}, {Smartt}, {Jerkstrand},
  {Kankare}, {Sim}, {Fraser}, {Inserra}, {Maguire}, {Leloudas}, {Magee},
  {Shingles}, {Smith}, {Young}, {Kotak}, {Gal-Yam}, {Lyman}, {Homan},
  {Agliozzo}, {Anderson}, {Angus}, {Ashall}, {Barbarino}, {Bauer}, {Berton},
  {Botticella}, {Bulla}, {Cannizzaro}, {Cartier}, {Cikota}, {Clark}, {De Cia},
  {Della Valle}, {Dennefeld}, {Dessart}, {Dimitriadis}, {Elias-Rosa}, {Firth},
  {Fl{\"o}rs}, {Frohmaier}, {Galbany}, {Gonz{\'a}lez-Gait{\'a}n}, {Gromadzki},
  {Guti{\'e}rrez}, {Hamanowicz}, {Harmanen}, {Heintz}, {Hernandez}, {Hodgkin},
  {Hook}, {Izzo}, {James}, {Jonker}, {Kerzendorf}, {Kostrzewa-Rutkowska},
  {Kromer}, {Kuncarayakti}, {Lawrence}, {Manulis}, {Mattila}, {McBrien},
  {M{\"u}ller}, {Nordin}, {O'Neill}, {Onori}, {Palmerio}, {Pastorello},
  {Patat}, {Pignata}, {Podsiadlowski}, {Razza}, {Reynolds}, {Roy}, {Ruiter},
  {Rybicki}, {Salmon}, {Pumo}, {Prentice}, {Seitenzahl}, {Smith}, {Sollerman},
  {Sullivan}, {Szegedi}, {Taddia}, {Taubenberger}, {Terreran}, {Van Soelen},
  {Vos}, {Walton}, {Wright}, {Wyrzykowski}, {Yaron}, {pre=''(''>ePESSTO},
  {Chen}, {Kr{\"u}hler}, {Schady}, {Wiseman}, {Greiner}, {Rau}, {Schweyer},
  {Klose}, {Nicuesa Guelbenzu}, {GROND}, {Palliyaguru}, {Tech University},
  {Shara}, {Williams}, {Vaisanen}, {Potter}, {Romero Colmenero}, {Crawford},
  {Buckley}, {Mao}, {SALT Group}, {D{\'\i}az}, {Macri}, {Garc{\'\i}a Lambas},
  {Mendes de Oliveira}, {Nilo Castell{\'o}n}, {Ribeiro}, {S{\'a}nchez},
  {Schoenell}, {Abramo}, {Akras}, {Alcaniz}, {Artola}, {Beroiz}, {Bonoli},
  {Cabral}, {Camuccio}, {Chavushyan}, {Coelho}, {Colazo}, {Costa-Duarte},
  {Cuevas Larenas}, {Dom{\'\i}nguez Romero}, {Dultzin}, {Fern{\'a}ndez},
  {Garc{\'\i}a}, {Girardini}, {Gon{\c{c}}alves}, {Gon{\c{c}}alves}, {Gurovich},
  {Jim{\'e}nez-Teja}, {Kanaan}, {Lares}, {Lopes de Oliveira}, {L{\'o}pez-Cruz},
  {Melia}, {Molino}, {Padilla}, {Pe{\~n}uela}, {Placco}, {Qui{\~n}ones},
  {Ram{\'\i}rez Rivera}, {Renzi}, {Riguccini}, {R{\'\i}os-L{\'o}pez},
  {Rodriguez}, {Sampedro}, {Schneiter}, {Sodr{\'e}}, {Starck}, {Torres-Flores},
  {Tornatore}, {Zadro{\.z}ny}, {Castillo}, {TOROS: Transient Robotic
  Observatory of South Collaboration}, {Castro-Tirado}, {Tello}, {Hu}, {Zhang},
  {Cunniffe}, {Castell{\'o}n}, {Hiriart}, {Caballero-Garc{\'\i}a},
  {Jel{\'\i}nek}, {Kub{\'a}nek}, {P{\'e}rez del Pulgar}, {Park}, {Jeong},
  {Castro Cer{\'o}n}, {Pandey}, {Yock}, {Querel}, {Fan}, {Wang}, {BOOTES
  Collaboration}, {Beardsley}, {Brown}, {Crosse}, {Emrich}, {Franzen},
  {Gaensler}, {Horsley}, {Johnston-Hollitt}, {Kenney}, {Morales}, {Pallot},
  {Sokolowski}, {Steele}, {Tingay}, {Trott}, {Walker}, {Wayth}, {Williams},
  {Wu}, {Murchison Widefield Array}, {Yoshida}, {Sakamoto}, {Kawakubo},
  {Yamaoka}, {Takahashi}, {Asaoka}, {Ozawa}, {Torii}, {Shimizu}, {Tamura},
  {Ishizaki}, {Cherry}, {Ricciarini}, {Penacchioni}, {Marrocchesi}, {CALET
  Collaboration}, {Pozanenko}, {Volnova}, {Mazaeva}, {Minaev}, {Krugov},
  {Kusakin}, {Reva}, {Moskvitin}, {Rumyantsev}, {Inasaridze}, {Klunko},
  {Tungalag}, {Schmalz}, {Burhonov}, {IKI-GW Follow-up Collaboration},
  {Abdalla}, {Abramowski}, {Aharonian}, {Ait Benkhali}, {Ang{\"u}ner},
  {Arakawa}, {Arrieta}, {Aubert}, {Backes}, {Balzer}, {Barnard}, {Becherini},
  {Becker Tjus}, {Berge}, {Bernhard}, {Bernl{\"o}hr}, {Blackwell},
  {B{\"o}ttcher}, {Boisson}, {Bolmont}, {Bonnefoy}, {Bordas}, {Bregeon},
  {Brun}, {Brun}, {Bryan}, {B{\"u}chele}, {Bulik}, {Capasso}, {Caroff},
  {Carosi}, {Casanova}, {Cerruti}, {Chakraborty}, {Chaves}, {Chen},
  {Chevalier}, {Colafrancesco}, {Condon}, {Conrad}, {Davids}, {Decock}, {Deil},
  {Devin}, {deWilt}, {Dirson}, {Djannati-Ata{\"\i}}, {Donath}, {O'C. Drury},
  {Dutson}, {Dyks}, {Edwards}, {Egberts}, {Emery}, {Ernenwein}, {Eschbach},
  {Farnier}, {Fegan}, {Fernandes}, {Fiasson}, {Fontaine}, {Funk},
  {F{\"u}ssling}, {Gabici}, {Gallant}, {Garrigoux}, {Gat{\'e}}, {Giavitto},
  {Giebels}, {Glawion}, {Glicenstein}, {Gottschall}, {Grondin}, {Hahn},
  {Haupt}, {Hawkes}, {Heinzelmann}, {Henri}, {Hermann}, {Hinton}, {Hofmann},
  {Hoischen}, {Holch}, {Holler}, {Horns}, {Ivascenko}, {Iwasaki},
  {Jacholkowska}, {Jamrozy}, {Jankowsky}, {Jankowsky}, {Jingo}, {Jouvin},
  {Jung-Richardt}, {Kastendieck}, {Katarzy{\'n}ski}, {Katsuragawa},
  {Kerszberg}, {Khangulyan}, {Kh{\'e}lifi}, {King}, {Klepser}, {Klochkov},
  {Klu{\'z}niak}, {Komin}, {Kosack}, {Krakau}, {Kraus}, {Kr{\"u}ger}, {Laffon},
  {Lamanna}, {Lau}, {Lees}, {Lefaucheur}, {Lemi{\`e}re}, {Lemoine-Goumard},
  {Lenain}, {Leser}, {Lohse}, {Lorentz}, {Liu}, {Lypova}, {Malyshev},
  {Marandon}, {Marcowith}, {Mariaud}, {Marx}, {Maurin}, {Maxted}, {Mayer},
  {Meintjes}, {Meyer}, {Mitchell}, {Moderski}, {Mohamed}, {Mohrmann},
  {Mor{\r{a}}}, {Moulin}, {Murach}, {Nakashima}, {de Naurois}, {Ndiyavala},
  {Niederwanger}, {Niemiec}, {Oakes}, {O'Brien}, {Odaka}, {Ohm}, {Ostrowski},
  {Oya}, {Padovani}, {Panter}, {Parsons}, {Pekeur}, {Pelletier}, {Perennes},
  {Petrucci}, {Peyaud}, {Piel}, {Pita}, {Poireau}, {Poon}, {Prokhorov},
  {Prokoph}, {P{\"u}hlhofer}, {Punch}, {Quirrenbach}, {Raab}, {Rauth},
  {Reimer}, {Reimer}, {Renaud}, {de los Reyes}, {Rieger}, {Rinchiuso},
  {Romoli}, {Rowell}, {Rudak}, {Rulten}, {Sahakian}, {Saito}, {Sanchez},
  {Santangelo}, {Sasaki}, {Schlickeiser}, {Sch{\"u}ssler}, {Schulz},
  {Schwanke}, {Schwemmer}, {Seglar-Arroyo}, {Settimo}, {Seyffert}, {Shafi},
  {Shilon}, {Shiningayamwe}, {Simoni}, {Sol}, {Spanier}, {Spir-Jacob},
  {Stawarz}, {Steenkamp}, {Stegmann}, {Steppa}, {Sushch}, {Takahashi},
  {Tavernet}, {Tavernier}, {Taylor}, {Terrier}, {Tibaldo}, {Tiziani},
  {Tluczykont}, {Trichard}, {Tsirou}, {Tsuji}, {Tuffs}, {Uchiyama}, {van der
  Walt}, {van Eldik}, {van Rensburg}, {van Soelen}, {Vasileiadis}, {Veh},
  {Venter}, {Viana}, {Vincent}, {Vink}, {Voisin}, {V{\"o}lk}, {Vuillaume},
  {Wadiasingh}, {Wagner}, {Wagner}, {Wagner}, {White}, {Wierzcholska},
  {Willmann}, {W{\"o}rnlein}, {Wouters}, {Yang}, {Zaborov}, {Zacharias},
  {Zanin}, {Zdziarski}, {Zech}, {Zefi}, {Ziegler}, {Zorn}, {{\.Z}ywucka},
  {H.~E.~S.~S. Collaboration}, {Fender}, {Broderick}, {Rowlinson}, {Wijers},
  {Stewart}, {ter Veen}, {Shulevski}, {LOFAR Collaboration}, {Kavic},
  {Simonetti}, {League}, {Tsai}, {Obenberger}, {Nathaniel}, {Taylor}, {Dowell},
  {Liebling}, {Estes}, {Lippert}, {Sharma}, {Vincent}, {Farella}, {Wavelength
  Array}, {Abeysekara}, {Albert}, {Alfaro}, {Alvarez}, {Arceo},
  {Arteaga-Vel{\'a}zquez}, {Avila Rojas}, {Ayala Solares}, {Barber}, {Becerra
  Gonzalez}, {Becerril}, {Belmont-Moreno}, {BenZvi}, {Berley}, {Bernal},
  {Braun}, {Brisbois}, {Caballero-Mora}, {Capistr{\'a}n}, {Carrami{\~n}ana},
  {Casanova}, {Castillo}, {Cotti}, {Cotzomi}, {Couti{\~n}o de Le{\'o}n}, {De
  Le{\'o}n}, {De la Fuente}, {Diaz Hernandez}, {Dichiara}, {Dingus},
  {DuVernois}, {D{\'\i}az-V{\'e}lez}, {Ellsworth}, {Engel},
  {Enr{\'\i}quez-Rivera}, {Fiorino}, {Fleischhack}, {Fraija},
  {Garc{\'\i}a-Gonz{\'a}lez}, {Garfias}, {Gerhardt}, {Gonz{\~o}lez Mu{\~n}oz},
  {Gonz{\'a}lez}, {Goodman}, {Hampel-Arias}, {Harding}, {Hernandez},
  {Hernandez-Almada}, {Hona}, {H{\"u}ntemeyer}, {Iriarte}, {Jardin-Blicq},
  {Joshi}, {Kaufmann}, {Kieda}, {Lara}, {Lauer}, {Lennarz}, {Le{\'o}n Vargas},
  {Linnemann}, {Longinotti}, {Raya}, {Luna-Garc{\'\i}a}, {L{\'o}pez-Coto},
  {Malone}, {Marinelli}, {Martinez}, {Martinez-Castellanos},
  {Mart{\'\i}nez-Castro}, {Mart{\'\i}nez-Huerta}, {Matthews},
  {Miranda-Romagnoli}, {Moreno}, {Mostaf{\'a}}, {Nellen}, {Newbold}, {Nisa},
  {Noriega-Papaqui}, {Pelayo}, {Pretz}, {P{\'e}rez-P{\'e}rez}, {Ren}, {Rho},
  {Rivi{\`e}re}, {Rosa-Gonz{\'a}lez}, {Rosenberg}, {Ruiz-Velasco}, {Salazar},
  {Salesa Greus}, {Sandoval}, {Schneider}, {Schoorlemmer}, {Sinnis}, {Smith},
  {Springer}, {Surajbali}, {Tibolla}, {Tollefson}, {Torres}, {Ukwatta},
  {Weisgarber}, {Westerhoff}, {Wisher}, {Wood}, {Yapici}, {Yodh}, {Younk},
  {Zhou}, {{\'A}lvarez}, {HAWC Collaboration}, {Aab}, {Abreu}, {Aglietta},
  {Albuquerque}, {Albury}, {Allekotte}, {Almela}, {Alvarez Castillo},
  {Alvarez-Mu{\~n}iz}, {Anastasi}, {Anchordoqui}, {Andrada}, {Andringa},
  {Aramo}, {Arsene}, {Asorey}, {Assis}, {Avila}, {Badescu}, {Balaceanu},
  {Barbato}, {Barreira Luz}, {Becker}, {Bellido}, {Berat}, {Bertaina},
  {Bertou}, {Biermann}, {Biteau}, {Blaess}, {Blanco}, {Blazek}, {Bleve},
  {Boh{\'a}{\v{c}}ov{\'a}}, {Bonifazi}, {Borodai}, {Botti}, {Brack}, {Brancus},
  {Bretz}, {Bridgeman}, {Briechle}, {Buchholz}, {Bueno}, {Buitink}, {Buscemi},
  {Caballero-Mora}, {Caccianiga}, {Cancio}, {Canfora}, {Caruso}, {Castellina},
  {Catalani}, {Cataldi}, {Cazon}, {Chavez}, {Chinellato}, {Chudoba}, {Clay},
  {Cobos Cerutti}, {Colalillo}, {Coleman}, {Collica}, {Coluccia},
  {Concei{\c{c}}{\~a}o}, {Consolati}, {Contreras}, {Cooper}, {Coutu},
  {Covault}, {Cronin}, {D'Amico}, {Daniel}, {Dasso}, {Daumiller}, {Dawson},
  {Day}, {de Almeida}, {de Jong}, {De Mauro}, {de Mello Neto}, {De Mitri}, {de
  Oliveira}, {de Souza}, {Debatin}, {Deligny}, {D{\'\i}az Castro}, {Diogo},
  {Dobrigkeit}, {D'Olivo}, {Dorosti}, {Dos Anjos}, {Dova}, {Dundovic}, {Ebr},
  {Engel}, {Erdmann}, {Erfani}, {Escobar}, {Espadanal}, {Etchegoyen}, {Falcke},
  {Farmer}, {Farrar}, {Fauth}, {Fazzini}, {Feldbusch}, {Fenu}, {Fick},
  {Figueira}, {Filip{\v{c}}i{\v{c}}}, {Freire}, {Fujii}, {Fuster},
  {Ga{\"\i}or}, {Garc{\'\i}a}, {Gat{\'e}}, {Gemmeke}, {Gherghel-Lascu}, {Ghia},
  {Giaccari}, {Giammarchi}, {Giller}, {G{\l}as}, {Glaser}, {Golup}, {G{\'o}mez
  Berisso}, {G{\'o}mez Vitale}, {Gonz{\'a}lez}, {Gorgi}, {Gottowik}, {Grillo},
  {Grubb}, {Guarino}, {Guedes}, {Halliday}, {Hampel}, {Hansen}, {Harari},
  {Harrison}, {Harvey}, {Haungs}, {Hebbeker}, {Heck}, {Heimann}, {Herve},
  {Hill}, {Hojvat}, {Holt}, {Homola}, {H{\"o}randel}, {Horvath},
  {Hrabovsk{\'y}}, {Huege}, {Hulsman}, {Insolia}, {Isar}, {Jandt}, {Johnsen},
  {Josebachuili}, {Jurysek}, {K{\"a}{\"a}p{\"a}}, {Kampert}, {Keilhauer},
  {Kemmerich}, {Kemp}, {Kieckhafer}, {Klages}, {Kleifges}, {Kleinfeller},
  {Krause}, {Krohm}, {Kuempel}, {Kukec Mezek}, {Kunka}, {Kuotb Awad}, {Lago},
  {LaHurd}, {Lang}, {Lauscher}, {Legumina}, {Leigui de Oliveira},
  {Letessier-Selvon}, {Lhenry-Yvon}, {Link}, {Lo Presti}, {Lopes}, {L{\'o}pez},
  {L{\'o}pez Casado}, {Lorek}, {Luce}, {Lucero}, {Malacari}, {Mallamaci},
  {Mandat}, {Mantsch}, {Mariazzi}, {Maris}, {Marsella}, {Martello}, {Martinez},
  {Mart{\'\i}nez Bravo}, {Mas{\'\i}as Meza}, {Mathes}, {Mathys}, {Matthews},
  {Matthiae}, {Mayotte}, {Mazur}, {Medina}, {Medina-Tanco}, {Melo},
  {Menshikov}, {Merenda}, {Michal}, {Micheletti}, {Middendorf}, {Miramonti},
  {Mitrica}, {Mockler}, {Mollerach}, {Montanet}, {Morello}, {Morlino},
  {M{\"u}ller}, {M{\"u}ller}, {Muller}, {M{\"u}ller}, {Mussa}, {Naranjo},
  {Nguyen}, {Niculescu-Oglinzanu}, {Niechciol}, {Niemietz}, {Niggemann},
  {Nitz}, {Nosek}, {Novotny}, {No{\v{z}}ka}, {N{\'u}{\~n}ez}, {Oikonomou},
  {Olinto}, {Palatka}, {Pallotta}, {Papenbreer}, {Parente}, {Parra}, {Paul},
  {Pech}, {Pedreira}, {P{\c{e}}kala}, {Pe{\~n}a-Rodriguez}, {Pereira},
  {Perlin}, {Perrone}, {Peters}, {Petrera}, {Phuntsok}, {Pierog}, {Pimenta},
  {Pirronello}, {Platino}, {Plum}, {Poh}, {Porowski}, {Prado}, {Privitera},
  {Prouza}, {Quel}, {Querchfeld}, {Quinn}, {Ramos-Pollan}, {Rautenberg},
  {Ravignani}, {Ridky}, {Riehn}, {Risse}, {Ristori}, {Rizi}, {Rodrigues de
  Carvalho}, {Rodriguez Fernandez}, {Rodriguez Rojo}, {Roncoroni}, {Roth},
  {Roulet}, {Rovero}, {Ruehl}, {Saffi}, {Saftoiu}, {Salamida}, {Salazar},
  {Saleh}, {Salina}, {S{\'a}nchez}, {Sanchez-Lucas}, {Santos}, {Santos},
  {Sarazin}, {Sarmento}, {Sarmiento-Cano}, {Sato}, {Schauer}, {Scherini},
  {Schieler}, {Schimp}, {Schmidt}, {Scholten}, {Schov{\'a}nek}, {Schr{\"o}der},
  {Schr{\"o}der}, {Schulz}, {Schumacher}, {Sciutto}, {Segreto}, {Shadkam},
  {Shellard}, {Sigl}, {Silli}, {{\v{S}}m{\'\i}da}, {Snow}, {Sommers},
  {Sonntag}, {Soriano}, {Squartini}, {Stanca}, {Stani{\v{c}}}, {Stasielak},
  {Stassi}, {Stolpovskiy}, {Strafella}, {Streich}, {Suarez},
  {Suarez-Dur{\'a}n}, {Sudholz}, {Suomij{\"a}rvi}, {Supanitsky},
  {{\v{S}}up{\'\i}k}, {Swain}, {Szadkowski}, {Taboada}, {Taborda},
  {Timmermans}, {Todero Peixoto}, {Tomankova}, {Tom{\'e}}, {Torralba Elipe},
  {Travnicek}, {Trini}, {Tueros}, {Ulrich}, {Unger}, {Urban}, {Vald{\'e}s
  Galicia}, {Vali{\~n}o}, {Valore}, {van Aar}, {van Bodegom}, {van den Berg},
  {van Vliet}, {Varela}, {Vargas C{\'a}rdenas}, {V{\'a}zquez}, {Veberi{\v{c}}},
  {Ventura}, {Vergara Quispe}, {Verzi}, {Vicha}, {Villase{\~n}or}, {Vorobiov},
  {Wahlberg}, {Wainberg}, {Walz}, {Watson}, {Weber}, {Weindl}, {Wiede{\'n}ski},
  {Wiencke}, {Wilczy{\'n}ski}, {Wirtz}, {Wittkowski}, {Wundheiler}, {Yang},
  {Yushkov}, {Zas}, {Zavrtanik}, {Zavrtanik}, {Zepeda}, {Zimmermann},
  {Ziolkowski}, {Zong}, {Zuccarello}, {Pierre Auger Collaboration}, {Kim},
  {Schulze}, {Bauer}, {Corral-Santana}, {de Gregorio-Monsalvo},
  {Gonz{\'a}lez-L{\'o}pez}, {Hartmann}, {Ishwara-Chandra}, {Mart{\'\i}n},
  {Mehner}, {Misra}, {Micha{\l}owski}, {Resmi}, {ALMA Collaboration}, {Paragi},
  {Agudo}, {An}, {Beswick}, {Casadio}, {Frey}, {Jonker}, {Kettenis}, {Marcote},
  {Moldon}, {Szomoru}, {van Langevelde}, {Yang}, {Euro VLBI Team}, {Cwiek},
  {Cwiok}, {Czyrkowski}, {Dabrowski}, {Kasprowicz}, {Mankiewicz}, {Nawrocki},
  {Opiela}, {Piotrowski}, {Wrochna}, {Zaremba}, {{\.Z}arnecki}, {Pi of the Sky
  Collaboration}, {Haggard}, {Nynka}, {Ruan}, {Chandra Team at McGill
  University}, {Bland}, {Booler}, {Devillepoix}, {de Gois}, {Hancock}, {Howie},
  {Paxman}, {Sansom}, {Towner}, {Desert Fireball Network}, {Tonry}, {Coughlin},
  {Stubbs}, {Denneau}, {Heinze}, {Stalder}, {Weiland}, {ATLAS}, {Eatough},
  {Kramer}, {Kraus}, {Time Resolution Universe Survey}, {Troja}, {Piro},
  {Becerra Gonz{\'a}lez}, {Butler}, {Fox}, {Khandrika}, {Kutyrev}, {Lee},
  {Ricci}, {Ryan}, {S{\'a}nchez-Ram{\'\i}rez}, {Veilleux}, {Watson},
  {Wieringa}, {Burgess}, {van Eerten}, {Fontes}, {Fryer}, {Korobkin},
  {Wollaeger}, {RIMAS}, {RATIR}, {Camilo}, {Foley}, {Goedhart}, {Makhathini},
  {Oozeer}, {Smirnov}, {Fender}, {Woudt}, \& {South
  Africa/MeerKAT}}]{KNGW170817}
{Abbott}, B.~P., {Abbott}, R., {Abbott}, T.~D., {et~al.} 2017, \apjl, 848, L12

\bibitem[{{Abbott} {et~al.}(2019){Abbott}, {Abbott}, {Abbott}, {Abraham},
  {Acernese}, {Ackley}, {Adams}, {Adhikari}, {Adya}, {Affeldt}, {Agathos},
  {Agatsuma}, {Aggarwal}, {Aguiar}, {Aiello}, {Ain}, {Ajith}, {Allen},
  {Allocca}, {Aloy}, {Altin}, {Amato}, {Ananyeva}, {Anderson}, {Anderson},
  {Angelova}, {Antier}, {Appert}, {Arai}, {Araya}, {Areeda}, {Ar{\`e}ne},
  {Arnaud}, {Arun}, {Ascenzi}, {Ashton}, {Aston}, {Astone}, {Aubin}, {Aufmuth},
  {AultONeal}, {Austin}, {Avendano}, {Avila-Alvarez}, {Babak}, {Bacon},
  {Badaracco}, {Bader}, {Bae}, {Baker}, {Baldaccini}, {Ballardin}, {Ballmer},
  {Banagiri}, {Barayoga}, {Barclay}, {Barish}, {Barker}, {Barkett}, {Barnum},
  {Barone}, {Barr}, {Barsotti}, {Barsuglia}, {Barta}, {Bartlett}, {Bartos},
  {Bassiri}, {Basti}, {Bawaj}, {Bayley}, {Bazzan}, {B{\'e}csy}, {Bejger},
  {Belahcene}, {Bell}, {Beniwal}, {Berger}, {Bergmann}, {Bernuzzi}, {Bero},
  {Berry}, {Bersanetti}, {Bertolini}, {Betzwieser}, {Bhandare}, {Bidler},
  {Bilenko}, {Bilgili}, {Billingsley}, {Birch}, {Birney}, {Birnholtz},
  {Biscans}, {Biscoveanu}, {Bisht}, {Bitossi}, {Bizouard}, {Blackburn},
  {Blackman}, {Blair}, {Blair}, {Blair}, {Bloemen}, {Bode}, {Boer}, {Boetzel},
  {Bogaert}, {Bondu}, {Bonilla}, {Bonnand}, {Booker}, {Boom}, {Booth}, {Bork},
  {Boschi}, {Bose}, {Bossie}, {Bossilkov}, {Bosveld}, {Bouffanais}, {Bozzi},
  {Bradaschia}, {Brady}, {Bramley}, {Branchesi}, {Brau}, {Briant}, {Briggs},
  {Brighenti}, {Brillet}, {Brinkmann}, {Brisson}, {Brockill}, {Brooks},
  {Brown}, {Brunett}, {Buikema}, {Bulik}, {Bulten}, {Buonanno}, {Buskulic},
  {Bustamante Rosell}, {Buy}, {Byer}, {Cabero}, {Cadonati}, {Cagnoli},
  {Cahillane}, {Calder{\'o}n Bustillo}, {Callister}, {Calloni}, {Camp},
  {Campbell}, {Canepa}, {Cannon}, {Cao}, {Cao}, {Capocasa}, {Carbognani},
  {Caride}, {Carney}, {Carullo}, {Casanueva Diaz}, {Casentini}, {Caudill},
  {Cavagli{\`a}}, {Cavalier}, {Cavalieri}, {Cella}, {Cerd{\'a}-Dur{\'a}n},
  {Cerretani}, {Cesarini}, {Chaibi}, {Chakravarti}, {Chamberlin}, {Chan},
  {Chao}, {Charlton}, {Chase}, {Chassande-Mottin}, {Chatterjee}, {Chaturvedi},
  {Chatziioannou}, {Cheeseboro}, {Chen}, {Chen}, {Chen}, {Cheng}, {Cheong},
  {Chia}, {Chincarini}, {Chiummo}, {Cho}, {Cho}, {Cho}, {Christensen}, {Chu},
  {Chua}, {Chung}, {Chung}, {Ciani}, {Ciobanu}, {Ciolfi}, {Cipriano}, {Cirone},
  {Clara}, {Clark}, {Clearwater}, {Cleva}, {Cocchieri}, {Coccia}, {Cohadon},
  {Cohen}, {Colgan}, {Colleoni}, {Collette}, {Collins}, {Cominsky},
  {Constancio}, {Conti}, {Cooper}, {Corban}, {Corbitt}, {Cordero-Carri{\'o}n},
  {Corley}, {Cornish}, {Corsi}, {Cortese}, {Costa}, {Cotesta}, {Coughlin},
  {Coughlin}, {Coulon}, {Countryman}, {Couvares}, {Covas}, {Cowan}, {Coward},
  {Cowart}, {Coyne}, {Coyne}, {Creighton}, {Creighton}, {Cripe}, {Croquette},
  {Crowder}, {Cullen}, {Cumming}, {Cunningham}, {Cuoco}, {Canton}, {D{\'a}lya},
  {Danilishin}, {D'Antonio}, {Danzmann}, {Dasgupta}, {Da Silva Costa},
  {Datrier}, {Dattilo}, {Dave}, {Davier}, {Davis}, {Daw}, {DeBra},
  {Deenadayalan}, {Degallaix}, {De Laurentis}, {Del{\'e}glise}, {Del Pozzo},
  {DeMarchi}, {Demos}, {Dent}, {De Pietri}, {Derby}, {De Rosa}, {De Rossi},
  {DeSalvo}, {de Varona}, {Dhurandhar}, {D{\'\i}az}, {Dietrich}, {Di Fiore},
  {Di Giovanni}, {Di Girolamo}, {Di Lieto}, {Ding}, {Di Pace}, {Di Palma}, {Di
  Renzo}, {Dmitriev}, {Doctor}, {Donovan}, {Dooley}, {Doravari}, {Dorrington},
  {Downes}, {Drago}, {Driggers}, {Du}, {Ducoin}, {Dupej}, {Dwyer}, {Easter},
  {Edo}, {Edwards}, {Effler}, {Ehrens}, {Eichholz}, {Eikenberry}, {Eisenmann},
  {Eisenstein}, {Essick}, {Estelles}, {Estevez}, {Etienne}, {Etzel}, {Evans},
  {Evans}, {Fafone}, {Fair}, {Fairhurst}, {Fan}, {Farinon}, {Farr}, {Farr},
  {Fauchon-Jones}, {Favata}, {Fays}, {Fazio}, {Fee}, {Feicht}, {Fejer}, {Feng},
  {Fernandez-Galiana}, {Ferrante}, {Ferreira}, {Ferreira}, {Ferrini},
  {Fidecaro}, {Fiori}, {Fiorucci}, {Fishbach}, {Fisher}, {Fishner},
  {Fitz-Axen}, {Flaminio}, {Fletcher}, {Flynn}, {Fong}, {Font}, {Forsyth},
  {Fournier}, {Frasca}, {Frasconi}, {Frei}, {Freise}, {Frey}, {Frey},
  {Fritschel}, {Frolov}, {Fulda}, {Fyffe}, {Gabbard}, {Gadre}, {Gaebel},
  {Gair}, {Gammaitoni}, {Ganija}, {Gaonkar}, {Garcia},
  {Garc{\'\i}a-Quir{\'o}s}, {Garufi}, {Gateley}, {Gaudio}, {Gaur}, {Gayathri},
  {Gemme}, {Genin}, {Gennai}, {George}, {George}, {Gergely}, {Germain},
  {Ghonge}, {Ghosh}, {Ghosh}, {Ghosh}, {Giacomazzo}, {Giaime}, {Giardina},
  {Giazotto}, {Gill}, {Giordano}, {Glover}, {Godwin}, {Goetz}, {Goetz},
  {Goncharov}, {Gonz{\'a}lez}, {Gonzalez Castro}, {Gopakumar}, {Gorodetsky},
  {Gossan}, {Gosselin}, {Gouaty}, {Grado}, {Graef}, {Granata}, {Grant}, {Gras},
  {Grassia}, {Gray}, {Gray}, {Greco}, {Green}, {Green}, {Gretarsson}, {Groot},
  {Grote}, {Grunewald}, {Gruning}, {Guidi}, {Gulati}, {Guo}, {Gupta}, {Gupta},
  {Gustafson}, {Gustafson}, {Haegel}, {Halim}, {Hall}, {Hall}, {Hamilton},
  {Hammond}, {Haney}, {Hanke}, {Hanks}, {Hanna}, {Hannam}, {Hannuksela},
  {Hanson}, {Hardwick}, {Haris}, {Harms}, {Harry}, {Harry}, {Haster},
  {Haughian}, {Hayes}, {Healy}, {Heidmann}, {Heintze}, {Heitmann}, {Hello},
  {Hemming}, {Hendry}, {Heng}, {Hennig}, {Heptonstall}, {Hernandez Vivanco},
  {Heurs}, {Hild}, {Hinderer}, {Hoak}, {Hochheim}, {Hofman}, {Holgado},
  {Holland}, {Holt}, {Holz}, {Hopkins}, {Horst}, {Hough}, {Howell}, {Hoy},
  {Hreibi}, {Huang}, {Huerta}, {Huet}, {Hughey}, {Hulko}, {Husa}, {Huttner},
  {Huynh-Dinh}, {Idzkowski}, {Iess}, {Ingram}, {Inta}, {Intini}, {Irwin},
  {Isa}, {Isac}, {Isi}, {Iyer}, {Izumi}, {Jacqmin}, {Jadhav}, {Jani},
  {Janthalur}, {Jaranowski}, {Jenkins}, {Jiang}, {Johnson}, {Johnson-McDaniel},
  {Jones}, {Jones}, {Jones}, {Jonker}, {Ju}, {Junker}, {Kalaghatgi},
  {Kalogera}, {Kamai}, {Kandhasamy}, {Kang}, {Kanner}, {Kapadia}, {Karki},
  {Karvinen}, {Kashyap}, {Kasprzack}, {Katsanevas}, {Katsavounidis}, {Katzman},
  {Kaufer}, {Kawabe}, {Keerthana}, {K{\'e}f{\'e}lian}, {Keitel}, {Kennedy},
  {Key}, {Khalili}, {Khan}, {Khan}, {Khan}, {Khan}, {Khazanov}, {Khursheed},
  {Kijbunchoo}, {Kim}, {Kim}, {Kim}, {Kim}, {Kim}, {Kim}, {Kimball}, {King},
  {King}, {Kinley-Hanlon}, {Kirchhoff}, {Kissel}, {Kleybolte}, {Klika},
  {Klimenko}, {Knowles}, {Koch}, {Koehlenbeck}, {Koekoek}, {Koley},
  {Kondrashov}, {Kontos}, {Koper}, {Korobko}, {Korth}, {Kowalska}, {Kozak},
  {Kringel}, {Krishnendu}, {Kr{\'o}lak}, {Kuehn}, {Kumar}, {Kumar}, {Kumar},
  {Kumar}, {Kuo}, {Kutynia}, {Kwang}, {Lackey}, {Lai}, {Lam}, {Landry}, {Lane},
  {Lang}, {Lange}, {Lantz}, {Lanza}, {Lartaux-Vollard}, {Lasky}, {Laxen},
  {Lazzarini}, {Lazzaro}, {Leaci}, {Leavey}, {Lecoeuche}, {Lee}, {Lee}, {Lee},
  {Lee}, {Lee}, {Lee}, {Lehmann}, {Lenon}, {Leroy}, {Letendre}, {Levin}, {Li},
  {Li}, {Li}, {Li}, {Lin}, {Linde}, {Linker}, {Littenberg}, {Liu}, {Liu}, {Lo},
  {Lockerbie}, {London}, {Longo}, {Lorenzini}, {Loriette}, {Lormand},
  {Losurdo}, {Lough}, {Lousto}, {Lovelace}, {Lower}, {L{\"u}ck}, {Lumaca},
  {Lundgren}, {Lynch}, {Ma}, {Macas}, {Macfoy}, {MacInnis}, {Macleod},
  {Macquet}, {Maga{\~n}a-Sandoval}, {Maga{\~n}a Zertuche}, {Magee}, {Majorana},
  {Maksimovic}, {Malik}, {Man}, {Mandic}, {Mangano}, {Mansell}, {Manske},
  {Mantovani}, {Marchesoni}, {Marion}, {M{\'a}rka}, {M{\'a}rka}, {Markakis},
  {Markosyan}, {Markowitz}, {Maros}, {Marquina}, {Marsat}, {Martelli},
  {Martin}, {Martin}, {Martynov}, {Mason}, {Massera}, {Masserot}, {Massinger},
  {Masso-Reid}, {Mastrogiovanni}, {Matas}, {Matichard}, {Matone}, {Mavalvala},
  {Mazumder}, {McCann}, {McCarthy}, {McClelland}, {McCormick}, {McCuller},
  {McGuire}, {McIver}, {McManus}, {McRae}, {McWilliams}, {Meacher}, {Meadors},
  {Mehmet}, {Mehta}, {Meidam}, {Melatos}, {Mendell}, {Mercer}, {Mereni},
  {Merilh}, {Merzougui}, {Meshkov}, {Messenger}, {Messick}, {Metzdorff},
  {Meyers}, {Miao}, {Michel}, {Middleton}, {Mikhailov}, {Milano}, {Miller},
  {Miller}, {Millhouse}, {Mills}, {Milovich-Goff}, {Minazzoli}, {Minenkov},
  {Mishkin}, {Mishra}, {Mistry}, {Mitra}, {Mitrofanov}, {Mitselmakher},
  {Mittleman}, {Mo}, {Moffa}, {Mogushi}, {Mohapatra}, {Montani}, {Moore},
  {Moraru}, {Moreno}, {Morisaki}, {Mours}, {Mow-Lowry}, {Mukherjee},
  {Mukherjee}, {Mukherjee}, {Mukund}, {Mullavey}, {Munch}, {Mu{\~n}iz},
  {Muratore}, {Murray}, {Nagar}, {Nardecchia}, {Naticchioni}, {Nayak},
  {Neilson}, {Nelemans}, {Nelson}, {Nery}, {Neunzert}, {Ng}, {Ng}, {Nguyen},
  {Nichols}, {Nielsen}, {Nissanke}, {Nitz}, {Nocera}, {North}, {Nuttall},
  {Obergaulinger}, {Oberling}, {O'Brien}, {O'Dea}, {Ogin}, {Oh}, {Oh}, {Ohme},
  {Ohta}, {Okada}, {Oliver}, {Oppermann}, {Oram}, {O'Reilly}, {Ormiston},
  {Ortega}, {O'Shaughnessy}, {Ossokine}, {Ottaway}, {Overmier}, {Owen}, {Pace},
  {Pagano}, {Page}, {Pai}, {Pai}, {Palamos}, {Palashov}, {Palomba},
  {Pal-Singh}, {Pan}, {Pang}, {Pang}, {Pankow}, {Pannarale}, {Pant},
  {Paoletti}, {Paoli}, {Papa}, {Parida}, {Parker}, {Pascucci}, {Pasqualetti},
  {Passaquieti}, {Passuello}, {Patil}, {Patricelli}, {Pearlstone}, {Pedersen},
  {Pedraza}, {Pedurand}, {Pele}, {Penn}, {Perego}, {Perez}, {Perreca},
  {Pfeiffer}, {Phelps}, {Phukon}, {Piccinni}, {Pichot}, {Piergiovanni},
  {Pillant}, {Pinard}, {Pirello}, {Pitkin}, {Poggiani}, {Pong}, {Ponrathnam},
  {Popolizio}, {Porter}, {Powell}, {Prajapati}, {Prasad}, {Prasai}, {Prasanna},
  {Pratten}, {Prestegard}, {Privitera}, {Prodi}, {Prokhorov}, {Puncken},
  {Punturo}, {Puppo}, {P{\"u}rrer}, {Qi}, {Quetschke}, {Quinonez}, {Quintero},
  {Quitzow-James}, {Raab}, {Radkins}, {Radulescu}, {Raffai}, {Raja}, {Rajan},
  {Rajbhandari}, {Rakhmanov}, {Ramirez}, {Ramos-Buades}, {Rana}, {Rao},
  {Rapagnani}, {Raymond}, {Razzano}, {Read}, {Regimbau}, {Rei}, {Reid},
  {Reitze}, {Ren}, {Ricci}, {Richardson}, {Richardson}, {Ricker},
  {Riemenschneider}, {Riles}, {Rizzo}, {Robertson}, {Robie}, {Robinet},
  {Rocchi}, {Rolland}, {Rollins}, {Roma}, {Romanelli}, {Romano}, {Romel},
  {Romie}, {Rose}, {Rosi{\'n}ska}, {Rosofsky}, {Ross}, {Rowan}, {R{\"u}diger},
  {Ruggi}, {Rutins}, {Ryan}, {Sachdev}, {Sadecki}, {Sakellariadou}, {Salafia},
  {Salconi}, {Saleem}, {Salemi}, {Samajdar}, {Sammut}, {Sanchez}, {Sanchez},
  {Sanchis-Gual}, {Sandberg}, {Sanders}, {Santiago}, {Sarin}, {Sassolas},
  {Sathyaprakash}, {Saulson}, {Sauter}, {Savage}, {Schale}, {Scheel},
  {Scheuer}, {Schmidt}, {Schnabel}, {Schofield}, {Sch{\"o}nbeck}, {Schreiber},
  {Schulte}, {Schutz}, {Schwalbe}, {Scott}, {Scott}, {Seidel}, {Sellers},
  {Sengupta}, {Sennett}, {Sentenac}, {Sequino}, {Sergeev}, {Setyawati},
  {Shaddock}, {Shaffer}, {Shahriar}, {Shaner}, {Shao}, {Sharma}, {Shawhan},
  {Shen}, {Shink}, {Shoemaker}, {Shoemaker}, {ShyamSundar}, {Siellez},
  {Sieniawska}, {Sigg}, {Silva}, {Singer}, {Singh}, {Singhal}, {Sintes},
  {Sitmukhambetov}, {Skliris}, {Slagmolen}, {Slaven-Blair}, {Smith}, {Smith},
  {Somala}, {Son}, {Sorazu}, {Sorrentino}, {Souradeep}, {Sowell}, {Spencer},
  {Srivastava}, {Srivastava}, {Staats}, {Stachie}, {Standke}, {Steer},
  {Steinke}, {Steinlechner}, {Steinlechner}, {Steinmeyer}, {Stevenson},
  {Stocks}, {Stone}, {Stops}, {Strain}, {Stratta}, {Strigin}, {Strunk},
  {Sturani}, {Stuver}, {Sudhir}, {Summerscales}, {Sun}, {Sunil}, {Suresh},
  {Sutton}, {Swinkels}, {Szczepa{\'n}czyk}, {Tacca}, {Tait}, {Talbot},
  {Talukder}, {Tanner}, {T{\'a}pai}, {Taracchini}, {Tasson}, {Taylor}, {Thies},
  {Thomas}, {Thomas}, {Thondapu}, {Thorne}, {Thrane}, {Tiwari}, {Tiwari},
  {Tiwari}, {Toland}, {Tonelli}, {Tornasi}, {Torres-Forn{\'e}}, {Torrie},
  {T{\"o}yr{\"a}}, {Travasso}, {Traylor}, {Tringali}, {Trovato}, {Trozzo},
  {Trudeau}, {Tsang}, {Tse}, {Tso}, {Tsukada}, {Tsuna}, {Tuyenbayev}, {Ueno},
  {Ugolini}, {Unnikrishnan}, {Urban}, {Usman}, {Vahlbruch}, {Vajente},
  {Valdes}, {van Bakel}, {van Beuzekom}, {van den Brand}, {Van Den Broeck},
  {Vander-Hyde}, {van Heijningen}, {van der Schaaf}, {van Veggel}, {Vardaro},
  {Varma}, {Vass}, {Vas{\'u}th}, {Vecchio}, {Vedovato}, {Veitch}, {Veitch},
  {Venkateswara}, {Venugopalan}, {Verkindt}, {Vetrano}, {Vicer{\'e}}, {Viets},
  {Vine}, {Vinet}, {Vitale}, {Vo}, {Vocca}, {Vorvick}, {Vyatchanin}, {Wade},
  {Wade}, {Wade}, {Walet}, {Walker}, {Wallace}, {Walsh}, {Wang}, {Wang},
  {Wang}, {Wang}, {Wang}, {Ward}, {Warden}, {Warner}, {Was}, {Watchi},
  {Weaver}, {Wei}, {Weinert}, {Weinstein}, {Weiss}, {Wellmann}, {Wen},
  {Wessel}, {We{\ss}els}, {Westhouse}, {Wette}, {Whelan}, {White}, {Whiting},
  {Whittle}, {Wilken}, {Williams}, {Williamson}, {Willis}, {Willke}, {Wimmer},
  {Winkler}, {Wipf}, {Wittel}, {Woan}, {Woehler}, {Wofford}, {Worden},
  {Wright}, {Wu}, {Wysocki}, {Xiao}, {Yamamoto}, {Yancey}, {Yang}, {Yap},
  {Yazback}, {Yeeles}, {Yu}, {Yu}, {Yuen}, {Yvert}, {Zadro{\.Z}ny}, {Zanolin},
  {Zappa}, {Zelenova}, {Zendri}, {Zevin}, {Zhang}, {Zhang}, {Zhang}, {Zhao},
  {Zhou}, {Zhou}, {Zhu}, {Zimmerman}, {Zlochower}, {Zucker}, {Zweizig}, {LIGO
  Scientific Collaboration}, \& {Virgo Collaboration}}]{GWTC1}
---. 2019, Physical Review X, 9, 031040

\bibitem[{Abbott {et~al.}(2020)}]{LVCGW190425}
Abbott, B.~P., {et~al.} 2020, arXiv:2001.01761

\bibitem[{{Abbott} {et~al.}(2020){Abbott}, {Abbott}, {Abraham}, {Acernese},
  {Ackley}, {Adams}, {Adams}, {Adhikari}, {Adya}, {Affeldt}, {Agathos},
  {Agatsuma}, {Aggarwal}, {Aguiar}, {Aiello}, {Ain}, {Ajith}, {Akcay}, {Allen},
  {Allocca}, {Altin}, {Amato}, {Anand}, {Ananyeva}, {Anderson}, {Anderson},
  {Angelova}, {Ansoldi}, {Antelis}, {Antier}, {Appert}, {Arai}, {Araya},
  {Areeda}, {Ar{\`e}ne}, {Arnaud}, {Aronson}, {Arun}, {Asali}, {Ascenzi},
  {Ashton}, {Aston}, {Astone}, {Aubin}, {Aufmuth}, {AultONeal}, {Austin},
  {Avendano}, {Babak}, {Badaracco}, {Bader}, {Bae}, {Baer}, {Bagnasco},
  {Baird}, {Ball}, {Ballardin}, {Ballmer}, {Bals}, {Balsamo}, {Baltus},
  {Banagiri}, {Bankar}, {Bankar}, {Barayoga}, {Barbieri}, {Barish}, {Barker},
  {Barneo}, {Barnum}, {Barone}, {Barr}, {Barsotti}, {Barsuglia}, {Barta},
  {Bartlett}, {Bartos}, {Bassiri}, {Basti}, {Bawaj}, {Bayley}, {Bazzan},
  {Becher}, {B{\'e}csy}, {Bedakihale}, {Bejger}, {Belahcene}, {Beniwal},
  {Benjamin}, {Bennett}, {Bentley}, {Bergamin}, {Berger}, {Bergmann},
  {Bernuzzi}, {Berry}, {Bersanetti}, {Bertolini}, {Betzwieser}, {Bhandare},
  {Bhandari}, {Bhattacharjee}, {Bidler}, {Bilenko}, {Billingsley}, {Birney},
  {Birnholtz}, {Biscans}, {Bischi}, {Biscoveanu}, {Bisht}, {Bitossi},
  {Bizouard}, {Blackburn}, {Blackman}, {Blair}, {Blair}, {Blair}, {Blanch},
  {Bobba}, {Bode}, {Boer}, {Boetzel}, {Bogaert}, {Boldrini}, {Bondu},
  {Bonnand}, {Bonilla}, {Booker}, {Boom}, {Bork}, {Boschi}, {Bose},
  {Bossilkov}, {Boudart}, {Bouffanais}, {Bozzi}, {Bradaschia}, {Brady},
  {Bramley}, {Branchesi}, {Brau}, {Breschi}, {Briant}, {Briggs}, {Brighenti},
  {Brillet}, {Brinkmann}, {Brockill}, {Brooks}, {Brooks}, {Brown}, {Brunett},
  {Bruno}, {Bruntz}, {Buikema}, {Bulik}, {Bulten}, {Buonanno}, {Buscicchio},
  {Buskulic}, {Byer}, {Cabero}, {Cadonati}, {Caesar}, {Cagnoli}, {Cahillane},
  {Calder{\'o}n Bustillo}, {Callaghan}, {Callister}, {Calloni}, {Camp},
  {Canepa}, {Cannon}, {Cao}, {Cao}, {Carapella}, {Carbognani}, {Carney},
  {Carpinelli}, {Carullo}, {Carver}, {Casanueva Diaz}, {Casentini}, {Caudill},
  {Cavagli{\`a}}, {Cavalier}, {Cavalieri}, {Cella}, {Cerd{\'a}-Dur{\'a}n},
  {Cesarini}, {Chaibi}, {Chakravarti}, {Chan}, {Chan}, {Chandra}, {Chanial},
  {Chao}, {Charlton}, {Chase}, {Chassande-Mottin}, {Chatterjee},
  {Chattopadhyay}, {Chaturvedi}, {Chatziioannou}, {Chen}, {Chen}, {Chen},
  {Chen}, {Cheng}, {Cheong}, {Chia}, {Chiadini}, {Chierici}, {Chincarini},
  {Chiummo}, {Cho}, {Cho}, {Cho}, {Choate}, {Christensen}, {Chu}, {Chua},
  {Chung}, {Chung}, {Ciani}, {Ciecielag}, {Cie{\'s}lar}, {Cifaldi}, {Ciobanu},
  {Ciolfi}, {Cipriano}, {Cirone}, {Clara}, {Clark}, {Clark}, {Clarke},
  {Clearwater}, {Clesse}, {Cleva}, {Coccia}, {Cohadon}, {Cohen}, {Colleoni},
  {Collette}, {Collins}, {Colpi}, {Constancio}, {Conti}, {Cooper}, {Corban},
  {Corbitt}, {Cordero-Carri{\'o}n}, {Corezzi}, {Corley}, {Cornish}, {Corre},
  {Corsi}, {Cortese}, {Costa}, {Cotesta}, {Coughlin}, {Coughlin}, {Coulon},
  {Countryman}, {Cousins}, {Couvares}, {Covas}, {Coward}, {Cowart}, {Coyne},
  {Coyne}, {Creighton}, {Creighton}, {Croquette}, {Crowder}, {Cudell},
  {Cullen}, {Cumming}, {Cummings}, {Cunningham}, {Cuoco}, {Curylo}, {Dal
  Canton}, {D{\'a}lya}, {Dana}, {DaneshgaranBajastani}, {D'Angelo}, {Danila},
  {Danilishin}, {D'Antonio}, {Danzmann}, {Darsow-Fromm}, {Dasgupta}, {Datrier},
  {Dattilo}, {Dave}, {Davier}, {Davies}, {Davis}, {Daw}, {Dean}, {DeBra},
  {Deenadayalan}, {Degallaix}, {De Laurentis}, {Del{\'e}glise}, {Del Favero},
  {De Lillo}, {De Lillo}, {Del Pozzo}, {DeMarchi}, {De Matteis}, {D'Emilio},
  {Demos}, {Denker}, {Dent}, {Depasse}, {De Pietri}, {De Rosa}, {De Rossi},
  {DeSalvo}, {de Varona}, {Dhurandhar}, {D{\'\i}az}, {Diaz-Ortiz}, {Didio},
  {Dietrich}, {Di Fiore}, {DiFronzo}, {Di Giorgio}, {Di Giovanni}, {Di
  Giovanni}, {Di Girolamo}, {Di Lieto}, {Ding}, {Di Pace}, {Di Palma}, {Di
  Renzo}, {Divakarla}, {Dmitriev}, {Doctor}, {D'Onofrio}, {Donovan}, {Dooley},
  {Doravari}, {Dorrington}, {Downes}, {Drago}, {Driggers}, {Du}, {Ducoin},
  {Dupej}, {Durante}, {D'Urso}, {Duverne}, {Dwyer}, {Easter}, {Eddolls},
  {Edelman}, {Edo}, {Edy}, {Effler}, {Eichholz}, {Eikenberry}, {Eisenmann},
  {Eisenstein}, {Ejlli}, {Errico}, {Essick}, {Estell{\'e}s}, {Estevez},
  {Etienne}, {Etzel}, {Evans}, {Evans}, {Ewing}, {Fafone}, {Fair}, {Fairhurst},
  {Fan}, {Farah}, {Farinon}, {Farr}, {Farr}, {Fauchon-Jones}, {Favata}, {Fays},
  {Fazio}, {Feicht}, {Fejer}, {Feng}, {Fenyvesi}, {Ferguson},
  {Fernandez-Galiana}, {Ferrante}, {Ferreira}, {Fidecaro}, {Figura}, {Fiori},
  {Fiorucci}, {Fishbach}, {Fisher}, {Fishner}, {Fittipaldi}, {Fitz-Axen},
  {Fiumara}, {Flaminio}, {Floden}, {Flynn}, {Fong}, {Font}, {Forsyth},
  {Fournier}, {Frasca}, {Frasconi}, {Frei}, {Freise}, {Frey}, {Frey},
  {Fritschel}, {Frolov}, {Fronz{\'e}}, {Fulda}, {Fyffe}, {Gabbard}, {Gadre},
  {Gaebel}, {Gair}, {Gais}, {Galaudage}, {Gamba}, {Ganapathy}, {Ganguly},
  {Gaonkar}, {Garaventa}, {Garc{\'\i}a-Quir{\'o}s}, {Garufi}, {Gateley},
  {Gaudio}, {Gayathri}, {Gemme}, {Gennai}, {George}, {George}, {George},
  {Gergely}, {Ghonge}, {Ghosh}, {Ghosh}, {Ghosh}, {Giacomazzo}, {Giacoppo},
  {Giaime}, {Giardina}, {Gibson}, {Gier}, {Gill}, {Giri}, {Glanzer}, {Gleckl},
  {Godwin}, {Goetz}, {Goetz}, {Gohlke}, {Goncharov}, {Gonz{\'a}lez},
  {Gopakumar}, {Gossan}, {Gosselin}, {Gouaty}, {Grace}, {Grado}, {Granata},
  {Granata}, {Grant}, {Gras}, {Grassia}, {Gray}, {Gray}, {Greco}, {Green},
  {Green}, {Gretarsson}, {Griggs}, {Grignani}, {Grimaldi}, {Grimes}, {Grimm},
  {Grote}, {Grunewald}, {Gruning}, {Guerrero}, {Guidi}, {Guimaraes},
  {Guix{\'e}}, {Gulati}, {Guo}, {Gupta}, {Gupta}, {Gupta}, {Gustafson},
  {Gustafson}, {Guzman}, {Haegel}, {Halim}, {Hall}, {Hamilton}, {Hammond},
  {Haney}, {Hanke}, {Hanks}, {Hanna}, {Hannam}, {Hannuksela}, {Hannuksela},
  {Hansen}, {Hansen}, {Hanson}, {Harder}, {Hardwick}, {Haris}, {Harms},
  {Harry}, {Harry}, {Hartwig}, {Hasskew}, {Haster}, {Haughian}, {Hayes},
  {Healy}, {Heidmann}, {Heintze}, {Heinze}, {Heinzel}, {Heitmann}, {Hellman},
  {Hello}, {Helmling-Cornell}, {Hemming}, {Hendry}, {Heng}, {Hennes}, {Hennig},
  {Hennig}, {Hernandez Vivanco}, {Heurs}, {Hild}, {Hill}, {Hines}, {Hochheim},
  {Hofgard}, {Hofman}, {Hohmann}, {Holgado}, {Holland}, {Hollows}, {Holmes},
  {Holt}, {Holz}, {Hopkins}, {Horst}, {Hough}, {Howell}, {Hoy}, {Hoyland},
  {Huang}, {H{\"u}bner}, {Huddart}, {Huerta}, {Hughey}, {Hui}, {Husa},
  {Huttner}, {Hutzler}, {Huxford}, {Huynh-Dinh}, {Idzkowski}, {Iess},
  {Imperato}, {Inchauspe}, {Ingram}, {Intini}, {Isi}, {Iyer},
  {JaberianHamedan}, {Jacqmin}, {Jadhav}, {Jadhav}, {James}, {Jani},
  {Janssens}, {Janthalur}, {Jaranowski}, {Jariwala}, {Jaume}, {Jenkins},
  {Jeunon}, {Jiang}, {Johns}, {Johnson-McDaniel}, {Jones}, {Jones}, {Jones},
  {Jones}, {Jones}, {Jonker}, {Ju}, {Junker}, {Kalaghatgi}, {Kalogera},
  {Kamai}, {Kandhasamy}, {Kang}, {Kanner}, {Kapadia}, {Kapasi}, {Karathanasis},
  {Karki}, {Kashyap}, {Kasprzack}, {Kastaun}, {Katsanevas}, {Katsavounidis},
  {Katzman}, {Kawabe}, {K{\'e}f{\'e}lian}, {Keitel}, {Key}, {Khadka},
  {Khalili}, {Khan}, {Khan}, {Khazanov}, {Khetan}, {Khursheed}, {Kijbunchoo},
  {Kim}, {Kim}, {Kim}, {Kim}, {Kim}, {Kim}, {Kimball}, {King}, {Kinley-Hanlon},
  {Kirchhoff}, {Kissel}, {Kleybolte}, {Klimenko}, {Knowles}, {Knyazev}, {Koch},
  {Koehlenbeck}, {Koekoek}, {Koley}, {Kolstein}, {Komori}, {Kondrashov},
  {Kontos}, {Koper}, {Korobko}, {Korth}, {Kovalam}, {Kozak}, {Kr{\"a}mer},
  {Kringel}, {Krishnendu}, {Kr{\'o}lak}, {Kuehn}, {Kumar}, {Kumar}, {Kumar},
  {Kumar}, {Kuns}, {Kwang}, {Lackey}, {Laghi}, {Lalande}, {Lam}, {Lamberts},
  {Landry}, {Lane}, {Lang}, {Lange}, {Lantz}, {Lanza}, {La Rosa},
  {Lartaux-Vollard}, {Lasky}, {Laxen}, {Lazzarini}, {Lazzaro}, {Leaci},
  {Leavey}, {Lecoeuche}, {Lee}, {Lee}, {Lee}, {Lee}, {Lehmann}, {Leon},
  {Leroy}, {Letendre}, {Levin}, {Li}, {Li}, {Li}, {Li}, {Li}, {Linde},
  {Linker}, {Linley}, {Littenberg}, {Liu}, {Liu}, {Llorens-Monteagudo}, {Lo},
  {Lockwood}, {London}, {Longo}, {Lorenzini}, {Loriette}, {Lormand}, {Losurdo},
  {Lough}, {Lousto}, {Lovelace}, {L{\"u}ck}, {Lumaca}, {Lundgren}, {Ma},
  {Macas}, {MacInnis}, {Macleod}, {MacMillan}, {Macquet}, {Maga{\~n}a
  Hernandez}, {Maga{\~n}a-Sandoval}, {Magazz{\`u}}, {Magee}, {Majorana},
  {Maksimovic}, {Maliakal}, {Malik}, {Man}, {Mandic}, {Mangano}, {Mansell},
  {Manske}, {Mantovani}, {Mapelli}, {Marchesoni}, {Marion}, {M{\'a}rka},
  {M{\'a}rka}, {Markakis}, {Markosyan}, {Markowitz}, {Maros}, {Marquina},
  {Marsat}, {Martelli}, {Martin}, {Martin}, {Martinez}, {Martinez}, {Martynov},
  {Masalehdan}, {Mason}, {Massera}, {Masserot}, {Massinger}, {Masso-Reid},
  {Mastrogiovanni}, {Matas}, {Mateu-Lucena}, {Matichard}, {Matiushechkina},
  {Mavalvala}, {Maynard}, {McCann}, {McCarthy}, {McClelland}, {McCormick},
  {McCuller}, {McGuire}, {McIsaac}, {McIver}, {McManus}, {McRae}, {McWilliams},
  {Meacher}, {Meadors}, {Mehmet}, {Mehta}, {Melatos}, {Melchor}, {Mendell},
  {Menendez-Vazquez}, {Mercer}, {Mereni}, {Merfeld}, {Merilh}, {Merritt},
  {Merzougui}, {Meshkov}, {Messenger}, {Messick}, {Metzdorff}, {Meyers},
  {Meylahn}, {Mhaske}, {Miani}, {Miao}, {Michaloliakos}, {Michel}, {Middleton},
  {Milano}, {Miller}, {Millhouse}, {Mills}, {Milotti}, {Milovich-Goff},
  {Minazzoli}, {Minenkov}, {Mir}, {Mishkin}, {Mishra}, {Mistry}, {Mitra},
  {Mitrofanov}, {Mitselmakher}, {Mittleman}, {Mo}, {Mogushi}, {Mohapatra},
  {Mohite}, {Molina}, {Molina-Ruiz}, {Mondin}, {Montani}, {Moore}, {Moraru},
  {Morawski}, {Moreno}, {Morisaki}, {Mours}, {Mow-Lowry}, {Mozzon},
  {Muciaccia}, {Mukherjee}, {Mukherjee}, {Mukherjee}, {Mukherjee}, {Mukund},
  {Mullavey}, {Munch}, {Mu{\~n}iz}, {Murray}, {Nadji}, {Nagar}, {Nardecchia},
  {Naticchioni}, {Nayak}, {Neil}, {Neilson}, {Nelemans}, {Nelson}, {Nery},
  {Neunzert}, {Nitz}, {Ng}, {Ng}, {Nguyen}, {Nguyen}, {Nguyen}, {Nichols},
  {Nissanke}, {Nocera}, {Noh}, {North}, {Nothard}, {Nuttall}, {Oberling},
  {O'Brien}, {O'Dell}, {Oganesyan}, {Ogin}, {Oh}, {Oh}, {Ohme}, {Ohta},
  {Okada}, {Olivetto}, {Oppermann}, {Oram}, {O'Reilly}, {Ormiston}, {Ortega},
  {O'Shaughnessy}, {Ossokine}, {Osthelder}, {Ottaway}, {Overmier}, {Owen},
  {Pace}, {Pagano}, {Page}, {Pagliaroli}, {Pai}, {Pai}, {Palamos}, {Palashov},
  {Palomba}, {Pan}, {Panda}, {Pang}, {Pankow}, {Pannarale}, {Pant}, {Paoletti},
  {Paoli}, {Paolone}, {Parker}, {Pascucci}, {Pasqualetti}, {Passaquieti},
  {Passuello}, {Patel}, {Patricelli}, {Payne}, {Pechsiri}, {Pedraza},
  {Pegoraro}, {Pele}, {Penn}, {Perego}, {Perez}, {P{\'e}rigois}, {Perreca},
  {Perri{\`e}s}, {Petermann}, {Petterson}, {Pfeiffer}, {Pham}, {Phukon},
  {Piccinni}, {Pichot}, {Piendibene}, {Piergiovanni}, {Pierini}, {Pierro},
  {Pillant}, {Pilo}, {Pinard}, {Pinto}, {Piotrzkowski}, {Pirello}, {Pitkin},
  {Placidi}, {Plastino}, {Pluchar}, {Poggiani}, {Polini}, {Pong}, {Ponrathnam},
  {Popolizio}, {Porter}, {Poverman}, {Powell}, {Pracchia}, {Prajapati},
  {Prasai}, {Prasanna}, {Pratten}, {Prestegard}, {Principe}, {Prodi},
  {Prokhorov}, {Prosposito}, {Prudenzi}, {Puecher}, {Punturo}, {Puosi},
  {Puppo}, {P{\"u}rrer}, {Qi}, {Quetschke}, {Quinonez}, {Quitzow-James},
  {Raab}, {Raaijmakers}, {Radkins}, {Radulesco}, {Raffai}, {Rafferty}, {Rail},
  {Raja}, {Rajan}, {Rajbhandari}, {Rakhmanov}, {Ramirez}, {Ramirez},
  {Ramos-Buades}, {Rana}, {Rao}, {Rapagnani}, {Rapol}, {Ratto}, {Raymond},
  {Razzano}, {Read}, {Regimbau}, {Rei}, {Reid}, {Reitze}, {Rettegno}, {Ricci},
  {Richardson}, {Richardson}, {Richardson}, {Ricker}, {Riemenschneider},
  {Riles}, {Rizzo}, {Robertson}, {Robinet}, {Rocchi}, {Rocha}, {Rodriguez},
  {Rodriguez-Soto}, {Rolland}, {Rollins}, {Roma}, {Romanelli}, {Romano},
  {Romel}, {Romero}, {Romero-Shaw}, {Romie}, {Ronchini}, {Rose}, {Rose},
  {Rose}, {Rosell}, {Rosi{\'n}ska}, {Rosofsky}, {Ross}, {Rowan}, {Rowlinson},
  {Roy}, {Roy}, {Ruggi}, {Ryan}, {Sachdev}, {Sadecki}, {Sadiq},
  {Sakellariadou}, {Salafia}, {Salconi}, {Saleem}, {Samajdar}, {Sanchez},
  {Sanchez}, {Sanchez}, {Sanchis-Gual}, {Sanders}, {Sandles}, {Santiago},
  {Santos}, {Saravanan}, {Sarin}, {Sassolas}, {Sathyaprakash}, {Sauter},
  {Savage}, {Savant}, {Sawant}, {Sayah}, {Schaetzl}, {Schale}, {Scheel},
  {Scheuer}, {Schindler-Tyka}, {Schmidt}, {Schnabel}, {Schofield},
  {Sch{\"o}nbeck}, {Schreiber}, {Schulte}, {Schutz}, {Schwarm}, {Schwartz},
  {Scott}, {Scott}, {Seglar-Arroyo}, {Seidel}, {Sellers}, {Sengupta},
  {Sennett}, {Sentenac}, {Sequino}, {Sergeev}, {Setyawati}, {Shaffer},
  {Shahriar}, {Sharifi}, {Sharma}, {Sharma}, {Shawhan}, {Shen}, {Shikauchi},
  {Shink}, {Shoemaker}, {Shoemaker}, {Shukla}, {ShyamSundar}, {Sieniawska},
  {Sigg}, {Singer}, {Singh}, {Singh}, {Singha}, {Singhal}, {Sintes}, {Sipala},
  {Skliris}, {Slagmolen}, {Slaven-Blair}, {Smetana}, {Smith}, {Smith},
  {Somala}, {Son}, {Soni}, {Soni}, {Sorazu}, {Sordini}, {Sorrentino},
  {Sorrentino}, {Soulard}, {Souradeep}, {Sowell}, {Spencer}, {Spera},
  {Srivastava}, {Srivastava}, {Staats}, {Stachie}, {Steer}, {Steinhoff},
  {Steinke}, {Steinlechner}, {Steinlechner}, {Steinmeyer}, {Stevenson},
  {Stolle-McAllister}, {Stops}, {Stover}, {Strain}, {Stratta}, {Strunk},
  {Sturani}, {Stuver}, {S{\"u}dbeck}, {Sudhagar}, {Sudhir}, {Suh},
  {Summerscales}, {Sun}, {Sun}, {Sunil}, {Sur}, {Suresh}, {Sutton}, {Swinkels},
  {Szczepa{\'n}czyk}, {Tacca}, {Tait}, {Talbot}, {Tanasijczuk}, {Tanner},
  {Tao}, {Tapia}, {Tapia San Martin}, {Tasson}, {Taylor}, {Tenorio},
  {Terkowski}, {Thirugnanasambandam}, {Thomas}, {Thomas}, {Thomas}, {Thompson},
  {Thondapu}, {Thorne}, {Thrane}, {Tiwari}, {Tiwari}, {Tiwari}, {Toland},
  {Tolley}, {Tonelli}, {Tornasi}, {Torres-Forn{\'e}}, {Torrie}, {Melo},
  {T{\"o}yr{\"a}}, {Tran}, {Trapananti}, {Travasso}, {Traylor}, {Tringali},
  {Tripathee}, {Trovato}, {Trudeau}, {Tsai}, {Tsang}, {Tse}, {Tso}, {Tsukada},
  {Tsuna}, {Tsutsui}, {Turconi}, {Ubhi}, {Udall}, {Ueno}, {Ugolini},
  {Unnikrishnan}, {Urban}, {Usman}, {Utina}, {Vahlbruch}, {Vajente}, {Vajpeyi},
  {Valdes}, {Valentini}, {Valsan}, {van Bakel}, {van Beuzekom}, {van den
  Brand}, {Van Den Broeck}, {Vander-Hyde}, {van der Schaaf}, {van Heijningen},
  {Vardaro}, {Vargas}, {Varma}, {Vass}, {Vas{\'u}th}, {Vecchio}, {Vedovato},
  {Veitch}, {Veitch}, {Venkateswara}, {Venneberg}, {Venugopalan}, {Verkindt},
  {Verma}, {Veske}, {Vetrano}, {Vicer{\'e}}, {Viets}, {Vijaykumar},
  {Villa-Ortega}, {Vinet}, {Vitale}, {Vo}, {Vocca}, {Vorvick}, {Vyatchanin},
  {Wade}, {Wade}, {Wade}, {Walet}, {Walker}, {Wallace}, {Wallace}, {Walsh},
  {Wang}, {Wang}, {Wang}, {Wang}, {Ward}, {Warner}, {Was}, {Washington},
  {Watchi}, {Weaver}, {Wei}, {Weinert}, {Weinstein}, {Weiss}, {Wellmann},
  {Wen}, {We{\ss}els}, {Westhouse}, {Wette}, {Whelan}, {White}, {White},
  {Whiting}, {Whittle}, {Wilken}, {Williams}, {Williams}, {Williamson},
  {Willis}, {Willke}, {Wilson}, {Wimmer}, {Winkler}, {Wipf}, {Woan}, {Woehler},
  {Wofford}, {Wong}, {Wrangel}, {Wright}, {Wu}, {Wysocki}, {Xiao}, {Yamamoto},
  {Yang}, {Yang}, {Yang}, {Yap}, {Yeeles}, {Yoon}, {Yu}, {Yu}, {Yuen},
  {Zadro{\.z}ny}, {Zanolin}, {Zelenova}, {Zendri}, {Zevin}, {Zhang}, {Zhang},
  {Zhang}, {Zhang}, {Zhao}, {Zhao}, {Zheng}, {Zhou}, {Zhou}, {Zhu},
  {Zimmerman}, {Zlochower}, {Zucker}, \& {Zweizig}}]{GWTC2}
{Abbott}, R., {Abbott}, T.~D., {Abraham}, S., {et~al.} 2020, arXiv e-prints,
  arXiv:2010.14527

\bibitem[{Acernese {et~al.}(2015)}]{aVirgo2015}
Acernese, F., {et~al.} 2015, Class. Quant. Grav., 32, 024001

\bibitem[{Anand {et~al.}(2020)Anand, Coughlin, Kasliwal, Bulla, Ahumada,
  Sagués~Carracedo, Almualla, Andreoni, Stein, Foucart, \&
  et~al.}]{AnandCoughlin_2020}
Anand, S., Coughlin, M.~W., Kasliwal, M.~M., {et~al.} 2020, Nature Astronomy,
  doi:10.1038/s41550-020-1183-3

\bibitem[{Andreoni {et~al.}(2020)Andreoni, Goldstein, {et~al.}}]{Andreoni2020}
Andreoni, I., Goldstein, D., {et~al.} 2020, Astrophys. J., 890, 131

\bibitem[{{Andreoni} {et~al.}(2017){Andreoni}, {Ackley}, {Cooke}, {Acharyya},
  {Allison}, {Anderson}, {Ashley}, {Baade}, {Bailes}, {Bannister}, {Beardsley},
  {Bessell}, {Bian}, {Bland}, {Boer}, {Booler}, {Brandeker}, {Brown},
  {Buckley}, {Chang}, {Coward}, {Crawford}, {Crisp}, {Crosse}, {Cucchiara},
  {Cup{\'a}k}, {de Gois}, {Deller}, {Devillepoix}, {Dobie}, {Elmer}, {Emrich},
  {Farah}, {Farrell}, {Franzen}, {Gaensler}, {Galloway}, {Gendre}, {Giblin},
  {Goobar}, {Green}, {Hancock}, {Hartig}, {Howell}, {Horsley}, {Hotan},
  {Howie}, {Hu}, {Hu}, {James}, {Johnston}, {Johnston-Hollitt}, {Kaplan},
  {Kasliwal}, {Keane}, {Kenney}, {Klotz}, {Lau}, {Laugier}, {Lenc}, {Li},
  {Liang}, {Lidman}, {Luvaul}, {Lynch}, {Ma}, {Macpherson}, {Mao},
  {McClelland}, {McCully}, {M{\"o}ller}, {Morales}, {Morris}, {Murphy},
  {Noysena}, {Onken}, {Orange}, {Os{\l}owski}, {Pallot}, {Paxman}, {Potter},
  {Pritchard}, {Raja}, {Ridden-Harper}, {Romero-Colmenero}, {Sadler}, {Sansom},
  {Scalzo}, {Schmidt}, {Scott}, {Seghouani}, {Shang}, {Shannon}, {Shao},
  {Shara}, {Sharp}, {Sokolowski}, {Sollerman}, {Staff}, {Steele}, {Sun},
  {Suntzeff}, {Tao}, {Tingay}, {Towner}, {Thierry}, {Trott}, {Tucker},
  {V{\"a}is{\"a}nen}, {Krishnan}, {Walker}, {Wang}, {Wang}, {Wayth}, {Whiting},
  {Williams}, {Williams}, {Wolf}, {Wu}, {Wu}, {Yang}, {Yuan}, {Zhang}, {Zhou},
  \& {Zovaro}}]{2017gfoAUS}
{Andreoni}, I., {Ackley}, K., {Cooke}, J., {et~al.} 2017, \pasa, 34, e069

\bibitem[{Andreoni {et~al.}(2021)Andreoni, Coughlin, Kool, Kasliwal, Kumar,
  Bhalerao, Carracedo, Ho, Pang, Saraogi, Sharma, Shenoy, Burns, Ahumada,
  Anand, Singer, Perley, De, Fremling, Bellm, Bulla, Crellin-Quick, Dietrich,
  Drake, Duev, Goobar, Graham, Kaplan, Kulkarni, Laher, Mahabal, Shupe,
  Sollerman, Walters, \& Yao}]{ZTFRest2021}
Andreoni, I., Coughlin, M.~W., Kool, E.~C., {et~al.} 2021, Fast-transient
  Searches in Real Time with ZTFReST: Identification of Three
  Optically-discovered Gamma-ray Burst Afterglows and New Constraints on the
  Kilonova Rate, , , arXiv:2104.06352

\bibitem[{{Andreoni} {et~al.}(2021){Andreoni}, {Margutti}, {Sharan Salafia},
  {Parazin}, {Villar}, {Coughlin}, {Yoachim}, {Mortensen}, {Brethauer},
  {Smartt}, {Kasliwal}, {Alexander}, {Anand}, {Berger}, {Grazia Bernardini},
  {Bianco}, {Blanchard}, {Bloom}, {Brocato}, {Cartier}, {Cenko}, {Chornock},
  {Copperwheat}, {Corsi}, {D'Ammando}, {D'Avanzo}, {Datrier}, {Foley},
  {Ghirlanda}, {Goobar}, {Grindlay}, {Hajela}, {Holz}, {Karambelkar}, {Kool},
  {Lamb}, {Laskar}, {Levan}, {Maguire}, {May}, {Melandri}, {Milisavljevic},
  {Miller}, {Nicholl}, {Palmese}, {Piranomonte}, {Rest}, {Sagues-Carracedo},
  {Siellez}, {Singer}, {Smith}, {Steeghs}, \& {Tanvir}}]{Andreoni2021_VRO}
{Andreoni}, I., {Margutti}, R., {Sharan Salafia}, O., {et~al.} 2021, arXiv
  e-prints, arXiv:2111.01945

\bibitem[{Antier {et~al.}(2020)Antier, Agayeva, AlMualla,
  {et~al.}}]{Grandma2020}
Antier, S., Agayeva, S., AlMualla, M., {et~al.} 2020, arXiv:2004.04277

\bibitem[{Arcavi(2018)}]{Arcavi2018}
Arcavi, I. 2018, The Astrophysical Journal, 855, L23

\bibitem[{Barbieri {et~al.}(2020)Barbieri, Salafia, Colpi, Ghirlanda, \&
  Perego}]{barbieri2020distinguishing}
Barbieri, C., Salafia, O.~S., Colpi, M., Ghirlanda, G., \& Perego, A. 2020,
  arXiv:2002.09395

\bibitem[{{Barnes} {et~al.}(2016){Barnes}, {Kasen}, {Wu}, \&
  {Mart{\'\i}nez-Pinedo}}]{Barnes2016}
{Barnes}, J., {Kasen}, D., {Wu}, M.-R., \& {Mart{\'\i}nez-Pinedo}, G. 2016,
  \apj, 829, 110

\bibitem[{{Barnes} {et~al.}(2020){Barnes}, {Zhu}, {Lund}, {Sprouse}, {Vassh},
  {McLaughlin}, {Mumpower}, \& {Surman}}]{Barnes2020}
{Barnes}, J., {Zhu}, Y.~L., {Lund}, K.~A., {et~al.} 2020, arXiv e-prints,
  arXiv:2010.11182

\bibitem[{Bauswein {et~al.}(2013)Bauswein, Baumgarte, \&
  Janka}]{BausBaumJan2013}
Bauswein, A., Baumgarte, T.~W., \& Janka, H.-T. 2013, Phys. Rev. Lett., 111,
  131101

\bibitem[{{Bellm} {et~al.}(2019){Bellm}, {Kulkarni}, {Graham}, {Dekany},
  {Smith}, {Riddle}, {Masci}, {Helou}, {Prince}, {Adams}, {Barbarino},
  {Barlow}, {Bauer}, {Beck}, {Belicki}, {Biswas}, {Blagorodnova}, {Bodewits},
  {Bolin}, {Brinnel}, {Brooke}, {Bue}, {Bulla}, {Burruss}, {Cenko}, {Chang},
  {Connolly}, {Coughlin}, {Cromer}, {Cunningham}, {De}, {Delacroix}, {Desai},
  {Duev}, {Eadie}, {Farnham}, {Feeney}, {Feindt}, {Flynn}, {Franckowiak},
  {Frederick}, {Fremling}, {Gal-Yam}, {Gezari}, {Giomi}, {Goldstein},
  {Golkhou}, {Goobar}, {Groom}, {Hacopians}, {Hale}, {Henning}, {Ho}, {Hover},
  {Howell}, {Hung}, {Huppenkothen}, {Imel}, {Ip}, {Ivezi{\'c}}, {Jackson},
  {Jones}, {Juric}, {Kasliwal}, {Kaspi}, {Kaye}, {Kelley}, {Kowalski},
  {Kramer}, {Kupfer}, {Landry}, {Laher}, {Lee}, {Lin}, {Lin}, {Lunnan},
  {Giomi}, {Mahabal}, {Mao}, {Miller}, {Monkewitz}, {Murphy}, {Ngeow},
  {Nordin}, {Nugent}, {Ofek}, {Patterson}, {Penprase}, {Porter}, {Rauch},
  {Rebbapragada}, {Reiley}, {Rigault}, {Rodriguez}, {van Roestel}, {Rusholme},
  {van Santen}, {Schulze}, {Shupe}, {Singer}, {Soumagnac}, {Stein}, {Surace},
  {Sollerman}, {Szkody}, {Taddia}, {Terek}, {Van Sistine}, {van Velzen},
  {Vestrand}, {Walters}, {Ward}, {Ye}, {Yu}, {Yan}, \&
  {Zolkower}}]{Bellm_2019ZTF}
{Bellm}, E.~C., {Kulkarni}, S.~R., {Graham}, M.~J., {et~al.} 2019, Publications
  of the Astronomical Society of the Pacific, 131, 018002

\bibitem[{{Blazek} {et~al.}(2019){Blazek}, {Christensen}, {Howell},
  {Vardosanidze}, {Boer}, {Eymar}, {Klotz}, {Laugier}, {Noysena}, {Antier},
  {Basa}, {Corre}, {Coughlin}, {Coward}, {Ducoin}, {Gendre}, {Hello},
  {Lachaud}, {Leroy}, {Turpin}, \& {Wang}}]{gcn24227}
{Blazek}, M., {Christensen}, N., {Howell}, E., {et~al.} 2019, GRB Coordinates
  Network, 24227, 1

\bibitem[{{Breschi} {et~al.}(2021){Breschi}, {Perego}, {Bernuzzi}, {Del Pozzo},
  {Nedora}, {Radice}, \& {Vescovi}}]{Breschi2021}
{Breschi}, M., {Perego}, A., {Bernuzzi}, S., {et~al.} 2021, arXiv e-prints,
  arXiv:2101.01201

\bibitem[{Bulla(2019)}]{Bulla_2019}
Bulla, M. 2019, Monthly Notices of the Royal Astronomical Society, 489,
  5037–5045

\bibitem[{{Chambers} {et~al.}(2016){Chambers}, {Magnier}, {Metcalfe},
  {Flewelling}, {Huber}, {Waters}, {Denneau}, {Draper}, {Farrow}, {Finkbeiner},
  {Holmberg}, {Koppenhoefer}, {Price}, {Rest}, {Saglia}, {Schlafly}, {Smartt},
  {Sweeney}, {Wainscoat}, {Burgett}, {Chastel}, {Grav}, {Heasley}, {Hodapp},
  {Jedicke}, {Kaiser}, {Kudritzki}, {Luppino}, {Lupton}, {Monet}, {Morgan},
  {Onaka}, {Shiao}, {Stubbs}, {Tonry}, {White}, {Ba{\~n}ados}, {Bell},
  {Bender}, {Bernard}, {Boegner}, {Boffi}, {Botticella}, {Calamida},
  {Casertano}, {Chen}, {Chen}, {Cole}, {Deacon}, {Frenk}, {Fitzsimmons},
  {Gezari}, {Gibbs}, {Goessl}, {Goggia}, {Gourgue}, {Goldman}, {Grant},
  {Grebel}, {Hambly}, {Hasinger}, {Heavens}, {Heckman}, {Henderson}, {Henning},
  {Holman}, {Hopp}, {Ip}, {Isani}, {Jackson}, {Keyes}, {Koekemoer}, {Kotak},
  {Le}, {Liska}, {Long}, {Lucey}, {Liu}, {Martin}, {Masci}, {McLean}, {Mindel},
  {Misra}, {Morganson}, {Murphy}, {Obaika}, {Narayan}, {Nieto-Santisteban},
  {Norberg}, {Peacock}, {Pier}, {Postman}, {Primak}, {Rae}, {Rai}, {Riess},
  {Riffeser}, {Rix}, {R{\"o}ser}, {Russel}, {Rutz}, {Schilbach}, {Schultz},
  {Scolnic}, {Strolger}, {Szalay}, {Seitz}, {Small}, {Smith}, {Soderblom},
  {Taylor}, {Thomson}, {Taylor}, {Thakar}, {Thiel}, {Thilker}, {Unger},
  {Urata}, {Valenti}, {Wagner}, {Walder}, {Walter}, {Watters}, {Werner},
  {Wood-Vasey}, \& {Wyse}}]{ChMa2016}
{Chambers}, K.~C., {Magnier}, E.~A., {Metcalfe}, N., {et~al.} 2016, arXiv
  e-prints, arXiv:1612.05560

\bibitem[{Chornock {et~al.}(2017)Chornock, Berger, Kasen, Cowperthwaite,
  Nicholl, Villar, Alexander, Blanchard, Eftekhari, Fong, Margutti, Williams,
  Annis, Brout, Brown, Chen, Drout, Farr, Foley, Frieman, Fryer, Herner, Holz,
  Kessler, Matheson, Metzger, Quataert, Rest, Sako, Scolnic, Smith, \&
  Soares-Santos}]{2017gfoGeminiS}
Chornock, R., Berger, E., Kasen, D., {et~al.} 2017, The Astrophysical Journal,
  848, L19

\bibitem[{{C{\^o}t{\'e}} {et~al.}(2018){C{\^o}t{\'e}}, {Fryer}, {Belczynski},
  {Korobkin}, {Chru{\'s}li{\'n}ska}, {Vassh}, {Mumpower}, {Lippuner},
  {Sprouse}, {Surman}, \& {Wollaeger}}]{Cote2018}
{C{\^o}t{\'e}}, B., {Fryer}, C.~L., {Belczynski}, K., {et~al.} 2018, \apj, 855,
  99

\bibitem[{Coughlin {et~al.}(2019{\natexlab{a}})Coughlin, Dietrich, Margalit, \&
  Metzger}]{Coughlin_2019b}
Coughlin, M.~W., Dietrich, T., Margalit, B., \& Metzger, B.~D.
  2019{\natexlab{a}}, Monthly Notices of the Royal Astronomical Society:
  Letters, 489, L91

\bibitem[{Coughlin {et~al.}(2018{\natexlab{a}})Coughlin, Dietrich, Doctor,
  Kasen, Coughlin, Jerkstrand, Leloudas, McBrien, Metzger, O’Shaughnessy, \&
  et~al.}]{Coughlin_2018}
Coughlin, M.~W., Dietrich, T., Doctor, Z., {et~al.} 2018{\natexlab{a}}, Monthly
  Notices of the Royal Astronomical Society, 480, 3871–3878

\bibitem[{Coughlin {et~al.}(2018{\natexlab{b}})Coughlin, Tao, Chan, Chatterjee,
  Christensen, Ghosh, Greco, Hu, Kapadia, Rana, \&
  et~al.}]{Coughlin2018gwemopt}
Coughlin, M.~W., Tao, D., Chan, M.~L., {et~al.} 2018{\natexlab{b}}, Monthly
  Notices of the Royal Astronomical Society, 478, 692–702

\bibitem[{Coughlin {et~al.}(2019{\natexlab{b}})Coughlin, Ahumada, Cenko,
  Cunningham, Ghosh, Singer, Bellm, Burns, De, Goldstein, Golkhou, Kaplan,
  Kasliwal, Perley, Sollerman, Bagdasaryan, Dekany, Duev, Feeney, Graham, Hale,
  Kulkarni, Kupfer, Laher, Mahabal, Masci, Miller, Neill, Patterson, Riddle,
  Rusholme, Smith, Tachibana, \& Walters}]{CoAh2019}
Coughlin, M.~W., Ahumada, T., Cenko, S.~B., {et~al.} 2019{\natexlab{b}},
  Publications of the Astronomical Society of the Pacific, 131, 048001

\bibitem[{Coughlin {et~al.}(2019{\natexlab{c}})Coughlin, Ahumada, Anand, De,
  Hankins, Kasliwal, Singer, Bellm, Andreoni, Cenko, Cooke, Copperwheat, Dugas,
  Jencson, Perley, Yu, Bhalerao, Kumar, Bloom, Anupama, Ashley, Bagdasaryan,
  Biswas, Buckley, Burdge, Cook, Cromer, Cunningham, D'A{\`{\i}}, Dekany,
  Delacroix, Dichiara, Duev, Dutta, Feeney, Frederick, Gatkine, Ghosh,
  Goldstein, Golkhou, Goobar, Graham, Hanayama, Horiuchi, Hung, Jha, Kong,
  Giomi, Kaplan, Karambelkar, Kowalski, Kulkarni, Kupfer, Masci, Mazzali,
  Moore, Mogotsi, Neill, Ngeow, Mart{\'{\i}}nez-Palomera, Parola, Pavana, Ofek,
  Patil, Riddle, Rigault, Rusholme, Serabyn, Shupe, Sharma, Singh, Sollerman,
  Soon, Staats, Taggart, Tan, Travouillon, Troja, Waratkar, \&
  Yatsu}]{ZTF_GW190425}
Coughlin, M.~W., Ahumada, T., Anand, S., {et~al.} 2019{\natexlab{c}}, The
  Astrophysical Journal, 885, L19

\bibitem[{Coughlin {et~al.}(2019{\natexlab{d}})Coughlin, Dietrich, Antier,
  Bulla, Foucart, Hotokezaka, Raaijmakers, Hinderer, \& Nissanke}]{CoDi2019}
Coughlin, M.~W., Dietrich, T., Antier, S., {et~al.} 2019{\natexlab{d}}, Mon.
  Not. Roy. Astron. Soc., 492, 863

\bibitem[{{Coulter} {et~al.}(2017){Coulter}, {Foley}, {Kilpatrick}, {Drout},
  {Piro}, {Shappee}, {Siebert}, {Simon}, {Ulloa}, {Kasen}, {Madore},
  {Murguia-Berthier}, {Pan}, {Prochaska}, {Ramirez-Ruiz}, {Rest}, \&
  {Rojas-Bravo}}]{2017gfoSwope}
{Coulter}, D.~A., {Foley}, R.~J., {Kilpatrick}, C.~D., {et~al.} 2017, Science,
  358, 1556

\bibitem[{{Cowperthwaite} {et~al.}(2017){Cowperthwaite}, {Berger}, {Villar},
  {Metzger}, {Nicholl}, {Chornock}, {Blanchard}, {Fong}, {Margutti},
  {Soares-Santos}, {Alexander}, {Allam}, {Annis}, {Brout}, {Brown}, {Butler},
  {Chen}, {Diehl}, {Doctor}, {Drout}, {Eftekhari}, {Farr}, {Finley}, {Foley},
  {Frieman}, {Fryer}, {Garc{\'\i}a-Bellido}, {Gill}, {Guillochon}, {Herner},
  {Holz}, {Kasen}, {Kessler}, {Marriner}, {Matheson}, {Neilsen}, {Quataert},
  {Palmese}, {Rest}, {Sako}, {Scolnic}, {Smith}, {Tucker}, {Williams},
  {Balbinot}, {Carlin}, {Cook}, {Durret}, {Li}, {Lopes}, {Louren{\c{c}}o},
  {Marshall}, {Medina}, {Muir}, {Mu{\~n}oz}, {Sauseda}, {Schlegel}, {Secco},
  {Vivas}, {Wester}, {Zenteno}, {Zhang}, {Abbott}, {Banerji}, {Bechtol},
  {Benoit-L{\'e}vy}, {Bertin}, {Buckley-Geer}, {Burke}, {Capozzi}, {Carnero
  Rosell}, {Carrasco Kind}, {Castander}, {Crocce}, {Cunha}, {D'Andrea}, {da
  Costa}, {Davis}, {DePoy}, {Desai}, {Dietrich}, {Drlica-Wagner}, {Eifler},
  {Evrard}, {Fernandez}, {Flaugher}, {Fosalba}, {Gaztanaga}, {Gerdes},
  {Giannantonio}, {Goldstein}, {Gruen}, {Gruendl}, {Gutierrez}, {Honscheid},
  {Jain}, {James}, {Jeltema}, {Johnson}, {Johnson}, {Kent}, {Krause}, {Kron},
  {Kuehn}, {Nuropatkin}, {Lahav}, {Lima}, {Lin}, {Maia}, {March}, {Martini},
  {McMahon}, {Menanteau}, {Miller}, {Miquel}, {Mohr}, {Neilsen}, {Nichol},
  {Ogando}, {Plazas}, {Roe}, {Romer}, {Roodman}, {Rykoff}, {Sanchez},
  {Scarpine}, {Schindler}, {Schubnell}, {Sevilla-Noarbe}, {Smith}, {Smith},
  {Sobreira}, {Suchyta}, {Swanson}, {Tarle}, {Thomas}, {Thomas}, {Troxel},
  {Vikram}, {Walker}, {Wechsler}, {Weller}, {Yanny}, \&
  {Zuntz}}]{2017gfoCowperthwaite}
{Cowperthwaite}, P.~S., {Berger}, E., {Villar}, V.~A., {et~al.} 2017, \apjl,
  848, L17

\bibitem[{{De} {et~al.}(2019){De}, {Adams}, {Coughlin}, {Kasliwal}, {Hankins},
  {Andreoni}, {Anand}, {Singer}, {Ahumada}, {Moore}, {Soon}, {Ashley}, \&
  {Travouillon}}]{gcn24187}
{De}, K., {Adams}, S.~M., {Coughlin}, M., {et~al.} 2019, GRB Coordinates
  Network, 24187, 1

\bibitem[{{Dekany} {et~al.}(2020){Dekany}, {Smith}, {Riddle}, {Feeney},
  {Porter}, {Hale}, {Zolkower}, {Belicki}, {Kaye}, {Henning}, {Walters},
  {Cromer}, {Delacroix}, {Rodriguez}, {Reiley}, {Mao}, {Hover}, {Murphy},
  {Burruss}, {Baker}, {Kowalski}, {Reif}, {Mueller}, {Bellm}, {Graham}, \&
  {Kulkarni}}]{Dekany_ZTF_2020}
{Dekany}, R., {Smith}, R.~M., {Riddle}, R., {et~al.} 2020, \pasp, 132, 038001

\bibitem[{{Dhawan} {et~al.}(2020){Dhawan}, {Bulla}, {Goobar}, {Sagu{\'e}s
  Carracedo}, \& {Setzer}}]{Dhawan2020}
{Dhawan}, S., {Bulla}, M., {Goobar}, A., {Sagu{\'e}s Carracedo}, A., \&
  {Setzer}, C.~N. 2020, \apj, 888, 67

\bibitem[{Dietrich {et~al.}(2020)Dietrich, Coughlin, Pang, Bulla, Heinzel,
  Issa, Tews, \& Antier}]{DiCo2020}
Dietrich, T., Coughlin, M.~W., Pang, P. T.~H., {et~al.} 2020, Science, 370,
  1450

\bibitem[{{Drout} {et~al.}(2017){Drout}, {Piro}, {Shappee}, {Kilpatrick},
  {Simon}, {Contreras}, {Coulter}, {Foley}, {Siebert}, {Morrell}, {Boutsia},
  {Di Mille}, {Holoien}, {Kasen}, {Kollmeier}, {Madore}, {Monson},
  {Murguia-Berthier}, {Pan}, {Prochaska}, {Ramirez-Ruiz}, {Rest}, {Adams},
  {Alatalo}, {Ba{\~n}ados}, {Baughman}, {Beers}, {Bernstein}, {Bitsakis},
  {Campillay}, {Hansen}, {Higgs}, {Ji}, {Maravelias}, {Marshall}, {Bidin},
  {Prieto}, {Rasmussen}, {Rojas-Bravo}, {Strom}, {Ulloa},
  {Vargas-Gonz{\'a}lez}, {Wan}, \& {Whitten}}]{2017gfoDrout}
{Drout}, M.~R., {Piro}, A.~L., {Shappee}, B.~J., {et~al.} 2017, Science, 358,
  1570

\bibitem[{{Dudi} {et~al.}(2021){Dudi}, {Adhikari}, {Br{\"u}gmann}, {Dietrich},
  {Hayashi}, {Kawaguchi}, {Kiuchi}, {Kyutoku}, {Shibata}, \&
  {Tichy}}]{Dudi2021}
{Dudi}, R., {Adhikari}, A., {Br{\"u}gmann}, B., {et~al.} 2021, arXiv e-prints,
  arXiv:2109.04063

\bibitem[{Etienne {et~al.}(2009)Etienne, Liu, Shapiro, \&
  Baumgarte}]{Etienne2009}
Etienne, Z.~B., Liu, Y.~T., Shapiro, S.~L., \& Baumgarte, T.~W. 2009, Phys.
  Rev. D, 79, 044024

\bibitem[{{Evans} {et~al.}(2017){Evans}, {Cenko}, {Kennea}, {Emery}, {Kuin},
  {Korobkin}, {Wollaeger}, {Fryer}, {Madsen}, {Harrison}, {Xu}, {Nakar},
  {Hotokezaka}, {Lien}, {Campana}, {Oates}, {Troja}, {Breeveld}, {Marshall},
  {Barthelmy}, {Beardmore}, {Burrows}, {Cusumano}, {Dai}, {D'Avanzo}, {D'Elia},
  {de Pasquale}, {Even}, {Fontes}, {Forster}, {Garcia}, {Giommi},
  {Grefenstette}, {Gronwall}, {Hartmann}, {Heida}, {Hungerford}, {Kasliwal},
  {Krimm}, {Levan}, {Malesani}, {Melandri}, {Miyasaka}, {Nousek}, {O'Brien},
  {Osborne}, {Pagani}, {Page}, {Palmer}, {Perri}, {Pike}, {Racusin}, {Rosswog},
  {Siegel}, {Sakamoto}, {Sbarufatti}, {Tagliaferri}, {Tanvir}, \&
  {Tohuvavohu}}]{2017gfoSwift}
{Evans}, P.~A., {Cenko}, S.~B., {Kennea}, J.~A., {et~al.} 2017, Science, 358,
  1565

\bibitem[{Farr {et~al.}(2015)Farr, Gair, Mandel, \& Cutler}]{FGMC2015}
Farr, W.~M., Gair, J.~R., Mandel, I., \& Cutler, C. 2015, Phys. Rev. D, 91,
  023005

\bibitem[{Feindt {et~al.}(2019)Feindt, Nordin, Rigault, Brinnel, Dhawan,
  Goobar, \& Kowalski}]{Feindt_2019}
Feindt, U., Nordin, J., Rigault, M., {et~al.} 2019, Journal of Cosmology and
  Astroparticle Physics, 2019, 005–005

\bibitem[{Foley {et~al.}(2020)Foley, Coulter, Kilpatrick, Piro, Ramirez-Ruiz,
  \& Schwab}]{Foley2020}
Foley, R.~J., Coulter, D.~A., Kilpatrick, C.~D., {et~al.} 2020, Monthly Notices
  of the Royal Astronomical Society, 494, 190

\bibitem[{Foucart {et~al.}(2018)Foucart, Hinderer, \& Nissanke}]{FoHiNi2018}
Foucart, F., Hinderer, T., \& Nissanke, S. 2018, Phys. Rev. D, 98, 081501

\bibitem[{{Foucart} {et~al.}(2021){Foucart}, {Moesta}, {Ramirez}, {Wright},
  {Darbha}, \& {Kasen}}]{Foucart2021}
{Foucart}, F., {Moesta}, P., {Ramirez}, T., {et~al.} 2021, arXiv e-prints,
  arXiv:2109.00565

\bibitem[{Gaebel {et~al.}(2019)Gaebel, Veitch, Dent, \& Farr}]{Gaebel2019}
Gaebel, S.~M., Veitch, J., Dent, T., \& Farr, W.~M. 2019, Monthly Notices of
  the Royal Astronomical Society, 484, 4008

\bibitem[{Ghosh {et~al.}(2017)Ghosh, Chatterjee, Kaplan, Brady, \&
  Sistine}]{Ghosh2017}
Ghosh, S., Chatterjee, D., Kaplan, D.~L., Brady, P.~R., \& Sistine, A.~V. 2017,
  Publications of the Astronomical Society of the Pacific, 129, 114503

\bibitem[{Gompertz {et~al.}(2020)}]{Goto2020}
Gompertz, B., {et~al.} 2020, arXiv:2004.00025

\bibitem[{{G{\'o}rski} {et~al.}(2005){G{\'o}rski}, {Hivon}, {Banday},
  {Wandelt}, {Hansen}, {Reinecke}, \& {Bartelmann}}]{Gorski2005}
{G{\'o}rski}, K.~M., {Hivon}, E., {Banday}, A.~J., {et~al.} 2005, \apj, 622,
  759

\bibitem[{Graham {et~al.}(2019)Graham, Kulkarni, Bellm, Adams, Barbarino,
  Blagorodnova, Bodewits, Bolin, Brady, Cenko, \& et~al.}]{Graham_2019}
Graham, M.~J., Kulkarni, S.~R., Bellm, E.~C., {et~al.} 2019, Publications of
  the Astronomical Society of the Pacific, 131, 078001

\bibitem[{{Green}(2018)}]{Dustmaps}
{Green}, G. 2018, The Journal of Open Source Software, 3, 695

\bibitem[{Harris {et~al.}(2020)Harris, Millman, van~der Walt, Gommers,
  Virtanen, Cournapeau, Wieser, Taylor, Berg, Smith, Kern, Picus, Hoyer, van
  Kerkwijk, Brett, Haldane, Fernández~del Río, Wiebe, Peterson,
  Gérard-Marchant, Sheppard, Reddy, Weckesser, Abbasi, Gohlke, \&
  Oliphant}]{numpy}
Harris, C.~R., Millman, K.~J., van~der Walt, S.~J., {et~al.} 2020, Nature, 585,
  357–362

\bibitem[{{Heinzel} {et~al.}(2021){Heinzel}, {Coughlin}, {Dietrich}, {Bulla},
  {Antier}, {Christensen}, {Coulter}, {Foley}, {Issa}, \&
  {Khetan}}]{Heinzel2021}
{Heinzel}, J., {Coughlin}, M.~W., {Dietrich}, T., {et~al.} 2021, \mnras,
  arXiv:2010.10746

\bibitem[{Hinderer {et~al.}(2019)Hinderer, Nissanke, Foucart, Hotokezaka,
  Vincent, Kasliwal, Schmidt, Williamson, Nichols, Duez, Kidder, Pfeiffer, \&
  Scheel}]{Hinderer2019}
Hinderer, T., Nissanke, S., Foucart, F., {et~al.} 2019, Phys. Rev. D, 100,
  063021

\bibitem[{{Hosseinzadeh} {et~al.}(2019){Hosseinzadeh}, {Cowperthwaite},
  {Gomez}, {Villar}, {Nicholl}, {Margutti}, {Berger}, {Chornock}, {Paterson},
  {Fong}, {Savchenko}, {Short}, {Alexander}, {Blanchard}, {Braga}, {Calkins},
  {Cartier}, {Coppejans}, {Eftekhari}, {Laskar}, {Ly}, {Patton}, {Pelisoli},
  {Reichart}, {Terreran}, \& {Williams}}]{Hosseinzadeh2019}
{Hosseinzadeh}, G., {Cowperthwaite}, P.~S., {Gomez}, S., {et~al.} 2019, \apjl,
  880, L4

\bibitem[{{Hotokezaka} {et~al.}(2018){Hotokezaka}, {Beniamini}, \&
  {Piran}}]{Hotokezaka2018}
{Hotokezaka}, K., {Beniamini}, P., \& {Piran}, T. 2018, International Journal
  of Modern Physics D, 27, 1842005

\bibitem[{Hotokezaka {et~al.}(2013)Hotokezaka, Kiuchi, Kyutoku, Okawa,
  Sekiguchi, Shibata, \& Taniguchi}]{Hotokezaka2013}
Hotokezaka, K., Kiuchi, K., Kyutoku, K., {et~al.} 2013, Phys. Rev. D, 87,
  024001

\bibitem[{{Hotokezaka} \& {Nakar}(2020)}]{Hotokezaka2019}
{Hotokezaka}, K., \& {Nakar}, E. 2020, \apj, 891, 152

\bibitem[{{Hotokezaka} {et~al.}(2019){Hotokezaka}, {Nakar}, {Gottlieb},
  {Nissanke}, {Masuda}, {Hallinan}, {Mooley}, \&
  {Deller}}]{Hotokezaka2018_HubbleConstant}
{Hotokezaka}, K., {Nakar}, E., {Gottlieb}, O., {et~al.} 2019, Nature Astronomy,
  3, 940

\bibitem[{Hunter(2007)}]{matplotlib}
Hunter, J.~D. 2007, Computing in Science Engineering, 9, 90

\bibitem[{{Just} {et~al.}(2021){Just}, {Kullmann}, {Goriely}, {Bauswein},
  {Janka}, \& {Collins}}]{Just2021}
{Just}, O., {Kullmann}, I., {Goriely}, S., {et~al.} 2021, arXiv e-prints,
  arXiv:2109.14617

\bibitem[{{Kasen} {et~al.}(2015){Kasen}, {Fern{\'a}ndez}, \&
  {Metzger}}]{Kasen2015}
{Kasen}, D., {Fern{\'a}ndez}, R., \& {Metzger}, B.~D. 2015, \mnras, 450, 1777

\bibitem[{{Kasen} {et~al.}(2017){Kasen}, {Metzger}, {Barnes}, {Quataert}, \&
  {Ramirez-Ruiz}}]{2017gfoKasen}
{Kasen}, D., {Metzger}, B., {Barnes}, J., {Quataert}, E., \& {Ramirez-Ruiz}, E.
  2017, \nat, 551, 80

\bibitem[{{Kasliwal}(2011)}]{KasliwalThesis2011}
{Kasliwal}, M.~M. 2011, PhD thesis, California Institute of Technology

\bibitem[{{Kasliwal} {et~al.}(2017){Kasliwal}, {Nakar}, {Singer}, {Kaplan},
  {Cook}, {Van Sistine}, {Lau}, {Fremling}, {Gottlieb}, {Jencson}, {Adams},
  {Feindt}, {Hotokezaka}, {Ghosh}, {Perley}, {Yu}, {Piran}, {Allison},
  {Anupama}, {Balasubramanian}, {Bannister}, {Bally}, {Barnes}, {Barway},
  {Bellm}, {Bhalerao}, {Bhattacharya}, {Blagorodnova}, {Bloom}, {Brady},
  {Cannella}, {Chatterjee}, {Cenko}, {Cobb}, {Copperwheat}, {Corsi}, {De},
  {Dobie}, {Emery}, {Evans}, {Fox}, {Frail}, {Frohmaier}, {Goobar}, {Hallinan},
  {Harrison}, {Helou}, {Hinderer}, {Ho}, {Horesh}, {Ip}, {Itoh}, {Kasen},
  {Kim}, {Kuin}, {Kupfer}, {Lynch}, {Madsen}, {Mazzali}, {Miller}, {Mooley},
  {Murphy}, {Ngeow}, {Nichols}, {Nissanke}, {Nugent}, {Ofek}, {Qi}, {Quimby},
  {Rosswog}, {Rusu}, {Sadler}, {Schmidt}, {Sollerman}, {Steele}, {Williamson},
  {Xu}, {Yan}, {Yatsu}, {Zhang}, \& {Zhao}}]{2017gfoKasliwal}
{Kasliwal}, M.~M., {Nakar}, E., {Singer}, L.~P., {et~al.} 2017, Science, 358,
  1559

\bibitem[{{Kasliwal} {et~al.}(2019{\natexlab{a}}){Kasliwal}, {Coughlin},
  {Bellm}, {Singer}, {de}, {Andreoni}, {Duev}, {Anand}, {Ahumada}, {Cenko},
  {Goldstein}, {Ho}, {Perley}, {Bhalerao}, {Kumar}, {Sharma}, {Goobar},
  {Kaplan}, {Sollerman}, {Bloom}, {Bulla}, {Kawai}, {Yatsu}, {Murata},
  {Hanayama}, {Horiuchi}, {Anupama}, {Rigault}, {Barbarino}, {Biswas}, {Cook},
  \& {Helou}}]{gcn24191}
{Kasliwal}, M.~M., {Coughlin}, M.~W., {Bellm}, E.~C., {et~al.}
  2019{\natexlab{a}}, GRB Coordinates Network, 24191, 1

\bibitem[{{Kasliwal} {et~al.}(2019{\natexlab{b}}){Kasliwal}, {Kasen}, {Lau},
  {Perley}, {Rosswog}, {Ofek}, {Hotokezaka}, {Chary}, {Sollerman}, {Goobar}, \&
  {Kaplan}}]{2017gfoSpitzer}
{Kasliwal}, M.~M., {Kasen}, D., {Lau}, R.~M., {et~al.} 2019{\natexlab{b}},
  \mnras, arXiv:1812.08708

\bibitem[{Kasliwal {et~al.}(2020)Kasliwal, Anand, Ahumada, Stein, Carracedo,
  Andreoni, Coughlin, Singer, Kool, De, Kumar, AlMualla, Yao, Bulla, Dobie,
  Reusch, Perley, Cenko, Bhalerao, Kaplan, Sollerman, Goobar, Copperwheat,
  Bellm, Anupama, Corsi, Nissanke, Agudo, Bagdasaryan, Barway, Belicki, Bloom,
  Bolin, Buckley, Burdge, Burruss, Caballero-Garcıa, Cannella, Castro-Tirado,
  Cook, Cooke, Cunningham, Dahiwale, Deshmukh, Dichiara, Duev, Dutta, Feeney,
  Franckowiak, Frederick, Fremling, Gal-Yam, Gatkine, Ghosh, Goldstein,
  Golkhou, Graham, Graham, Hankins, Helou, Hu, Ip, Jaodand, Karambelkar, Kong,
  Kowalski, Khandagale, Kulkarni, Kumar, Laher, Li, Mahabal, Masci, Miller,
  Mogotsi, Mohite, Mooley, Mroz, Newman, Ngeow, Oates, Patil, Pandey, Pavana,
  Pian, Riddle, Sanchez-Ramırez, Sharma, Singh, Smith, Soumagnac, Taggart,
  Tan, Tzanidakis, Troja, Valeev, Walters, Waratkar, Webb, Yu, Zhang, Zhou, \&
  Zolkower}]{kasliwal2020kilonova}
Kasliwal, M.~M., Anand, S., Ahumada, T., {et~al.} 2020, Kilonova Luminosity
  Function Constraints based on Zwicky Transient Facility Searches for 13
  Neutron Star Mergers, , , arXiv:2006.11306

\bibitem[{Kawaguchi {et~al.}(2016)Kawaguchi, Kyutoku, Shibata, \&
  Tanaka}]{Kawaguchi_2016}
Kawaguchi, K., Kyutoku, K., Shibata, M., \& Tanaka, M. 2016, The Astrophysical
  Journal, 825, 52

\bibitem[{{Kawaguchi} {et~al.}(2020){Kawaguchi}, {Shibata}, \&
  {Tanaka}}]{Kawaguchi2020}
{Kawaguchi}, K., {Shibata}, M., \& {Tanaka}, M. 2020, \apj, 889, 171

\bibitem[{{Kilpatrick} {et~al.}(2017){Kilpatrick}, {Foley}, {Kasen},
  {Murguia-Berthier}, {Ramirez-Ruiz}, {Coulter}, {Drout}, {Piro}, {Shappee},
  {Boutsia}, {Contreras}, {Di Mille}, {Madore}, {Morrell}, {Pan}, {Prochaska},
  {Rest}, {Rojas-Bravo}, {Siebert}, {Simon}, \& {Ulloa}}]{2017gfoKilpatrick}
{Kilpatrick}, C.~D., {Foley}, R.~J., {Kasen}, D., {et~al.} 2017, Science, 358,
  1583

\bibitem[{Kiuchi {et~al.}(2019)Kiuchi, Kyutoku, Shibata, \&
  Taniguchi}]{Kiuchi_2019}
Kiuchi, K., Kyutoku, K., Shibata, M., \& Taniguchi, K. 2019, The Astrophysical
  Journal, 876, L31

\bibitem[{Kluyver {et~al.}(2016)Kluyver, Ragan-Kelley, P{\'e}rez, Granger,
  Bussonnier, Frederic, Kelley, Hamrick, Grout, Corlay, Ivanov, Avila, Abdalla,
  Willing, \& development team}]{jupyter}
Kluyver, T., Ragan-Kelley, B., P{\'e}rez, F., {et~al.} 2016, in Positioning and
  Power in Academic Publishing: Players, Agents and Agendas, ed. F.~Loizides \&
  B.~Scmidt (Netherlands: IOS Press), 87--90

\bibitem[{{Korobkin} {et~al.}(2020){Korobkin}, {Wollaeger}, {Fryer},
  {Hungerford}, {Rosswog}, {Fontes}, {Mumpower}, {Chase}, {Even}, {Miller},
  {Misch}, \& {Lippuner}}]{Korobkin2020}
{Korobkin}, O., {Wollaeger}, R., {Fryer}, C., {et~al.} 2020, arXiv e-prints,
  arXiv:2004.00102

\bibitem[{{Kr{\"u}ger} \& {Foucart}(2020)}]{KrugerFoucart2020}
{Kr{\"u}ger}, C.~J., \& {Foucart}, F. 2020, \prd, 101, 103002

\bibitem[{{Kullmann} {et~al.}(2021){Kullmann}, {Goriely}, {Just},
  {Ardevol-Pulpillo}, {Bauswein}, \& {Janka}}]{Kullmann2021}
{Kullmann}, I., {Goriely}, S., {Just}, O., {et~al.} 2021, arXiv e-prints,
  arXiv:2109.02509

\bibitem[{{Kyutoku} {et~al.}(2020){Kyutoku}, {Fujibayashi}, {Hayashi},
  {Kawaguchi}, {Kiuchi}, {Shibata}, \& {Tanaka}}]{Kyutoku2020}
{Kyutoku}, K., {Fujibayashi}, S., {Hayashi}, K., {et~al.} 2020, \apjl, 890, L4

\bibitem[{Kyutoku {et~al.}(2015)Kyutoku, Ioka, Okawa, Shibata, \&
  Taniguchi}]{Kyutoko2015}
Kyutoku, K., Ioka, K., Okawa, H., Shibata, M., \& Taniguchi, K. 2015, Phys.
  Rev. D, 92, 044028

\bibitem[{Köppel {et~al.}(2019)Köppel, Bovard, \& Rezzolla}]{Koppel_2019}
Köppel, S., Bovard, L., \& Rezzolla, L. 2019, The Astrophysical Journal, 872,
  L16

\bibitem[{Levan(2020)}]{Engrave2020}
Levan, A. 2020, PoS, Asterics2019, 044

\bibitem[{{Li} {et~al.}(2019){Li}, {Xu}, {Zhou}, \& {Lu}}]{gcn24285}
{Li}, B., {Xu}, D., {Zhou}, X., \& {Lu}, H. 2019, GRB Coordinates Network,
  24285, 1

\bibitem[{Li \& Paczyński(1998)}]{Li_1998}
Li, L.-X., \& Paczyński, B. 1998, The Astrophysical Journal, 507, L59–L62

\bibitem[{{Lipunov} {et~al.}(2019){Lipunov}, {Gorbovskoy}, {Kornilov},
  {Kuvshinov}, {Tyurina}, {Balanutsa}, {Kuznetsov}, {Chazov}, {Vlasenko},
  {Tlatov}, {Senik}, {Parhomenko}, {Dormidontov}, {Ivanov}, {Gres}, {Budnev},
  {Yazev}, {Chuvalaev}, {Poleshchuk}, {Yurkov}, {Gabovich}, {Sergienko},
  {Podesta}, {Lopez}, {Podesta}, {Levato}, {Saffe}, {Rebolo}, {Serra},
  {Lodieu}, {Israelian}, {Suarez-Andres}, {Buckley}, {Potter}, {Kniazev}, \&
  {Kotze}}]{gcn24167}
{Lipunov}, V., {Gorbovskoy}, E., {Kornilov}, V., {et~al.} 2019, GRB Coordinates
  Network, 24167, 1

\bibitem[{{Lipunov} {et~al.}(2017){Lipunov}, {Gorbovskoy}, {Kornilov}, {.
  Tyurina}, {Balanutsa}, {Kuznetsov}, {Vlasenko}, {Kuvshinov}, {Gorbunov},
  {Buckley}, {Krylov}, {Podesta}, {Lopez}, {Podesta}, {Levato}, {Saffe},
  {Mallamachi}, {Potter}, {Budnev}, {Gress}, {Ishmuhametova}, {Vladimirov},
  {Zimnukhov}, {Yurkov}, {Sergienko}, {Gabovich}, {Rebolo}, {Serra-Ricart},
  {Israelyan}, {Chazov}, {Wang}, {Tlatov}, \& {Panchenko}}]{2017gfoMASTER}
{Lipunov}, V.~M., {Gorbovskoy}, E., {Kornilov}, V.~G., {et~al.} 2017, \apjl,
  850, L1

\bibitem[{{Lundquist} {et~al.}(2019{\natexlab{a}}){Lundquist}, {Paterson},
  {Sand}, {Valenti}, {Yang}, {Wyatt}, {Christensen}, {Gibbs}, {Shelly}, \&
  {Andrews}}]{gcn24172}
{Lundquist}, M.~J., {Paterson}, K., {Sand}, D.~J., {et~al.} 2019{\natexlab{a}},
  GRB Coordinates Network, 24172, 1

\bibitem[{{Lundquist} {et~al.}(2019{\natexlab{b}}){Lundquist}, {Paterson},
  {Fong}, {Sand}, {Andrews}, {Shivaei}, {Daly}, {Valenti}, {Yang},
  {Christensen}, {Gibbs}, {Shelly}, {Wyatt}, {Eskandari}, {Kuhn}, {Amaro},
  {Arcavi}, {Behroozi}, {Butler}, {Chomiuk}, {Corsi}, {Drout}, {Egami}, {Fan},
  {Foley}, {Frye}, {Gabor}, {Green}, {Grier}, {Guzman}, {Hamden}, {Howell},
  {Jannuzi}, {Kelly}, {Milne}, {Moe}, {Nugent}, {Olszewski}, {Palazzi},
  {Paschalidis}, {Psaltis}, {Reichart}, {Rest}, {Rossi}, {Schroeder}, {Smith},
  {Smith}, {Spekkens}, {Strader}, {Stark}, {Trilling}, {Veillet}, {Wagner},
  {Weiner}, {Wheeler}, {Williams}, \& {Zabludoff}}]{Lundquist2019}
{Lundquist}, M.~J., {Paterson}, K., {Fong}, W., {et~al.} 2019{\natexlab{b}},
  \apjl, 881, L26

\bibitem[{Mandel {et~al.}(2019)Mandel, Farr, \& Gair}]{Mandel2019}
Mandel, I., Farr, W.~M., \& Gair, J.~R. 2019, Monthly Notices of the Royal
  Astronomical Society, 486, 1086

\bibitem[{{Masci} {et~al.}(2019){Masci}, {Laher}, {Rusholme}, {Shupe}, {Groom},
  {Surace}, {Jackson}, {Monkewitz}, {Beck}, {Flynn}, {Terek}, {Landry},
  {Hacopians}, {Desai}, {Howell}, {Brooke}, {Imel}, {Wachter}, {Ye}, {Lin},
  {Cenko}, {Cunningham}, {Rebbapragada}, {Bue}, {Miller}, {Mahabal}, {Bellm},
  {Patterson}, {Juri{\'c}}, {Golkhou}, {Ofek}, {Walters}, {Graham}, {Kasliwal},
  {Dekany}, {Kupfer}, {Burdge}, {Cannella}, {Barlow}, {Van Sistine}, {Giomi},
  {Fremling}, {Blagorodnova}, {Levitan}, {Riddle}, {Smith}, {Helou}, {Prince},
  \& {Kulkarni}}]{Masci_2019ZTF}
{Masci}, F.~J., {Laher}, R.~R., {Rusholme}, B., {et~al.} 2019, Publications of
  the Astronomical Society of the Pacific, 131, 018003

\bibitem[{{McBrien} {et~al.}(2019){McBrien}, {Smartt}, {Smith}, {Young},
  {Denneau}, {Flewelling}, {Heinze}, {Tonry}, {Weiland}, {Gillanders},
  {Srivastav}, {O'Neil}, {Clark}, {Sim}, {Rest}, {Stalder}, {Stubbs},
  {Magnier}, {Schultz}, {Huber}, \& {Chambers}}]{gcn24197}
{McBrien}, O., {Smartt}, S., {Smith}, K.~W., {et~al.} 2019, GRB Coordinates
  Network, 24197, 1

\bibitem[{{McCully} {et~al.}(2017){McCully}, {Hiramatsu}, {Howell},
  {Hosseinzadeh}, {Arcavi}, {Kasen}, {Barnes}, {Shara}, {Williams},
  {V{\"a}is{\"a}nen}, {Potter}, {Romero-Colmenero}, {Crawford}, {Buckley},
  {Cooke}, {Andreoni}, {Pritchard}, {Mao}, {Gromadzki}, \&
  {Burke}}]{2017gfoMcCully}
{McCully}, C., {Hiramatsu}, D., {Howell}, D.~A., {et~al.} 2017, \apjl, 848, L32

\bibitem[{Metzger(2019)}]{Metzger_2019}
Metzger, B.~D. 2019, Living Reviews in Relativity, 23, arXiv:1910.01617

\bibitem[{Metzger {et~al.}(2010)Metzger, Martínez-Pinedo, Darbha, Quataert,
  Arcones, Kasen, Thomas, Nugent, Panov, \& Zinner}]{Metzger_2010}
Metzger, B.~D., Martínez-Pinedo, G., Darbha, S., {et~al.} 2010, Monthly
  Notices of the Royal Astronomical Society, 406, 2650

\bibitem[{{Mohite}(2021)}]{nimbus_zenodo}
{Mohite}, S.~R. 2021, {nimbus : A Bayesian inference framework to constrain
  kilonova models.}, v.v1.0.0,  Zenodo, doi:10.5281/zenodo.5648468

\bibitem[{{Morgan} {et~al.}(2020){Morgan}, {Soares-Santos}, {Annis}, {Herner},
  {Garcia}, {Palmese}, {Drlica-Wagner}, {Kessler}, {Garc{\'\i}a-Bellido},
  {Bachmann}, {Sherman}, {Allam}, {Bechtol}, {Bom}, {Brout}, {Butler},
  {Butner}, {Cartier}, {Chen}, {Conselice}, {Cook}, {Davis}, {Doctor}, {Farr},
  {Figueiredo}, {Finley}, {Foley}, {Galarza}, {Gill}, {Gruendl}, {Holz},
  {Kuropatkin}, {Lidman}, {Lin}, {Malik}, {Mann}, {Marriner}, {Marshall},
  {Mart{\'\i}nez-V{\'a}zquez}, {Meza}, {Neilsen}, {Nicolaou}, {Olivares E.},
  {Paz-Chinch{\'o}n}, {Points}, {Quirola-V{\'a}squez}, {Rodriguez}, {Sako},
  {Scolnic}, {Smith}, {Sobreira}, {Tucker}, {Vivas}, {Wiesner}, {Wood},
  {Yanny}, {Zenteno}, {Abbott}, {Aguena}, {Avila}, {Bertin}, {Bhargava},
  {Brooks}, {Burke}, {Rosell}, {Kind}, {Carretero}, {Costa}, {Costanzi}, {De
  Vicente}, {Desai}, {Diehl}, {Doel}, {Eifler}, {Everett}, {Flaugher},
  {Frieman}, {Gaztanaga}, {Gerdes}, {Gruen}, {Gschwend}, {Gutierrez},
  {Hartley}, {Hinton}, {Hollowood}, {Honscheid}, {James}, {Kuehn}, {Lahav},
  {Lima}, {Maia}, {March}, {Miquel}, {Ogando}, {Plazas}, {Roodman}, {Sanchez},
  {Scarpine}, {Schubnell}, {Serrano}, {Sevilla-Noarbe}, {Suchyta}, \&
  {Tarle}}]{MoSo2020}
{Morgan}, R., {Soares-Santos}, M., {Annis}, J., {et~al.} 2020, \apj, 901, 83

\bibitem[{Nakar(2019)}]{nakar2019electromagnetic}
Nakar, E. 2019, arXiv:1912.05659

\bibitem[{Nicholl {et~al.}(2021)Nicholl, Margalit, Schmidt, Smith, Ridley, \&
  Nuttall}]{Nicholl2021}
Nicholl, M., Margalit, B., Schmidt, P., {et~al.} 2021, Monthly Notices of the
  Royal Astronomical Society, 505, 3016

\bibitem[{{Nicholl} {et~al.}(2017){Nicholl}, {Berger}, {Kasen}, {Metzger},
  {Elias}, {Brice{\~n}o}, {Alexander}, {Blanchard}, {Chornock},
  {Cowperthwaite}, {Eftekhari}, {Fong}, {Margutti}, {Villar}, {Williams},
  {Brown}, {Annis}, {Bahramian}, {Brout}, {Brown}, {Chen}, {Clemens},
  {Dennihy}, {Dunlap}, {Holz}, {Marchesini}, {Massaro}, {Moskowitz},
  {Pelisoli}, {Rest}, {Ricci}, {Sako}, {Soares-Santos}, \&
  {Strader}}]{2017gfoNicholl}
{Nicholl}, M., {Berger}, E., {Kasen}, D., {et~al.} 2017, \apjl, 848, L18

\bibitem[{Pedregosa {et~al.}(2011)Pedregosa, Varoquaux, Gramfort, Michel,
  Thirion, Grisel, Blondel, Prettenhofer, Weiss, Dubourg, Vanderplas, Passos,
  Cournapeau, Brucher, Perrot, \& Duchesnay}]{scikit-learn}
Pedregosa, F., Varoquaux, G., Gramfort, A., {et~al.} 2011, Journal of Machine
  Learning Research, 12, 2825

\bibitem[{{Perego} {et~al.}(2017){Perego}, {Radice}, \&
  {Bernuzzi}}]{Perego2017}
{Perego}, A., {Radice}, D., \& {Bernuzzi}, S. 2017, \apjl, 850, L37

\bibitem[{P\'erez \& Granger(2007)}]{ipython}
P\'erez, F., \& Granger, B.~E. 2007, Computing in Science and Engineering, 9,
  21

\bibitem[{{Petrov} {et~al.}(2021){Petrov}, {Singer}, {Coughlin}, {Kumar},
  {Almualla}, {Anand}, {Bulla}, {Dietrich}, {Foucart}, \&
  {Guessoum}}]{PetrovSinger2021}
{Petrov}, P., {Singer}, L.~P., {Coughlin}, M.~W., {et~al.} 2021, arXiv
  e-prints, arXiv:2108.07277

\bibitem[{{Pian} {et~al.}(2017){Pian}, {D'Avanzo}, {Benetti}, {Branchesi},
  {Brocato}, {Campana}, {Cappellaro}, {Covino}, {D'Elia}, {Fynbo}, {Getman},
  {Ghirlanda}, {Ghisellini}, {Grado}, {Greco}, {Hjorth}, {Kouveliotou},
  {Levan}, {Limatola}, {Malesani}, {Mazzali}, {Melandri}, {M{\o}ller},
  {Nicastro}, {Palazzi}, {Piranomonte}, {Rossi}, {Salafia}, {Selsing},
  {Stratta}, {Tanaka}, {Tanvir}, {Tomasella}, {Watson}, {Yang}, {Amati},
  {Antonelli}, {Ascenzi}, {Bernardini}, {Bo{\"e}r}, {Bufano}, {Bulgarelli},
  {Capaccioli}, {Casella}, {Castro-Tirado}, {Chassande-Mottin}, {Ciolfi},
  {Copperwheat}, {Dadina}, {De Cesare}, {di Paola}, {Fan}, {Gendre},
  {Giuffrida}, {Giunta}, {Hunt}, {Israel}, {Jin}, {Kasliwal}, {Klose}, {Lisi},
  {Longo}, {Maiorano}, {Mapelli}, {Masetti}, {Nava}, {Patricelli}, {Perley},
  {Pescalli}, {Piran}, {Possenti}, {Pulone}, {Razzano}, {Salvaterra},
  {Schipani}, {Spera}, {Stamerra}, {Stella}, {Tagliaferri}, {Testa}, {Troja},
  {Turatto}, {Vergani}, \& {Vergani}}]{2017gfoPian}
{Pian}, E., {D'Avanzo}, P., {Benetti}, S., {et~al.} 2017, \nat, 551, 67

\bibitem[{Raaijmakers {et~al.}(2021)Raaijmakers, Nissanke, Foucart, Kasliwal,
  Bulla, Fernandez, Henkel, Hinderer, Hotokezaka, Lukošiūtė, Venumadhav,
  Antier, Coughlin, Dietrich, \& Edwards}]{raaijmakers2021challenges}
Raaijmakers, G., Nissanke, S., Foucart, F., {et~al.} 2021, arXiv:2102.11569

\bibitem[{{Radice} \& {Dai}(2019)}]{RadiceDai2019}
{Radice}, D., \& {Dai}, L. 2019, European Physical Journal A, 55, 50

\bibitem[{{Radice} {et~al.}(2018){Radice}, {Perego}, {Hotokezaka}, {Fromm},
  {Bernuzzi}, \& {Roberts}}]{Radice2018}
{Radice}, D., {Perego}, A., {Hotokezaka}, K., {et~al.} 2018, \apj, 869, 130

\bibitem[{{Roberts} {et~al.}(2011){Roberts}, {Kasen}, {Lee}, \&
  {Ramirez-Ruiz}}]{Roberts2011}
{Roberts}, L.~F., {Kasen}, D., {Lee}, W.~H., \& {Ramirez-Ruiz}, E. 2011, \apjl,
  736, L21

\bibitem[{Rosswog(2005)}]{Rosswog2005}
Rosswog, S. 2005, The Astrophysical Journal, 634, 1202

\bibitem[{{Rosswog} {et~al.}(2017){Rosswog}, {Feindt}, {Korobkin}, {Wu},
  {Sollerman}, {Goobar}, \& {Martinez-Pinedo}}]{Rosswog2017}
{Rosswog}, S., {Feindt}, U., {Korobkin}, O., {et~al.} 2017, Classical and
  Quantum Gravity, 34, 104001

\bibitem[{{Sagu{\'e}s Carracedo} {et~al.}(2021){Sagu{\'e}s Carracedo}, {Bulla},
  {Feindt}, \& {Goobar}}]{Carracedo2021}
{Sagu{\'e}s Carracedo}, A., {Bulla}, M., {Feindt}, U., \& {Goobar}, A. 2021,
  \mnras, 504, 1294

\bibitem[{{Sasada} {et~al.}(2021){Sasada}, {Utsumi}, {Itoh}, {Tominaga},
  {Tanaka}, {Morokuma}, {Yanagisawa}, {Kawabata}, {Ohgami}, {Yoshida}, {Abe},
  {Adachi}, {Akitaya}, {Chong}, {Daikuhara}, {Hamasaki}, {Honda}, {Hosokawa},
  {Iida}, {Imazato}, {Ishioka}, {Iwasaki}, {Jian}, {Kamei}, {Kanai}, {Kaneda},
  {Kaneko}, {Katoh}, {Kawai}, {Kubota}, {Kubota}, {Mamiya}, {Matsubayashi},
  {Morihana}, {Murata}, {Nagayama}, {Nakamura}, {Nakaoka}, {Niino},
  {Nishinaka}, {Niwano}, {Nogami}, {Oasa}, {Oeda}, {Ogawa}, {Ohsawa}, {Ohta},
  {Oide}, {Onozato}, {Sako}, {Saito}, {Sekiguchi}, {Shigeyama}, {Shigeyoshi},
  {Shikauchi}, {Shiraishi}, {Suzuki}, {Takagi}, {Takahashi}, {Takarada},
  {Takayama}, {Takeuchi}, {Tamura}, {Tanaka}, {Toma}, {Tozuka}, {Uchida},
  {Uzawa}, {Yamanaka}, {Yasuda}, \& {Yatsu}}]{Sasada2021_JGEM}
{Sasada}, M., {Utsumi}, Y., {Itoh}, R., {et~al.} 2021, Progress of Theoretical
  and Experimental Physics, 2021, 05A104

\bibitem[{Schlafly \& Finkbeiner(2011)}]{SFD2011}
Schlafly, E.~F., \& Finkbeiner, D.~P. 2011, The Astrophysical Journal, 737, 103

\bibitem[{{Shappee} {et~al.}(2017){Shappee}, {Simon}, {Drout}, {Piro},
  {Morrell}, {Prieto}, {Kasen}, {Holoien}, {Kollmeier}, {Kelson}, {Coulter},
  {Foley}, {Kilpatrick}, {Siebert}, {Madore}, {Murguia-Berthier}, {Pan},
  {Prochaska}, {Ramirez-Ruiz}, {Rest}, {Adams}, {Alatalo}, {Ba{\~n}ados},
  {Baughman}, {Bernstein}, {Bitsakis}, {Boutsia}, {Bravo}, {Di Mille}, {Higgs},
  {Ji}, {Maravelias}, {Marshall}, {Placco}, {Prieto}, \&
  {Wan}}]{2017gfoShappee}
{Shappee}, B.~J., {Simon}, J.~D., {Drout}, M.~R., {et~al.} 2017, Science, 358,
  1574

\bibitem[{{Siegel}(2019)}]{Siegel2019}
{Siegel}, D.~M. 2019, European Physical Journal A, 55, 203

\bibitem[{Singer \& Price(2016)}]{SiPr2016bayestar}
Singer, L.~P., \& Price, L.~R. 2016, Phys. Rev. D, 93, 024013

\bibitem[{Singer {et~al.}(2016{\natexlab{a}})Singer, Chen, Holz, Farr, Price,
  Raymond, Cenko, Gehrels, Cannizzo, Kasliwal, Nissanke, Coughlin, Farr, Urban,
  Vitale, Veitch, Graff, Berry, Mohapatra, \& Mandel}]{Singer2016}
Singer, L.~P., Chen, H.-Y., Holz, D.~E., {et~al.} 2016{\natexlab{a}}, The
  Astrophysical Journal, 829, L15

\bibitem[{Singer {et~al.}(2016{\natexlab{b}})Singer, Chen, Holz, Farr, Price,
  Raymond, Cenko, Gehrels, Cannizzo, Kasliwal, Nissanke, Coughlin, Farr, Urban,
  Vitale, Veitch, Graff, Berry, Mohapatra, \& Mandel}]{Singer2016Supp}
---. 2016{\natexlab{b}}, The Astrophysical Journal Supplement Series, 226, 10

\bibitem[{{Smartt} {et~al.}(2017){Smartt}, {Chen}, {Jerkstrand}, {Coughlin},
  {Kankare}, {Sim}, {Fraser}, {Inserra}, {Maguire}, {Chambers}, {Huber},
  {Kr{\"u}hler}, {Leloudas}, {Magee}, {Shingles}, {Smith}, {Young}, {Tonry},
  {Kotak}, {Gal-Yam}, {Lyman}, {Homan}, {Agliozzo}, {Anderson}, {Angus},
  {Ashall}, {Barbarino}, {Bauer}, {Berton}, {Botticella}, {Bulla}, {Bulger},
  {Cannizzaro}, {Cano}, {Cartier}, {Cikota}, {Clark}, {De Cia}, {Della Valle},
  {Denneau}, {Dennefeld}, {Dessart}, {Dimitriadis}, {Elias-Rosa}, {Firth},
  {Flewelling}, {Fl{\"o}rs}, {Franckowiak}, {Frohmaier}, {Galbany},
  {Gonz{\'a}lez-Gait{\'a}n}, {Greiner}, {Gromadzki}, {Guelbenzu},
  {Guti{\'e}rrez}, {Hamanowicz}, {Hanlon}, {Harmanen}, {Heintz}, {Heinze},
  {Hernandez}, {Hodgkin}, {Hook}, {Izzo}, {James}, {Jonker}, {Kerzendorf},
  {Klose}, {Kostrzewa-Rutkowska}, {Kowalski}, {Kromer}, {Kuncarayakti},
  {Lawrence}, {Lowe}, {Magnier}, {Manulis}, {Martin-Carrillo}, {Mattila},
  {McBrien}, {M{\"u}ller}, {Nordin}, {O'Neill}, {Onori}, {Palmerio},
  {Pastorello}, {Patat}, {Pignata}, {Podsiadlowski}, {Pumo}, {Prentice}, {Rau},
  {Razza}, {Rest}, {Reynolds}, {Roy}, {Ruiter}, {Rybicki}, {Salmon}, {Schady},
  {Schultz}, {Schweyer}, {Seitenzahl}, {Smith}, {Sollerman}, {Stalder},
  {Stubbs}, {Sullivan}, {Szegedi}, {Taddia}, {Taubenberger}, {Terreran}, {van
  Soelen}, {Vos}, {Wainscoat}, {Walton}, {Waters}, {Weiland}, {Willman},
  {Wiseman}, {Wright}, {Wyrzykowski}, \& {Yaron}}]{2017gfoSmartt}
{Smartt}, S.~J., {Chen}, T.~W., {Jerkstrand}, A., {et~al.} 2017, \nat, 551, 75

\bibitem[{{Smith} {et~al.}(2019){Smith}, {Young}, {McBrien},
  {et~al.}}]{gcn24210}
{Smith}, K., {Young}, D., {McBrien}, O., {et~al.} 2019, GRB Coordinates
  Network, 24210, 1

\bibitem[{Soares-Santos {et~al.}(2017)Soares-Santos, Holz, Annis, Chornock, \&
  Herner}]{2017gfoDECam}
Soares-Santos, M., Holz, D., Annis, J., Chornock, R., \& Herner, K. 2017,
  Astrophys. J. Lett., 848, L16

\bibitem[{{Steeghs} {et~al.}(2019){Steeghs}, {Dyer}, {Galloway}, {Dhillon},
  {O'Brien}, {Ramsay}, {Pollacco}, {Thrane}, {Poshyachinda}, {Palle},
  {Ulaczyk}, {Cutter}, {Stanway}, {Ackley}, {Obradovic}, {Mong}, {Casey},
  {Brown}, {Rol}, {Mullaney}, {Littlefair}, {Makrygianni}, {Daw}, {Maund},
  {Starling}, {Eyles}, {Sawangwit}, {Mkrtichian}, {Awiphan},
  {Aukkaravittayapun}, {Irawati}, {Kennedy}, {Breton}, {Mata-Sanchez},
  {Heikkila}, \& {Kotak}}]{gcn24224}
{Steeghs}, D., {Dyer}, M., {Galloway}, D., {et~al.} 2019, GRB Coordinates
  Network, 24224, 1

\bibitem[{{Tanaka}(2016)}]{Tanaka2016}
{Tanaka}, M. 2016, Advances in Astronomy, 2016, 634197

\bibitem[{Tanaka \& Hotokezaka(2013)}]{TaHo2013}
Tanaka, M., \& Hotokezaka, K. 2013, The Astrophysical Journal, 775, 113

\bibitem[{{Tanaka} {et~al.}(2017){Tanaka}, {Utsumi}, {Mazzali}, {Tominaga},
  {Yoshida}, {Sekiguchi}, {Morokuma}, {Motohara}, {Ohta}, {Kawabata}, {Abe},
  {Aoki}, {Asakura}, {Baar}, {Barway}, {Bond}, {Doi}, {Fujiyoshi}, {Furusawa},
  {Honda}, {Itoh}, {Kawabata}, {Kawai}, {Kim}, {Lee}, {Miyazaki}, {Morihana},
  {Nagashima}, {Nagayama}, {Nakaoka}, {Nakata}, {Ohsawa}, {Ohshima}, {Okita},
  {Saito}, {Sumi}, {Tajitsu}, {Takahashi}, {Takayama}, {Tamura}, {Tanaka},
  {Terai}, {Tristram}, {Yasuda}, \& {Zenko}}]{Tanaka2017}
{Tanaka}, M., {Utsumi}, Y., {Mazzali}, P.~A., {et~al.} 2017, \pasj, 69, 102

\bibitem[{Tanaka {et~al.}(2018)Tanaka, Kato, Gaigalas, Rynkun,
  Rad{\v{z}}i{\={u}}t{\.{e}}, Wanajo, Sekiguchi, Nakamura, Tanuma, Murakami, \&
  Sakaue}]{Tanaka2018}
Tanaka, M., Kato, D., Gaigalas, G., {et~al.} 2018, The Astrophysical Journal,
  852, 109

\bibitem[{{Tanvir} {et~al.}(2017){Tanvir}, {Levan},
  {Gonz{\'a}lez-Fern{\'a}ndez}, {Korobkin}, {Mandel}, {Rosswog}, {Hjorth},
  {D'Avanzo}, {Fruchter}, {Fryer}, {Kangas}, {Milvang-Jensen}, {Rosetti},
  {Steeghs}, {Wollaeger}, {Cano}, {Copperwheat}, {Covino}, {D'Elia}, {de Ugarte
  Postigo}, {Evans}, {Even}, {Fairhurst}, {Figuera Jaimes}, {Fontes}, {Fujii},
  {Fynbo}, {Gompertz}, {Greiner}, {Hodosan}, {Irwin}, {Jakobsson},
  {J{\o}rgensen}, {Kann}, {Lyman}, {Malesani}, {McMahon}, {Melandri},
  {O'Brien}, {Osborne}, {Palazzi}, {Perley}, {Pian}, {Piranomonte}, {Rabus},
  {Rol}, {Rowlinson}, {Schulze}, {Sutton}, {Th{\"o}ne}, {Ulaczyk}, {Watson},
  {Wiersema}, \& {Wijers}}]{2017gfolanthanides}
{Tanvir}, N.~R., {Levan}, A.~J., {Gonz{\'a}lez-Fern{\'a}ndez}, C., {et~al.}
  2017, \apjl, 848, L27

\bibitem[{Tonry {et~al.}(2018)Tonry, Denneau, Heinze, Stalder, Smith, Smartt,
  Stubbs, Weiland, \& Rest}]{ToDe2018}
Tonry, J.~L., Denneau, L., Heinze, A.~N., {et~al.} 2018, Publications of the
  Astronomical Society of the Pacific, 130, 064505

\bibitem[{{Utsumi} {et~al.}(2017){Utsumi}, {Tanaka}, {Tominaga}, {Yoshida},
  {Barway}, {Nagayama}, {Zenko}, {Aoki}, {Fujiyoshi}, {Furusawa}, {Kawabata},
  {Koshida}, {Lee}, {Morokuma}, {Motohara}, {Nakata}, {Ohsawa}, {Ohta},
  {Okita}, {Tajitsu}, {Tanaka}, {Terai}, {Yasuda}, {Abe}, {Asakura}, {Bond},
  {Miyazaki}, {Sumi}, {Tristram}, {Honda}, {Itoh}, {Itoh}, {Kawabata},
  {Morihana}, {Nagashima}, {Nakaoka}, {Ohshima}, {Takahashi}, {Takayama},
  {Aoki}, {Baar}, {Doi}, {Finet}, {Kanda}, {Kawai}, {Kim}, {Kuroda}, {Liu},
  {Matsubayashi}, {Murata}, {Nagai}, {Saito}, {Saito}, {Sako}, {Sekiguchi},
  {Tamura}, {Tanaka}, {Uemura}, \& {Yamaguchi}}]{2017gfoJGEM}
{Utsumi}, Y., {Tanaka}, M., {Tominaga}, N., {et~al.} 2017, \pasj, 69, 101

\bibitem[{Van~Rossum \& Drake(2009)}]{python3}
Van~Rossum, G., \& Drake, F.~L. 2009, Python 3 Reference Manual (Scotts Valley,
  CA: CreateSpace)

\bibitem[{{Veitch} {et~al.}(2015){Veitch}, {Raymond}, {Farr}, {Farr}, {Graff},
  {Vitale}, {Aylott}, {Blackburn}, {Christensen}, {Coughlin}, {Del Pozzo},
  {Feroz}, {Gair}, {Haster}, {Kalogera}, {Littenberg}, {Mandel},
  {O'Shaughnessy}, {Pitkin}, {Rodriguez}, {R{\"o}ver}, {Sidery}, {Smith}, {Van
  Der Sluys}, {Vecchio}, {Vousden}, \& {Wade}}]{Veitch2015}
{Veitch}, J., {Raymond}, V., {Farr}, B., {et~al.} 2015, \prd, 91, 042003

\bibitem[{Virtanen {et~al.}(2020)Virtanen, Gommers, Oliphant, Haberland, Reddy,
  Cournapeau, Burovski, Peterson, Weckesser, Bright, {van der Walt}, Brett,
  Wilson, Millman, Mayorov, Nelson, Jones, Kern, Larson, Carey, Polat, Feng,
  Moore, {VanderPlas}, Laxalde, Perktold, Cimrman, Henriksen, Quintero, Harris,
  Archibald, Ribeiro, Pedregosa, {van Mulbregt}, \& {SciPy 1.0
  Contributors}}]{scipy}
Virtanen, P., Gommers, R., Oliphant, T.~E., {et~al.} 2020, Nature Methods, 17,
  261

\bibitem[{{Waxman} {et~al.}(2018){Waxman}, {Ofek}, {Kushnir}, \&
  {Gal-Yam}}]{Waxman2017}
{Waxman}, E., {Ofek}, E.~O., {Kushnir}, D., \& {Gal-Yam}, A. 2018, \mnras, 481,
  3423

\bibitem[{{Wollaeger} {et~al.}(2018){Wollaeger}, {Korobkin}, {Fontes},
  {Rosswog}, {Even}, {Fryer}, {Sollerman}, {Hungerford}, {van Rossum}, \&
  {Wollaber}}]{Wollaeger2018}
{Wollaeger}, R.~T., {Korobkin}, O., {Fontes}, C.~J., {et~al.} 2018, \mnras,
  478, 3298

\bibitem[{Wu {et~al.}(2019)Wu, Barnes, Mart\'{\i}nez-Pinedo, \&
  Metzger}]{Wu2019}
Wu, M.-R., Barnes, J., Mart\'{\i}nez-Pinedo, G., \& Metzger, B.~D. 2019, Phys.
  Rev. Lett., 122, 062701

\bibitem[{{Xu} {et~al.}(2019){Xu}, {Zhu}, {Yu}, {Zhang}, {Zhou}, {Teng}, {Liu},
  {Guan}, {Yang}, {Zhao}, {Li}, {Liu}, {Niu}, {Liu}, {Zhang}, {Mao}, {Bai}, \&
  {Gao}}]{gcn24190}
{Xu}, D., {Zhu}, Z.~P., {Yu}, B.~Y., {et~al.} 2019, GRB Coordinates Network,
  24190, 1

\bibitem[{{Zhu} {et~al.}(2021){Zhu}, {Yang}, {Zhang}, {Gao}, \& {Yu}}]{Zhu2021}
{Zhu}, J.-P., {Yang}, Y.-P., {Zhang}, B., {Gao}, H., \& {Yu}, Y.-W. 2021, arXiv
  e-prints, arXiv:2110.10468

\bibitem[{{Zhu} {et~al.}(2020){Zhu}, {Wu}, {Yang}, {Zhang}, {Gao}, {Yu}, {Li},
  {Cao}, {Liu}, {Huang}, \& {Zhang}}]{Zhu2020NSBH}
{Zhu}, J.-P., {Wu}, S., {Yang}, Y.-P., {et~al.} 2020, arXiv e-prints,
  arXiv:2011.02717

\bibitem[{Zhu {et~al.}(2018)Zhu, Wollaeger, Vassh, Surman, Sprouse, Mumpower,
  Möller, McLaughlin, Korobkin, Kawano, Jaffke, Holmbeck, Fryer, Even,
  Couture, \& Barnes}]{Zhu2018}
Zhu, Y., Wollaeger, R.~T., Vassh, N., {et~al.} 2018, The Astrophysical Journal,
  863, L23

\end{thebibliography}
\end{document}